\documentclass[]{aa}
\usepackage{graphicx}
\usepackage{txfonts}
\usepackage{rotating}

\newcommand{\mini}{\mbox{$M_{\rm i}$}}

\newcommand{\diff}{\mbox{${\rm d}$}}

\newcommand{\jk}{\mbox{$J\!-\!K$}}
\newcommand{\jks}{\mbox{$J\!-\!K_{\rm s}$}}
\newcommand{\ks}{\mbox{$K_{\rm s}$}}

\newcommand{\mh}{\mbox{\rm [{\rm M}/{\rm H}]}}
\newcommand{\Msun}{\mbox{$M_{\odot}$}}
\newcommand{\Teff}{\mbox{$T_{\rm eff}$}}

\newcommand{\co}{\mbox{${\rm C/O}$}}

\newcommand{\comment}[1]{}
\newcommand{\beq}{\begin{equation}}
\newcommand{\eeq}{\end{equation}}
\newcommand{\beqa}{\begin{eqnarray}}
\newcommand{\eeqa}{\end{eqnarray}}
        \def\smallskip{\vskip 2pt}

\begin{document}
\title{Evolution of asymptotic giant branch stars}
\subtitle{
II. Optical to far-infrared isochrones with improved
TP-AGB models}

\author{Paola Marigo$^1$ \and L\'eo~Girardi$^2$ \and
 Alessandro~Bressan$^{2,3}$ \\ Martin A.T. Groenewegen$^4$ \and
 Laura Silva$^5$ \and Gian Luigi Granato$^2$}
\institute{
 Dipartimento di Astronomia, Universit\`a di Padova,
	Vicolo dell'Osservatorio 2, I-35122 Padova, Italy \and
 Osservatorio Astronomico di Padova -- INAF,
	Vicolo dell'Osservatorio 5, I-35122 Padova, Italy \and
 INAOE, Luis Enrique Erro 1, 72840 Tonantzintla, Puebla, Mexico \and
 Instituut voor Sterrenkunde, Katholieke Universiteit Leuven,
	Celestijnenlaan 200 D, B-3001 Leuven, Belgium \and
 Osservatorio Astronomico di Trieste -- INAF,
	Via Tiepolo 11, I-34131 Trieste, Italy
}

\offprints{P. Marigo \\ e-mail: paola.marigo@unipd.it}

\date{Received 2007 August 13 / Accepted 2008 January 29 }

\abstract{
We present a large set of theoretical isochrones, whose distinctive
features mostly reside on the greatly improved treatment of the
thermally pulsing asymptotic giant branch (TP-AGB) phase. Essentially,
we have coupled the TP-AGB tracks described in Paper I, at their
stages of pre-flash quiescent H-shell burning, with the evolutionary
tracks for the previous evolutionary phases from Girardi et
al. (2000). Theoretical isochrones for any intermediate value of age
and metallicity are then derived by interpolation in the grids. We
take care that the isochrones keep, to a good level of detail, the
several peculiarities present in these TP-AGB tracks -- 
e.g. the cool
tails of C-type stars owing to the use of 
proper molecular opacities as convective dredge up occurs 
along the TP-AGB, the bell-shaped
sequences in the Hertzsprung-Russell (HR) diagram for stars with
hot-bottom burning, the changes of pulsation mode between fundamental
and first overtone, the sudden changes of mean mass-loss rates as the
surface chemistry changes from M- to C-type, etc. Theoretical
isochrones are then converted to about 20 different photometric
systems -- including traditional ground-based systems, and those of
recent major wide-field surveys such as SDSS, OGLE, DENIS, 2MASS,
UKIDSS, etc., -- by means of synthetic photometry applied to an
updated library of stellar spectra, suitably extended to include
C-type stars. Finally, we correct the predicted photometry by the
effect of circumstellar dust during the mass-losing stages of
the AGB evolution, which allows us to improve the results for the
optical-to-infrared systems, and to simulate mid- and far-IR systems
such as those of Spitzer and AKARI. We illustrate the most striking
properties of these isochrones by means of basic comparisons with
observational data for the Milky Way disk and the Magellanic
Clouds. Access to the data is provided both via a web repository of
static tables ({\tt http://stev.oapd.inaf.it/dustyAGB07} and
CDS), and via an interactive web interface ({\tt
http://stev.oapd.inaf.it/cmd}) that provides tables for any
intermediate value of age and metallicity, for several photometric
systems, and for different choices of dust properties. }

\authorrunning{Marigo et al.}
\titlerunning{Optical to far-IR isochrones}
\maketitle

\section{Introduction}
\label{intro}

Theoretical isochrones find a wide range of applications in modern
astrophysics, going from the simple age-dating of star clusters, to
more complex problems like the derivation of star formation histories
of resolved galaxies, and, since Charlot \& Bruzual (1991), the
modelling of their spectrophotometric evolution across the Hubble
time.

In fact, many different groups are publishing large grids of stellar
isochrones, covering wide enough a range of stellar parameters (age,
mass, metal content, alpha-enhancement, etc.) to become useful in the
stellar population synthesis of galaxies. These isochrones normally
comprehend a detailed description of all evolutionary phases from the
zero-age main sequence (ZAMS) up to either carbon ignition or to the
onset of the thermally-pulsing asymptotic giant branch phase
(TP-AGB). The coverage of the TP-AGB evolution, however, has always
been far from detailed and physically accurate, as we will illustrate
below.  The poor modelling of this phase greatly contrasts with its
importance to the integrated light of stellar populations with ages
from a few $10^8$~yr up to the Hubble time. According to Frogel et
al. (1990), the TP-AGB contribution to the total luminosity of
single-burst stellar populations reaches a maximum of about 40 percent
at ages from 1 to 3 Gyr, falling to less than 10 percent at 10
Gyr. Obviously, TP-AGB stars have a great impact on the rest-frame
near-infrared spectra of gala\-xies. Moreover, TP-AGB stars account for
most of the bright-infrared (IR) objects in resolved galaxies, as
clearly demonstrated by DENIS, 2MASS, SAGE and S$^3$MC data for the
Magellanic Clouds (Cioni et al. 1999; Nikolaev \& Weinberg 2000; Blum
et al. 2006; Bolatto et al. 2007).

\begin{table*}
\small
\caption{Available stellar isochrones including the TP-AGB phase.}
\label{tab_isoc}
\begin{tabular}{lcccl}
\hline
Reference & $3^{\rm rd}$DU and HBB$^{(\rm a)}$ & $T_{\rm eff}$$^{(\rm b)}$ & Dust$^{(\rm c)}$ &  public www address for retrieval \\
\hline
Bertelli et al. (1994)\comment{, A\&AS, 106, 275}    &  NO  &  solar-scaled comp.  &  NO  &   http://pleiadi.pd.astro.it \\
Bressan et al. (1998)\comment{, A\&A, 332, 135}  & NO  &  solar scaled comp. &   YES &               upon request \\
Girardi et al. (2000)\comment{, A\&AS, 141, 371}  &     NO  &  solar-scaled  comp. &  NO    &        http://pleiadi.pd.astro.it\\
Marigo \& Girardi (2001)\comment{, A\&A, 377, 132}   &   YES &  solar-scaled  comp. &  NO   &         http://pleiadi.pd.astro.it \\
Mouhcine (2002)\comment{, A\&A, 394, 125}  & YES    &  solar scaled comp. &    YES  &               upon request \\
Piovan et al. (2003)\comment{, A\&A, 408, 559} &  NO  &  solar scaled comp. &   YES  &                upon request \\
Marigo et al. (2003a)\comment{, A\&A, 403, 225}  & YES  & O-/C-rich comp. &    NO  &          upon request  \\
Cioni et al. (2006)\comment{, A\&A, 448, 77}  & YES  & O-/C-rich comp. &    NO  &          http://pleiadi.pd.astro.it  \\
Cordier et al. (2007)\comment{, AJ, 133,468}    &  NO & solar scaled comp.  &   NO    & http://www.oa-teramo.inaf.it/BASTI \\
This work    &  YES & O-/C-rich comp.  &   YES    & http://stev.oapd.inaf.it/cmd \\
\hline
\end{tabular}
\\
$^{(\rm a)}$ Explicit inclusion or not of AGB nucleosynthesis,
i.e. the third dredge-up and hot-bottom burning, in the underlying AGB
models \\ $^{(\rm b)}$ Choice of low-temperature opacities affecting
the predicted effective temperature, i.e. for solar-scaled or variable
(O- or C-rich) chemical mixtures \\ $^{(\rm c)}$ Inclusion or not of
circumstellar dust in the computation of the emitted spectrum \\
\end{table*}

The situation regarding the TP-AGB phase in published sets of
isochrones is the following: Whereas early isochrones sets do not even
include the TP-AGB phase, more recent ones do it, but still in a very
approximate way, ignoring crucial aspects of the TP-AGB evolution.  A
summary table of available isochrones is provided 
in Marigo (2007) and updated in
Table~\ref{tab_isoc} herewith. In short, Bertelli et al. (1994) and
Girardi et al. (2000; hereafter Gi00) adopt a very simple TP-AGB
synthetic model that does not include the third dredge-up and the
over-luminosity above the core mass--luminosity relation caused by hot
bottom burning (HBB). A luminosity correction for stars undergoing HBB
is instead included by Cordier et al. (2007), but in this case the
third dredge up -- together with its most remarkable consequence,
i.e. the formation of C stars -- and the HBB nucleosynthesis are
completely missing. Marigo \& Girardi (2001) present the first
isochrones based on tracks (Marigo et al. 1999) with the third dredge
up included, and moreover its efficiency is calibrated to reproduce
the C-star luminosity functions in both Magellanic Clouds
(MC)\footnote{Many of the improvements in Marigo et al. (1999)
TP-AGB tracks were already introduced by Groenewegen \& de Jong (1993,
1994abc), who computed extended sets of tracks and performed
population synthesis of resolved AGB stellar populations. Groenewegen
\& de Jong, however, do not provide theoretical isochrones from their
tracks.}. Another distinctive aspect of Marigo et al. (1999) tracks,
is that the HBB efficiency, together with the model effective
temperatures, are derived from numerical integrations of
the envelope structure.

Besides the above mentioned works, the literature abounds in
descriptions of evolutionary population synthesis models which have
included the complete TP-AGB evolution (e.g. Charlot \& Bruzual 2003;
Raimondo et al. 2005; V\'azquez \& Leitherer 2005; only to mention a
few). However, in all these cases, critical processes of TP-AGB
evolution -- again the third dredge-up, HBB, or superwind-like mass
loss -- are missing in the models. In some cases (e.g. V\'azquez \&
Leitherer 2005; Charlot \& Bruzual 2003), the tracks for the TP-AGB
and previous evolutionary phases come from different sources; 
this implies that basic stellar properties like the core mass and
envelope chemical composition are discontinuous along the composite
tracks. As demonstrated by Marigo \& Girardi (2001), discontinuities
in these quantities also produce an energetic inconsistency, in the
sense that they violate the  strict coupling between stellar  
integrated light and chemical yields implied by the condition
of energy conservation.

A few authors have computed isochrones aimed at the modelling of
spectral evolution of galaxies which consider the effect of
circumstellar dust around AGB stars (Bressan et al. 1998; Mouhcine
2002; Piovan et al. 2003), but without improving the treatment of the
underlying TP-AGB evolution, with respect to those already mentioned.

In all the previously-mentioned cases, moreover, the effect of
variable opacities along the TP-AGB evolution is completely ignored:
The basic stellar properties, and in particular the effective
temperature \Teff, are derived using opacities valid for scaled-solar
compositions, which are highly inadequate to describe the chemically
evolved envelopes of TP-AGB stars. The proper consideration of
variable opacities is now known to be crucial (see Marigo 2002; Marigo
et al. 2003a), especially for stars undergoing the transition to the
C-rich phase -- i.e. comprising, for MC metallicities, all stars with
initial masses between about 1.5 and 3.5~\Msun, and all isochrones
with ages between about 0.5 and 5 Gyr. This situation is partially
remedied by Cioni et al. (2006), who has made publicly available the
first isochrone sets in which variable molecular opacities were
considered along the TP-AGB evolution\footnote{Marigo et al. (2003a)
also computed isochrones containing essentially the same kind of
TP-AGB tracks as in Cioni et al. (2006), but for a more limited range
of metallicities, and for exploratory choices of model
parameters. These isochrones were provided a few times upon request,
and were superseded first by Cioni et al.'s and now by our new
isochrones.}. In fact, the C-star phase in these isochrones is
characterized by much lower temperatures (reaching
\Teff\ as low as 2300~K) than former ones. Those specific TP-AGB
models, however, are previous to the extensive calibration of dredge
up efficiency  performed in Marigo \& Girardi (2007, hereafter
Paper~I). In this latter work, the dredge up parameters are
tuned so that the models reproduce the observed the C-star luminosity
functions in the Magellanic Clouds, and the TP-AGB lifetimes of M and
C stars in Magellanic Cloud star clusters (Girardi \& Marigo 2007a).

{\em The goal of the present paper is to remedy this highly
unsatisfactory situation, providing the first set of isochrones in
which the TP-AGB evolution is treated with a satisfactory level of
completeness and detail -- i.e. considering the crucial effects of
third dredge up, hot bottom burning, and variable molecular opacities
-- and calibrated at the same time -- i.e. able to reproduce the basic
observables of AGB stars in the MCs. Moreover, we wish to provide
these results in many photometric systems and including the
reprocessing of radiation by circumstellar dust in mass-losing stars,
which is crucial for the modelling of the infrared photometry.}

The properties of our stellar evolutionary tracks are briefly recalled
in Sect.~\ref{sec_tracks}, which discusses the basic aspects of
constructing isochrones using the new TP-AGB tracks of Paper~I -- including
new variables like the C/O ratio, pulsation periods, pulsation modes,
etc. Section~\ref{sec_bc} deals with the transformation of the
isochrones to many photometric systems, which require updating and
extending the libraries of stellar spectra previously
in use. The inclusion of dust
reprocessing in the predicted photometry is illustrated 
in Sect.~\ref{sec_dust}. A preliminary comparison
with observational data is done in Sect.~\ref{sec_obs}. 
Section~\ref{sec_conclu} describes how to access the data tables --
containing isochrones, bolometric corrections, dust attenuation
curves, integrated magnitudes of single-burst populations, etc. -- and
summarizes both the main novelties and caveats of our new isochrones.

\section{From evolutionary tracks to isochrones}
\label{sec_tracks}

\subsection{Putting tracks together}
\label{sec_trackstogether}

The evolutionary tracks used in this work have already been thoroughly
presented in previous papers. The evolution from the ZAMS up to the
first thermal pulse on the AGB is taken from Gi00 for the initial
metallicities $Z=0.0004$, $0.001$, $0.004$, $0.008$, $0.019$ (the
solar value), $0.03$, and from Girardi (2001, unpublished; see {\tt
http://pleiadi.oapd.inaf.it}) for $Z=0.0001$.  The range of initial
masses of these tracks goes from $\mini=0.15$ to 7~\Msun, out of which
all models between $\mini\sim0.5$ and 5~\Msun\ are considered to
undergo the TP-AGB phase.

The TP-AGB phase is then computed via a synthetic code as described in
Paper~I, starting from the physical conditions at the first thermal
pulse, and up to the complete loss of the envelope via
stellar winds. The model takes into account many critical aspects of the
TP-AGB evolution, like the third dredge-up events, HBB over-luminosity
and nucleosynthesis, pulse cycle luminosity and \Teff\ variations,
opacity variations driven by the changing chemical composition in the
envelope, long-period variability (LPV) in both first overtone and
fundamental modes, and their modulation of the mass-loss rates
$\dot{M}$ via changes in the pulsation period $P$.

Especially relevant for the present paper are the formalisms
adopted for the mass loss: besides affecting the AGB
lifetimes, high mass-loss rates determine the formation of
circumstellar dust and hence the obscuration of AGB stars in the
optical and their brightening in the mid- and far-IR (see
Sect.~\ref{sec_dust}). As detailed in Paper~I, for O-rich AGB stars we
use the results of dynamical atmospheres including dust for
long-period variables calculated by Bowen (1988) and Bowen \& Willson
(1991), taking into account whether pulsation occurs on the first
overtone or on the fundamental mode. For C-rich stars, we assume two
different regimes of mass loss: initially stars follow the
semi-empirical relation from Schr\"oder et al. (2003), then switching,
above a critical luminosity, to the $\dot{M}$ values derived from the
pulsating dust-driven wind models by Winters et al. (2000, 2003). In
both cases, mass-loss rates depend on the pulsation period, which is
predicted  along the evolutionary tracks according to the 
period-mass-radius relations derived
by Ostlie \& Cox (1986), Fox \& Wood (1982), Wood et al. (1983).

The most massive TP-AGB tracks ($\mini\ge3.0$~\Msun) of Paper~I have 
been replaced by new calculations, accounting for a better description of the
transition luminosity between the first-overtone and the fundamental-mode
pulsation. This aspect is thoroughly described in the Appendix.

The synthetic TP-AGB evolution of Paper~I has been followed up to
the complete ejection of the stellar envelopes, hence including
a good fraction of the post-AGB phase prior to the planetary nebula
regime, 
i.e. the fast terminal stages 
during which the star evolves at almost constant luminosity, 
while \Teff\ rapidly increases due to the combined effect 
of the outward
displacement of the H-burning shell, and the stripping of the residual
envelope by stellar winds. These final parts 
of the tracks may not be reliable since some relations used in
the synthetic TP-AGB code become inaccurate. Moreover, the mass-loss
rates are critical in determining the duration of this post-AGB
transition phase, but they are still poorly known.

Anyway, this  transition phase turns out to be
very short-living ($\la10^4$~yr for low-mass stars, and even shorter
for intermediate-mass ones) if compared to the previous TP-AGB
evolution. Therefore, although this post-AGB phase is mostly included in
the new isochrones, its overall impact on them is
almost negligible.

Another important detail is that before building the isochrones,
TP-AGB tracks are ``cleaned'' of their luminosity and \Teff\
variations caused by thermal pulse cycles. Just the stages of
quiescent H-shell burning previous to He-shell flashes, along with the
stage of coolest \Teff, are kept for the isochrone
interpolation. This is a necessary approximation since the large 
luminosity and \Teff\ variations between confining tracks would lead
to wild variations in the interpolated isochrones. Anyway, the 
detailed evolution of luminosity and \Teff\ over the 
thermal pulse cycles  can be easily
introduced in the tabulated isochrones a posteriori, 
since the shape of thermal pulse cycles is a
known function of other quantities that are also interpolated and
stored along the isochrones, like the core mass and surface chemical
composition (see e.g. Wagenhuber \& Groenewegen 1998).

The Gi00 and Paper~I sets of evolutionary tracks differ in their
source of low-temperature ($T\la10^4$~K) opacities: while Gi00 uses
tables for scaled-solar chemical compositions taken from Alexander \&
Ferguson (1994), in Paper~I the opacities are computed ``on-the-fly''
using the proper chemical mixture of the evolving TP-AGB envelopes, by
means of the methods and approximations described in Marigo (2002).
There is no dramatic change in opacities, however, in passing from the
scaled-solar tables of Alexander \& Ferguson (1994) to Marigo's (2002)
formalism, provided that the passage occurs at C/O ratios much smaller
than 1, as indeed is the case at the first thermal pulse of all TP-AGB
tracks. The molecular opacities depart from those in Alexander \&
Ferguson (1994) only later in the TP-AGB evolution, when the third dredge
up sets in and the C/O ratio increases above 1. Therefore, the
concatenation of both sets of tracks at the stage of first thermal
pulse produces just modest discontinuities in the tracks, i.e. small
jumps in \Teff\ that are caused by the small differences in opacities
for O-rich mixtures\footnote{These jumps in \Teff\ can be appreciated
in Fig.~\ref{fig_tracks}, at the points where the TP-AGB phase begins
(i.e. in passing from green to blue lines).}. For other quantities,
like the luminosity, core mass, and surface chemical composition, the
inspection of isochrones does not reveal any discontinuity.

Although the reader may feel uncomfortable with the small
discontinuities in \Teff\ {at the stage of first thermal pulse in} the
isochrones, we remark that they are anyway dramatically less important
than any discontinuity that one may find when amalgamating tracks from
completely different sources. The simple fact that the core mass and
envelope chemical composition vary in a smooth and continuous way, for
instance, ensures that the total emitted light derived from a given
set of isochrones will be fully consistent with the chemical yields
and total remnant masses derived from the same sets of tracks. As
illustrated by Marigo \& Girardi (2001), this is an important,
although largely ignored, consistency requirement to evolutionary
population synthesis models of galaxies.

\subsection{Resulting isochrones}
\label{sec_isochrones}

\begin{figure*}
\begin{minipage}{0.33\textwidth}
	\resizebox{\hsize}{!}{\includegraphics{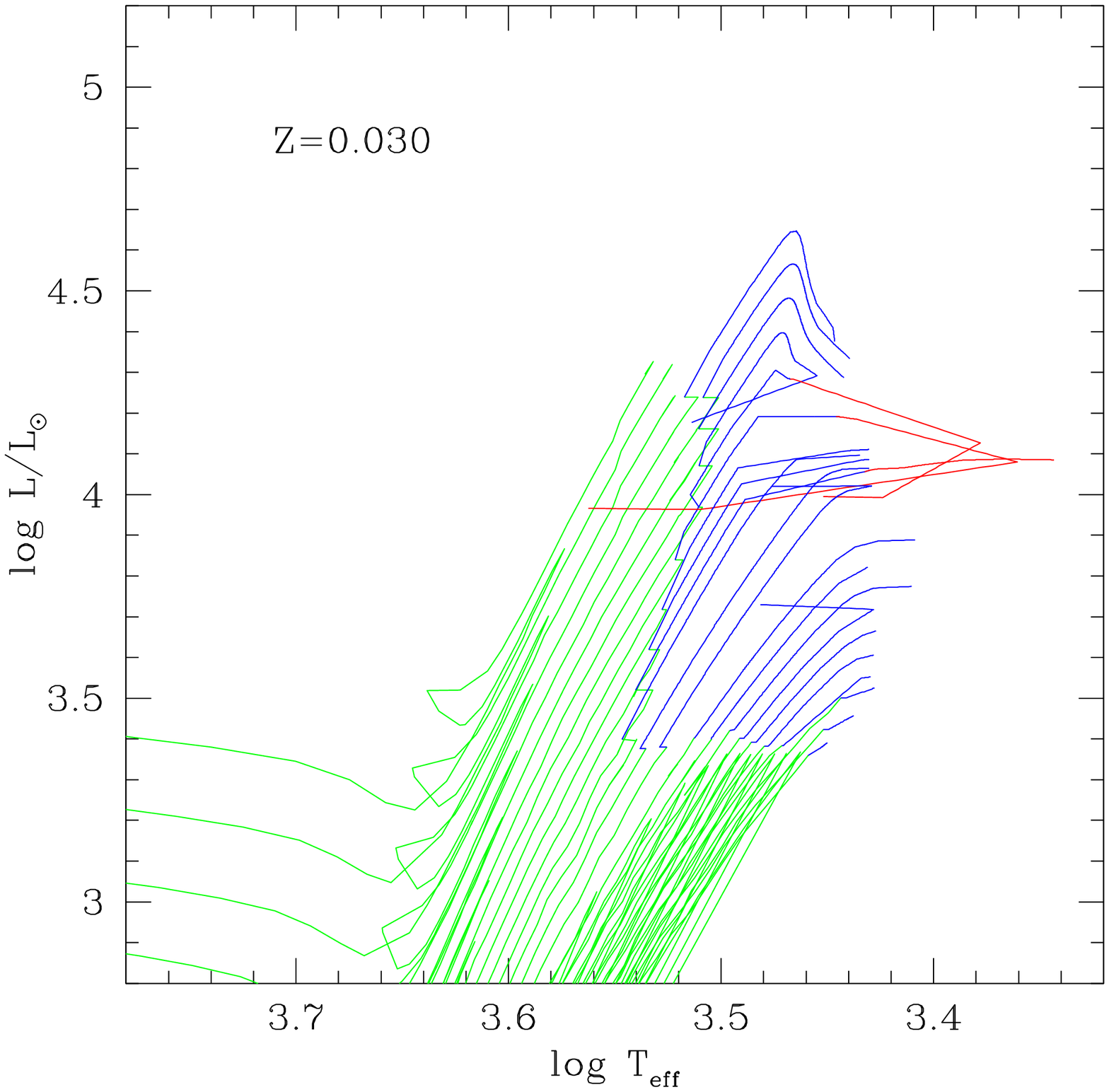}}
\end{minipage}
\hfill
\begin{minipage}{0.33\textwidth}
	\resizebox{\hsize}{!}{\includegraphics{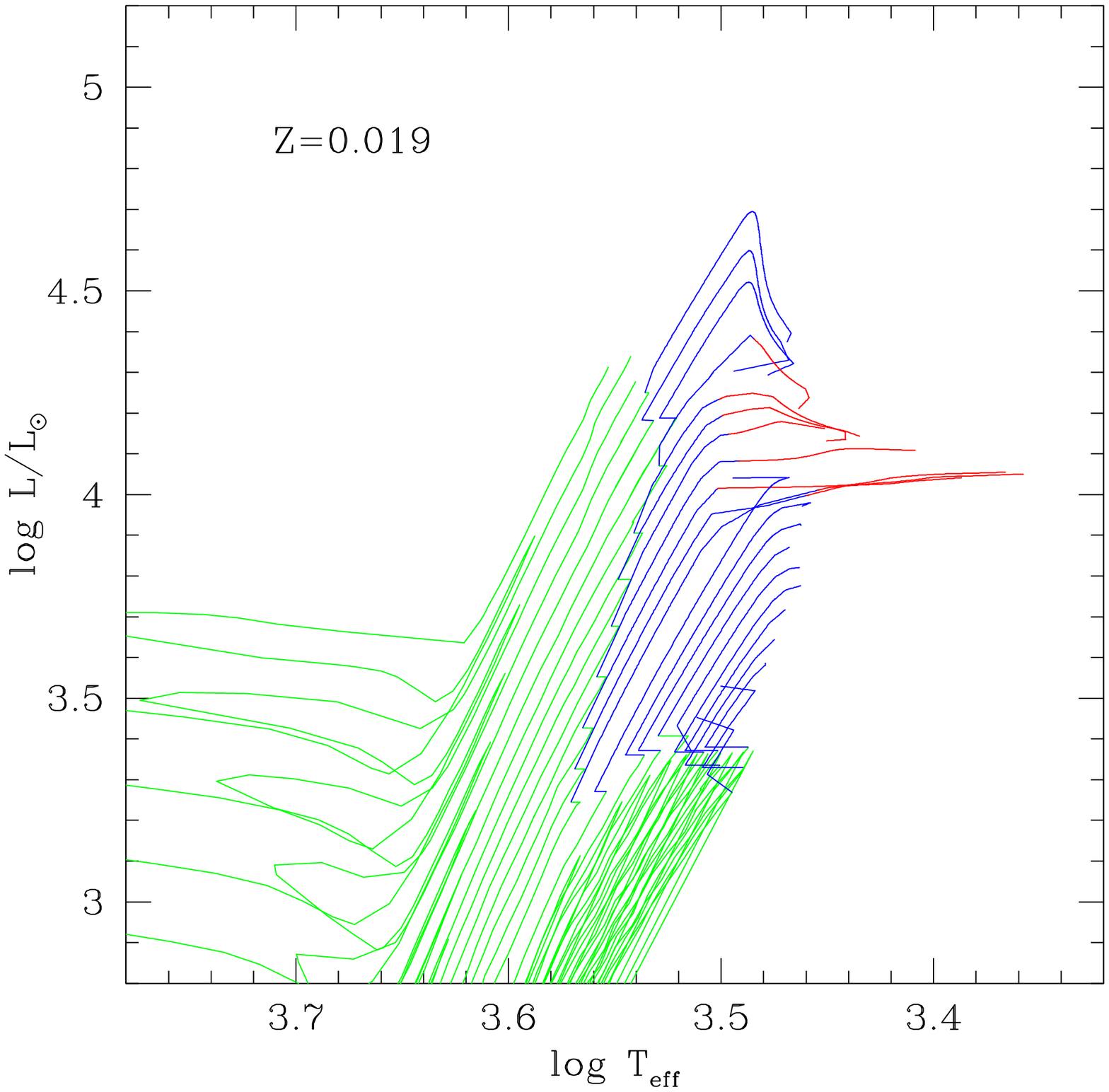}}
\end{minipage}
\hfill
\begin{minipage}{0.33\textwidth}
	\resizebox{\hsize}{!}{\includegraphics{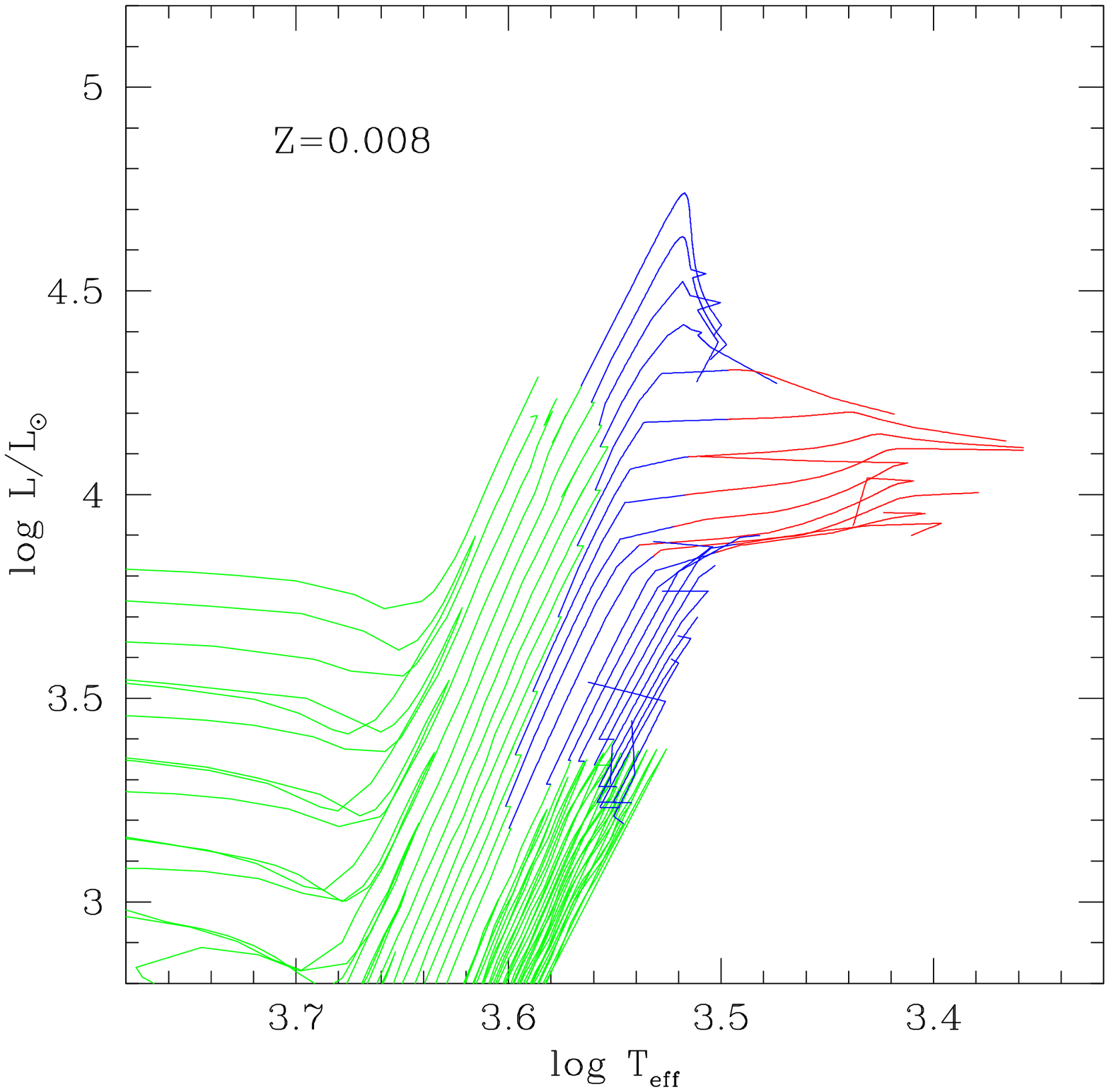}}
\end{minipage}
\\
\begin{minipage}{0.33\textwidth}
	\resizebox{\hsize}{!}{\includegraphics{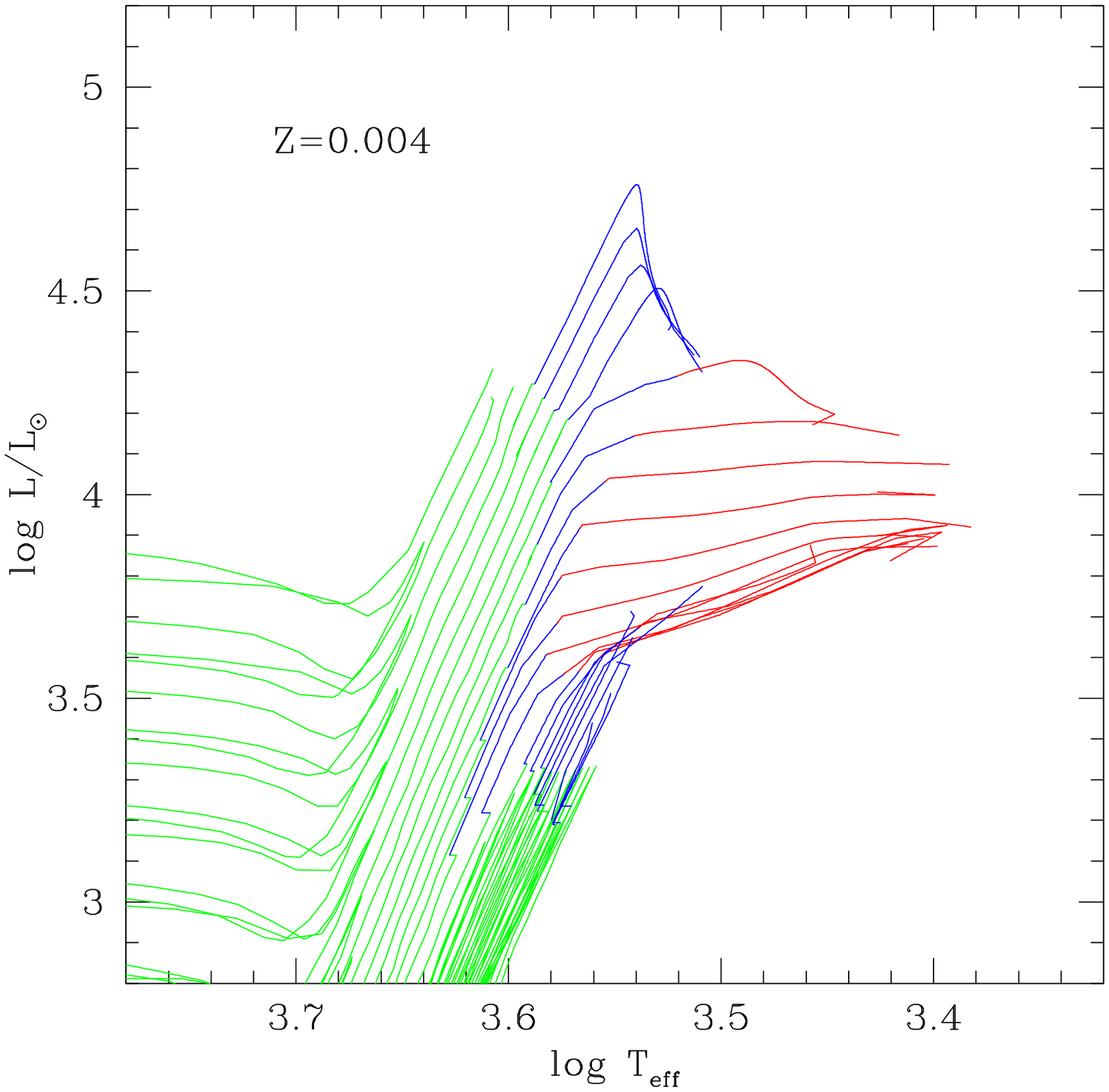}}
\end{minipage}
\hfill
\begin{minipage}{0.33\textwidth}
	\resizebox{\hsize}{!}{\includegraphics{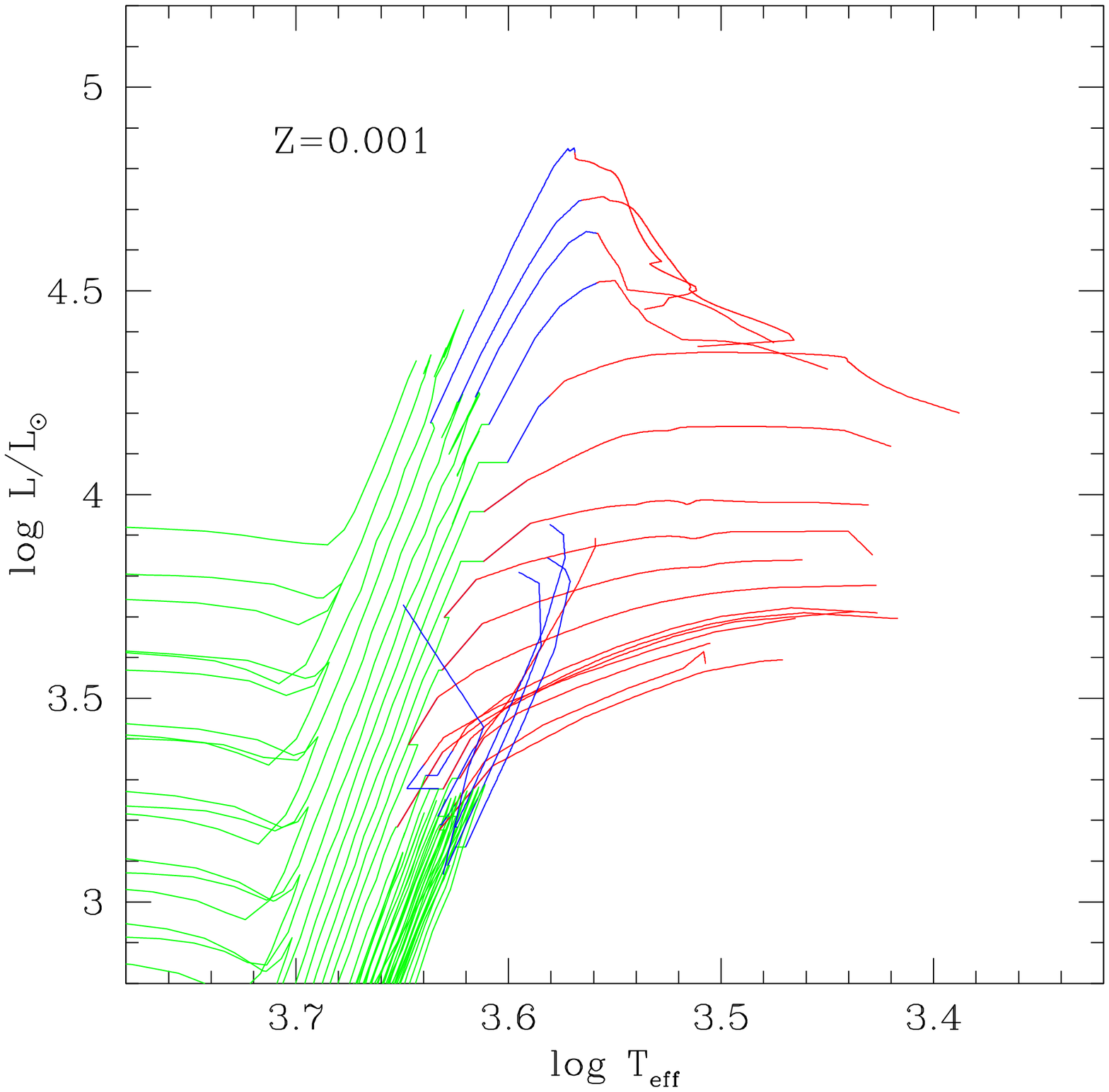}}
\end{minipage}
\hfill
\begin{minipage}{0.33\textwidth}
	\resizebox{\hsize}{!}{\includegraphics{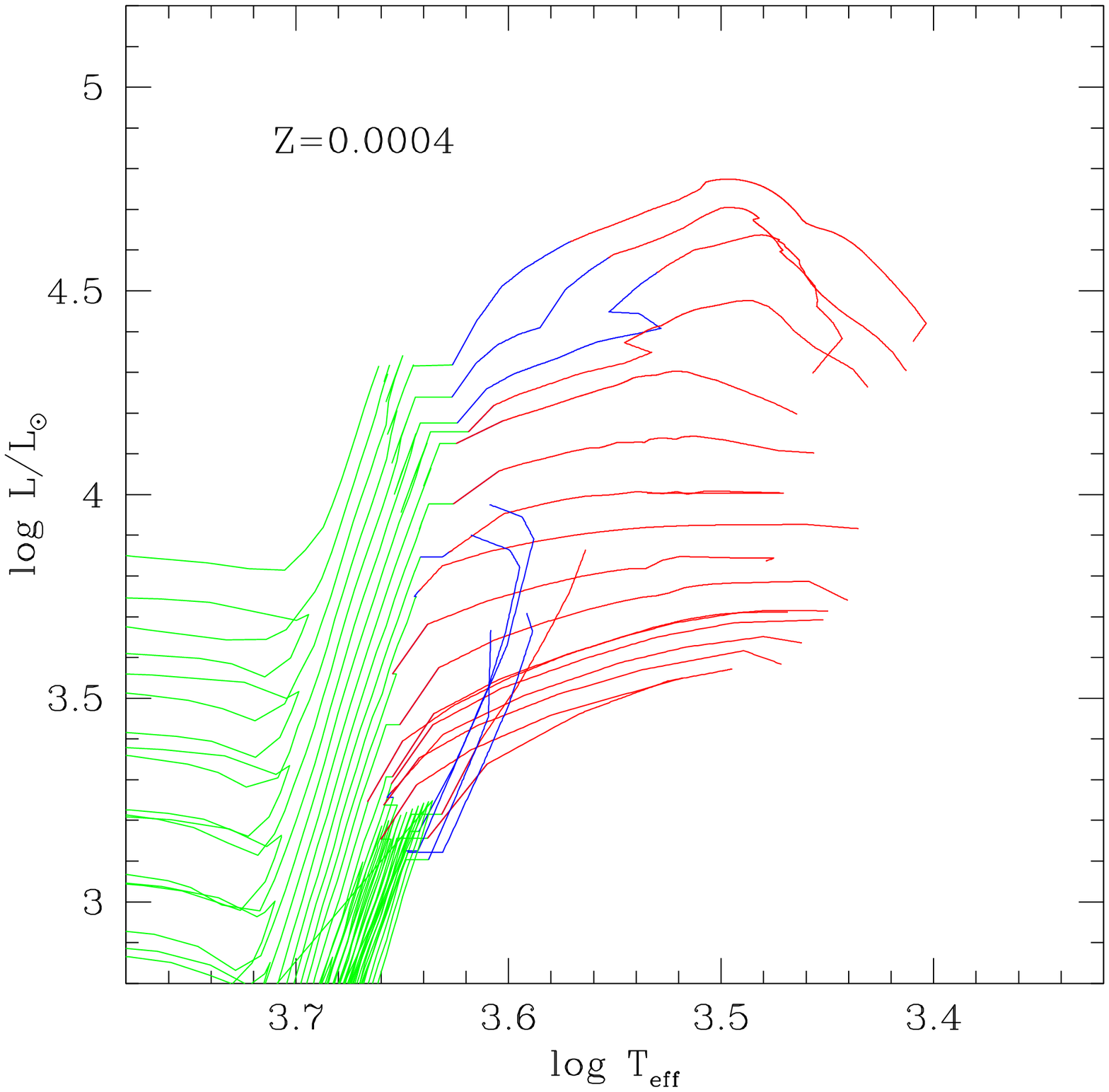}}
\end{minipage}
\\
\begin{minipage}{0.33\textwidth}
	\resizebox{\hsize}{!}{\includegraphics{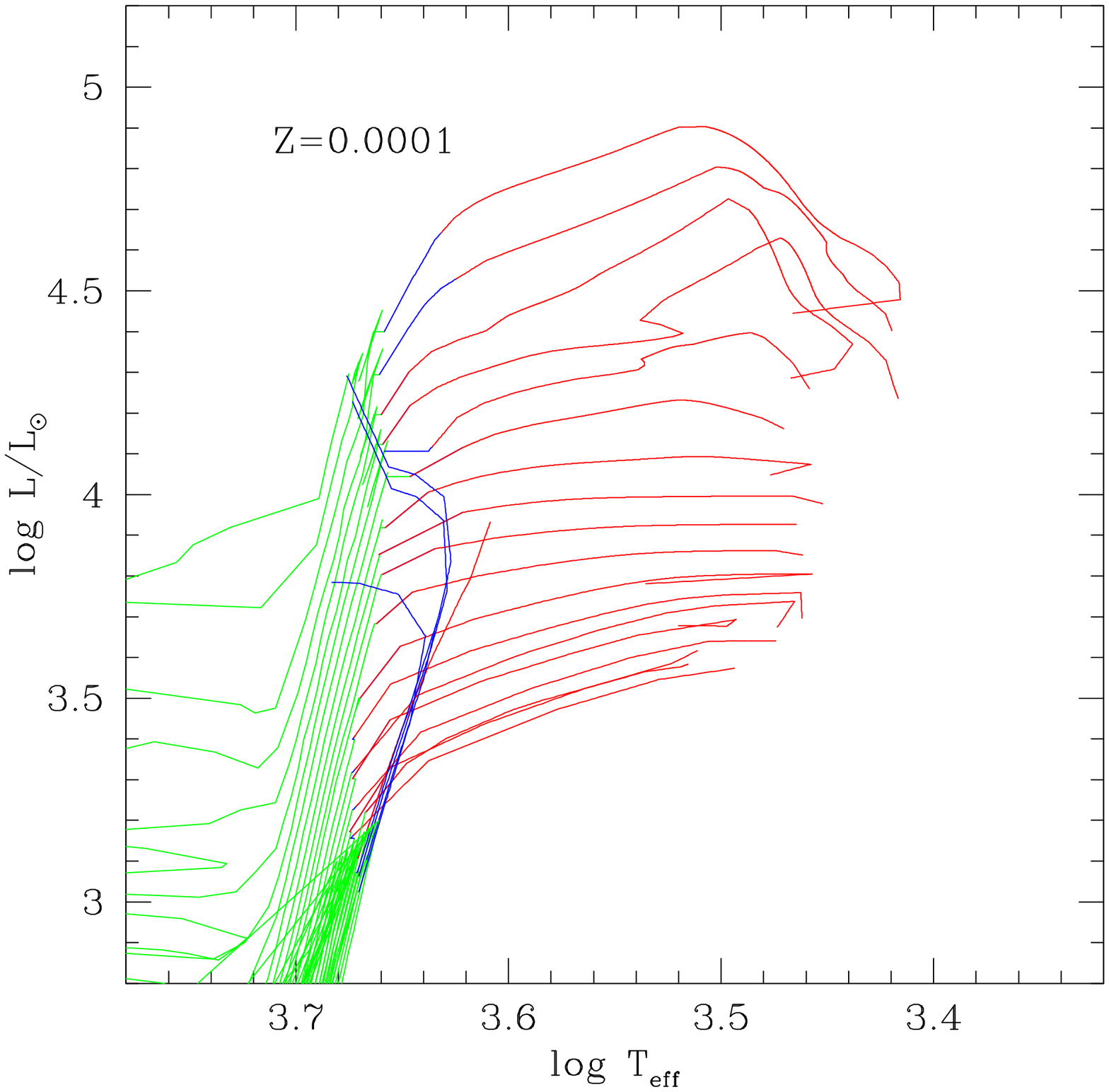}}
\end{minipage}
\hfill
\begin{minipage}{0.66\textwidth}
\caption{Isochrones in the H-R diagram ,
zooming on the region of the AGB, for the 7 metallicity values of the
original tracks. The isochrones are plotted with ages $\log(t/{\rm
yr})=7.8$ to $10.2$ (the youngest are at the top/left) with a spacing
of 0.1~dex in $\log t$. In the electronic version of the paper, the
pre-TP-AGB phases are drawn in green, whereas on the TP-AGB the
surface O-rich and C-rich configurations are distinguished with blue
and red lines, respectively. Along the TP-AGB, the luminosity and
effective temperature correspond to the quiescent values just before
the predicted occurrence of thermal pulses.}
\label{fig_tracks}
\end{minipage}
\end{figure*}

\begin{figure*}
\begin{minipage}{0.33\textwidth}
	\resizebox{\hsize}{!}{\includegraphics{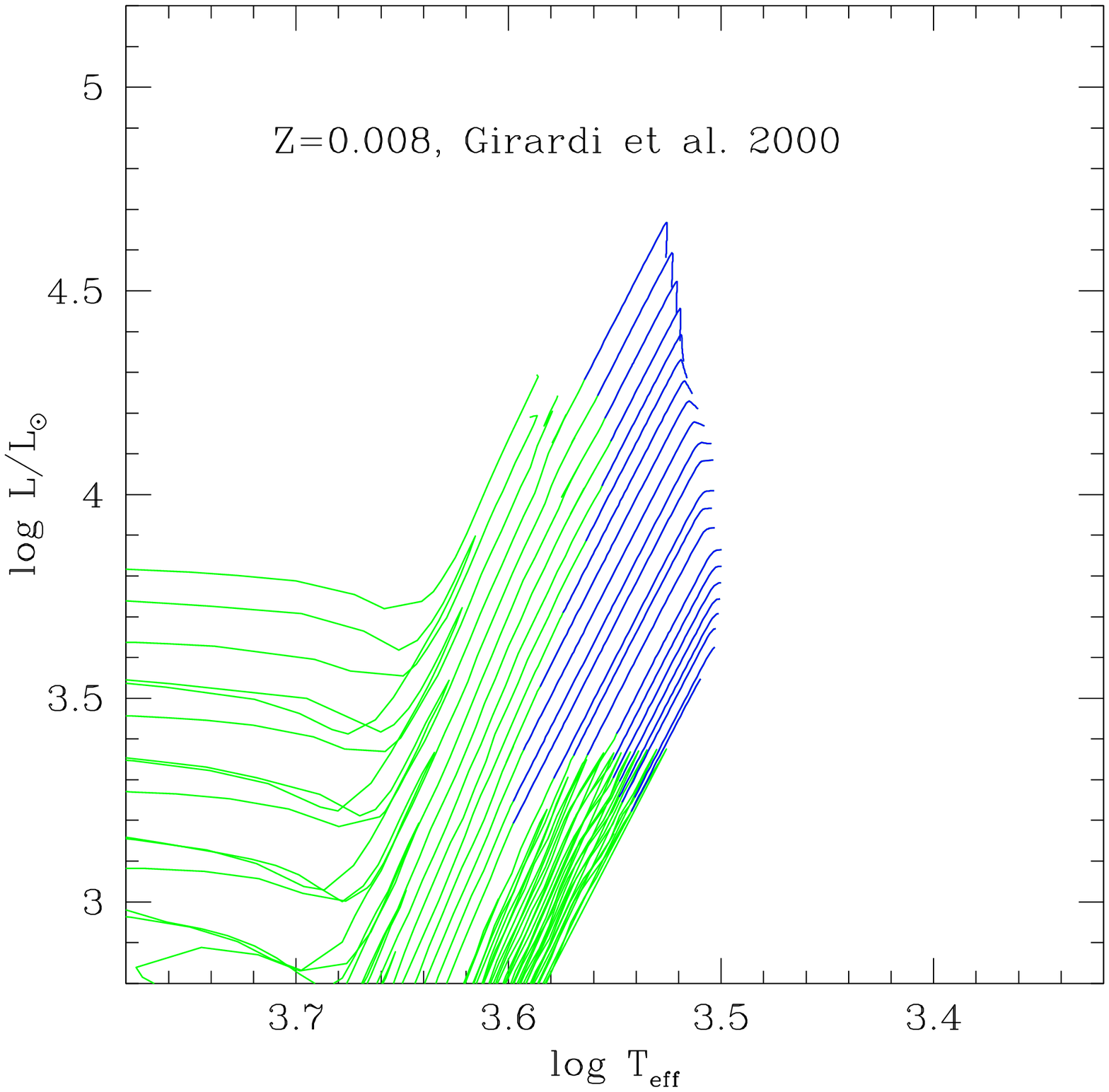}}
\end{minipage}
\hfill
\begin{minipage}{0.33\textwidth}
	\resizebox{\hsize}{!}{\includegraphics{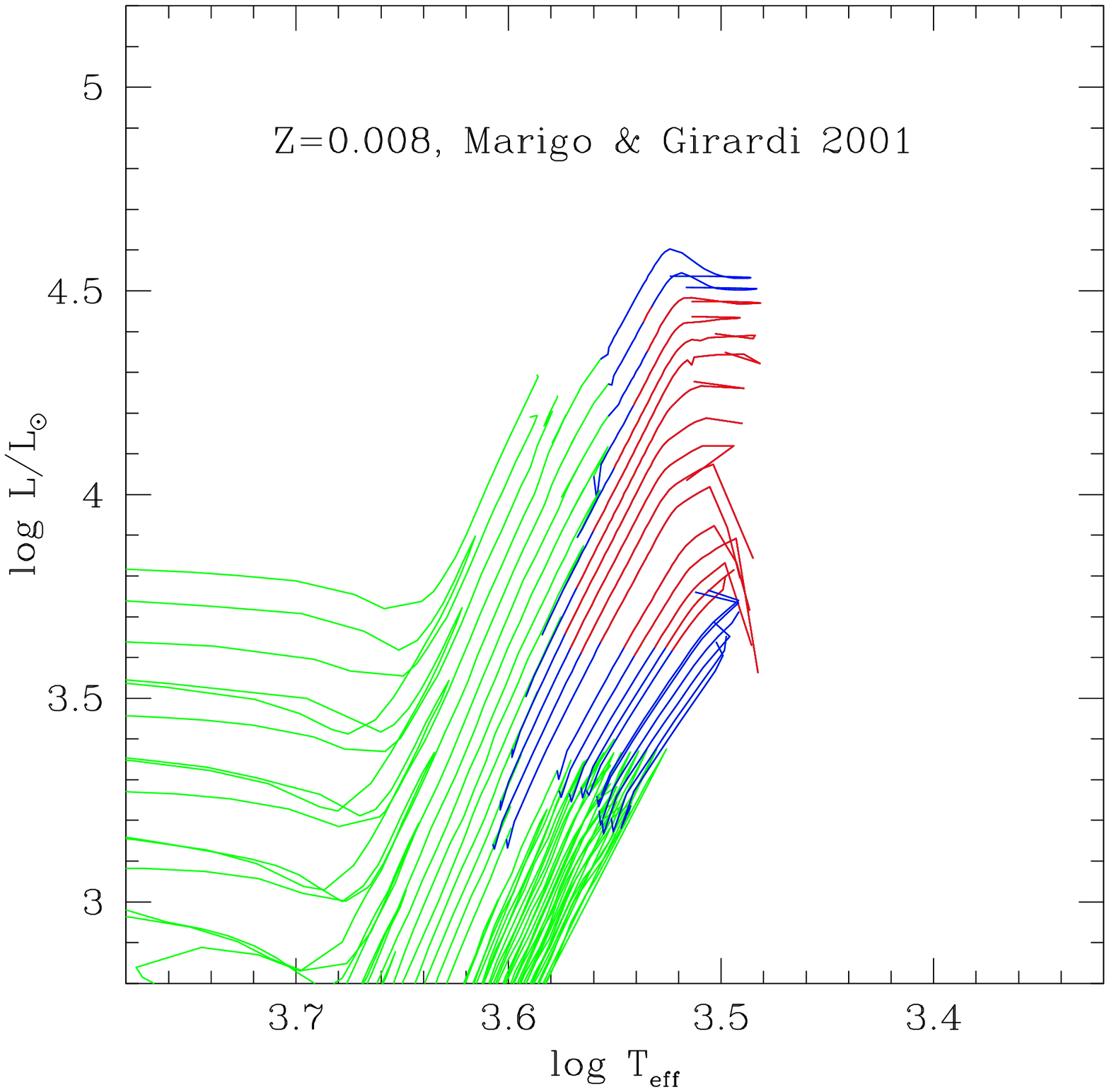}}
\end{minipage}
\hfill
\begin{minipage}{0.33\textwidth}
	\resizebox{\hsize}{!}{\includegraphics{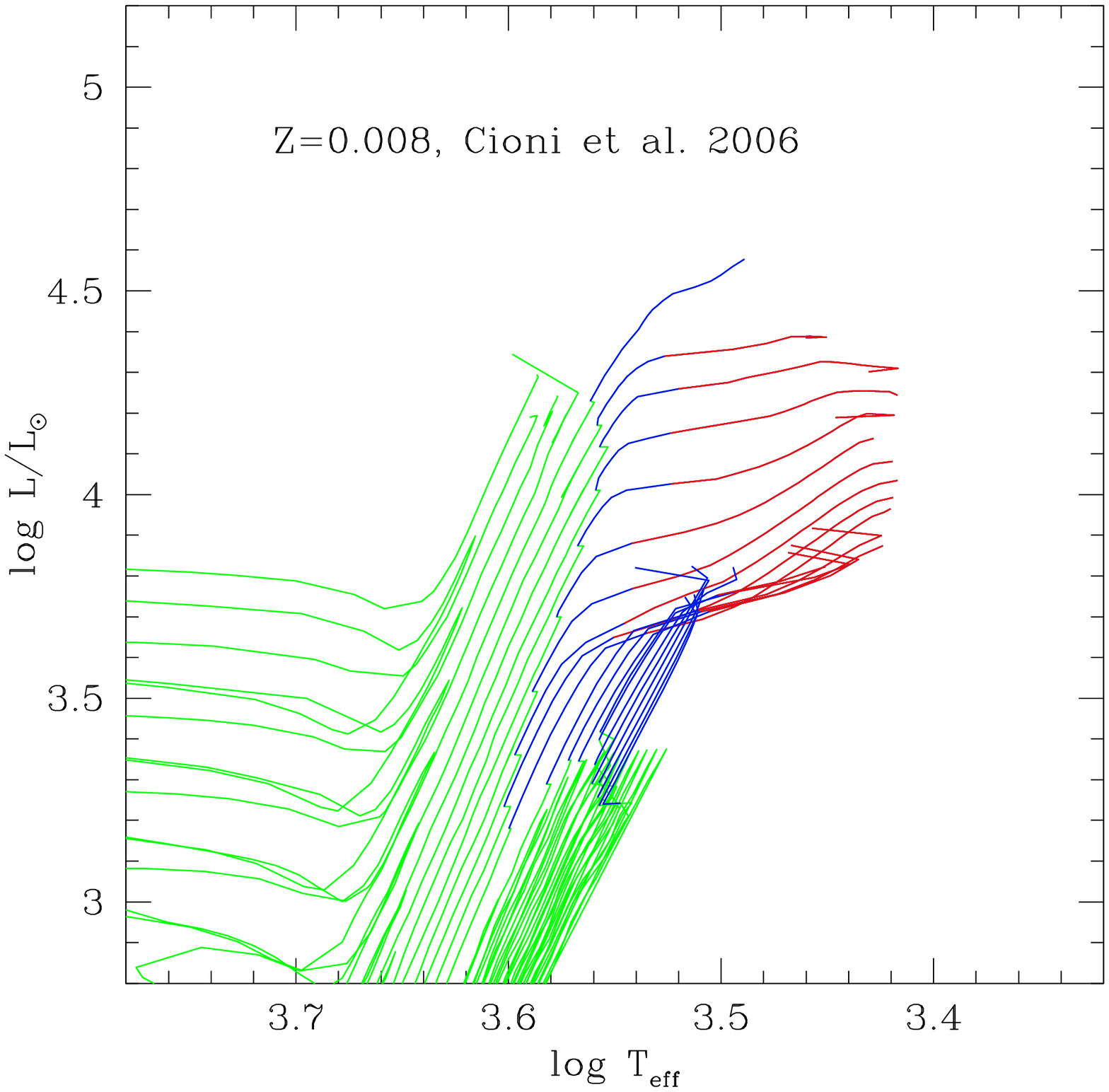}}
\end{minipage}
\\
\caption{The same as in Fig.~\ref{fig_tracks}, but for a sequence that
illustrates the evolution of the TP-AGB phase since the release
of Gi00 isochrones. From left to right, isochrones are from Gi00,
Marigo \& Girardi (2001), and Cioni et al. (2006). This sequence would
be completed with the third panel of Fig.~\ref{fig_tracks}.}
\label{fig_isoc}
\end{figure*}

Theoretical isochrones are built via simple linear interpolations
inside the grids of evolutionary tracks, performed between
``equivalent evolutionary points''. The final isochrone, for a given
age $t$, is made of a sample of such points conveniently distributed
on the HR diagram. For each point,  several stellar parameters
are stored. In fact, the newest TP-AGB tracks take into account
various  aspects of evolution that have been ignored while making the
Gi00 isochrones, but that are essential components of the TP-AGB
models of Paper~I, namely: surface chemical composition (especially
the C/O ratio), stellar core and envelope mass, mass-loss
rates, pulsation modes and periods. These parameters can, for
instance, allow an easy identification of C-type and dust-obscured
stars, and are essential to the simulation of variations in stellar
properties due to thermal pulse cycles and long-period
variability. Therefore, our isochrone-making routines have been
extended to take explicit care of these properties.

When an isochrone is built out of our tabulated tracks, 
mass loss and LPV pulsation are taken into detailed
account only after the transition from  the early-AGB to the TP-AGB
regime. Variables such as the mass-loss rates, pulsation periods, and
circumstellar dust emission, pass from ``non-defined'' values to well
defined ones, corresponding to the occurrence of 
the first thermal pulse.  In order to avoid
discontinuities in these variables, we have a posteriori evaluated the
mass-loss rates and pulsation properties of stars in earlier phases,
even if these have been computed at constant mass. 
The procedure is straightforward: for each star in the pre-TP-AGB
section of the isochrones, the basic stellar properties ($L$, \Teff,
\mini) are used to determine pulsation periods and mass-loss rates in
the same way as for O-rich AGB stars. Notice that in most cases mass
loss prior to the TP-AGB phase is expected to be very modest and not
to have evolutionary implications, so that this procedure is well
justified. The only exception is for low-mass models close to the tip
of the red giant branch (TRGB), where the use of Reimers (1975)
mass-loss formula with a multiplicative factor of $\eta=0.4$, leads to
the loss of $\Delta M^{\rm RGB}\la0.2$~\Msun\ for low-mass
stars\footnote{The behaviour of $\Delta M^{\rm RGB}$ as a function of
age can be appreciated in figure 6 of Girardi (1999), top
panel.}. This amount of mass lost is necessary to explain the HB
morphologies observed in galactic globular clusters (see Renzini \&
Fusi Pecci 1988). The use of the Reimers mass-loss formula for
the RGB shall be seen as just a ``conservative choice'' since, 
although alternative formulas do exist (see Catelan 2000, 2007),  
at the present time none can be clearly favoured over the others. 
We take the RGB mass
loss into consideration during the construction of stellar isochrones
by simply jumping from a TRGB track of given mass $M$ to a ZAHB track
of mass $M-\Delta M^{\rm RGB}$. The procedure is somewhat detailed in
Bertelli et al. (1994) and in Gi00.

Extended sets of isochrones are depicted in the HR diagrams of
Fig.~\ref{fig_tracks}.  Since the main novelty of the present paper is
in the detailed description of the TP-AGB phase, the HR diagrams are
zoomed on this phase. In other parts of the HR, isochrones are simply
identical to those previously computed by Gi00.

It is evident from the figure that our new isochrones reflect all
of the peculiar behaviours of the TP-AGB tracks they derive from (see
figure 20 in Paper~I). In particular, one can notice the following:

\paragraph{At moderate-to-high metallicities, i.e. for $Z\ge0.004$}: 
In the youngest isochrones, the combination of HBB and high mass-loss
rates gives origin to bell-shaped TP-AGB curves. The luminosity
first rapidly grows above the core-mass luminosity relation because of
the occurrence of HBB, then higher mass-loss rates are attained,
drastically reducing the stellar envelope mass so that HBB and the
related over-luminosity are eventually extinguished.

These stars usually have C/O$<1$, although a late transition to
the C-rich phase may take place due to final dredge up events when HBB
has already ceased (see Frost et al. 1998) in agreement with
observations (e.g. van Loon et al. 1998, 1999).

The C-star phase is completely developed in isochrones of intermediate
ages, and it happens at very low \Teff\ (reaching $\Teff\simeq
2500$~K), in a sort of ``red tail'' which is neatly separated from the
sequences of O-rich stars. This separation naturally results from
the very different low-temperature opacities of C-rich envelopes
(Marigo 2002).

In the oldest isochrones, dredge up is not operating and the TP-AGB
sections of the isochrones simply align above the previous early-AGB
phase. They steadily increase in luminosity and slightly bend to
the left at their upper end, as a result of the progressive thinning
of stellar envelopes.

\paragraph{At low metallicities, i.e. for $Z<0.004$:} Some of the
previous trends change. One important difference is  that,
despite the high efficiency of HBB, a
C-star phase is present even at younger ages. This is caused by
the activation of the ON cycle  
which converts O into N, so that the C/O ratio is predicted to 
increase above unity 
(Ventura et al. 2002; and Paper~I). Compared to the tracks of higher
metallicities, the C-star sequences start at higher effective
temperatures, as expected from the lower molecular concentrations
in their atmospheres. 
The oldest O-rich sequences evolve to
larger luminosities while bending to the left at their high ends, just
as in the tracks for higher metallicities. This feature is essentially
related to the
outward displacement of the H-burning shell, which is found to be more  
efficient than the stellar winds in reducing the envelope mass of
low-mass AGB stars ($M_{\rm i} \la 1.\, M_{\odot}$).

A more detailed discussion of the above-mentioned behaviours is
presented in Paper~I.

In order to better illustrate the peculiarities of the TP-AGB models
presently available, Fig.~\ref{fig_isoc} presents a series of
$Z=0.008$ isochrones derived from Gi00 tracks connected with different
extensions of the TP-AGB phase. In the first one (Gi00) the TP-AGB
phase is very simplified and just extends the previous evolution
straight above the RGB and early-AGB phases, without any C-star
sequence and without any signature of HBB. Then, the Marigo \& Girardi
(2001) isochrones do introduce dredge-up and HBB in the TP-AGB, and
present a well-defined C-star region, which is still aligned on top of
the RGB and early-AGB, as expected from models using opacities for
scaled-solar compositions.  Cioni et al. (2006) introduce variable
molecular opacities and then the C-star sequence clearly changes its
shape on the HR diagram (as explained by Marigo 2002). Finally, the
sequence of Fig.~\ref{fig_isoc} could be completed with the third
panel of Fig.~\ref{fig_tracks}, which represents the present state of
the $Z=0.008$ isochrones, differing from the Cioni et al. (2006) in
that the C-star luminosities and lifetimes are now calibrated (see
Paper~I).

\section{Transformations to several photometric systems}
\label{sec_bc}

The new isochrones are transformed to several photometric systems,
following the formalism described in Girardi et al. (2002) to derive
bolometric corrections from a library of stellar spectra. Although
this argument will be more thoroughly discussed in a forthcoming paper
(Girardi et al. 2007, in preparation), it is worth recalling the main
steps we follow to deal with the novelties of the present release.

\subsection{Bolometric corrections and extinction coefficients}
\label{sec_bolom}

Given the spectral flux at the stellar surface, $F_\lambda$,
bolometric corrections for any set of filter transmission curves
$S_\lambda$ are given by (see Girardi et al. 2002 for details)
	\beqa
BC_{S_\lambda} & = & M_{\rm bol, \odot}
	- 2.5\,\log \left[
		4\pi (10\,{\rm pc})^2 F_{\rm bol}/L_\odot
		\right] \label{eq_bcfinal}
	\\ \nonumber
	&& + 2.5\,\log\left(
	\frac { \int_{\lambda_1}^{\lambda_2}
		\lambda F_\lambda \, 10^{-0.4A_\lambda}
		S_\lambda \diff\lambda }
		{ \int_{\lambda_1}^{\lambda_2}
		\lambda f^0_\lambda S_\lambda \diff\lambda }
	\right)
	- m_{S_\lambda}^0 \,\,\,\,\,\,,
	\eeqa
where $f^0_\lambda$ represents a reference spectrum (at the Earth)
that produces a known apparent magnitude $m^0_{S_\lambda}$, and
$F_{\rm bol} = \int_0^\infty F_\lambda \diff\lambda =
\sigma T_{\rm eff}^4 $ is the total emerging flux at the stellar
surface. Once $BC_{S_\lambda}$ are computed, stellar absolute
magnitudes follow from
	\beq
M_{S_\lambda} = M_{\rm bol} - BC_{S_\lambda} \,\,\, ,
\label{eq_absolmag}
	\eeq
where
	\beqa
M_{\rm bol} & = & M_{\rm bol, \odot} - 2.5\, \log(L/L_\odot)
	\label{eq_mbol}
	 \\ \nonumber
	 & = & M_{\rm bol, \odot} - 2.5\, \log(
		4\pi R^2 F_{\rm bol} /L_\odot) \,\,\,\, .
	\eeqa
As in Girardi et al. (2002), we adopt $M_{\rm bol,
\odot}=4.77$, and $L_\odot=3.844\times10^{33}\,{\rm erg\,s^{-1}}$
(Bahcall et al.\ 1995).

Notice that Eq.~(\ref{eq_bcfinal}) uses photon count integration instead
of energy integration, as adequate to most modern photometric
systems which use photon-counting devices instead of energy-amplifiers.

The equation also includes the interstellar extinction curve
$A_\lambda$, that we set to $A_\lambda=0$. For the optical and
near-IR, we provide the extinction coefficients $A_\lambda/A_V$ as
computed for a yellow dwarf of solar metallicity (the Sun) using the
Cardelli et al. (1989) $R_V=3.1$ extinction curve. The small
variations of such coefficients with spectral type and extinction (see
Grebel \& Roberts 1995) will be discussed in a separate paper (Girardi
et al., in prep.) which will detail the transformations to WFPC2 and
ACS systems, and describe a way to consistently include the
interstellar extinction into the isochrones.

\subsection{Assembling a library of spectral fluxes}
\label{sec_spectra}

Girardi et al. (2002) describes the assembly of a large library of
spectral fluxes, covering wide ranges of initial metallicities (\mh,
from $-2.5$ to $+0.5$), \Teff\ (from 600 to 50\,000~K), and $\log g$
(from $-2$ to $5$)\footnote{In this paper, the surface gravity $g$ is
expressed in c.g.s. units.}. This grid is wide enough to include most
of the spectral types that constitute the bulk of observed samples. It
is composed of ATLAS9 ``NOVER'' spectra from Castelli et al. (1997;
see also Bessell et al. 1998) for most of the stellar types, Allard et
al. (2000; see also Chabrier et al. 2000) for M, L and T
dwarfs\footnote{Although brown dwarfs are not the subject of this
paper, the tables of bolometric corrections we are going to distribute
do include these objects. More details can be found in the original
paper by Allard et al.\ (and to a smaller extent also Girardi et al.
2002).}, Fluks et al. (1994) empirical spectra for M giants, and
finally pure black-body spectra for stars exceeding 50\,000~K.

Since then this library  been revised. We have replaced the ATLAS9
spectra by the newest ``ODFNEW'' ones from Castelli \& Kurucz (2003),
which incorporate several corrections to the atomic and molecular
data, especially regarding the molecular line lists of CN, OH, SiO,
H$_2$O and TiO, and the inclusion of H{\sc I}-H+ and H{\sc I}-H+
quasi-molecular absorption in the UV. These new models are
considerably better than previous ones in the UV region of the
spectrum, and over most of the wavelength range for $\Teff$ between
$\sim4500$ and $3500$~K (see figs. 1 to 3 in Castelli \& Kurucz 2003).

Moreover, the library has been extended with the Loidl et al. (2001)
synthetic C-star spectra for $3600\ga(\Teff/{\rm K})\ga2600$, derived
from hydrostatic model atmospheres computed with MARCS (see also
H\"ofner et al. 2003 for more details). For C-stars cooler than
2600~K, we simply adopt the BC values derived from 2600~K
spectra. These Loidl et al. (2001) spectra are computed for C/O=1.1
for all \Teff, and for both C/O=1.1 and 1.4 for $\Teff\le3200$~K. We
have verified that bolometric corrections for all broad bands redder
than $V$ are quite insensitive to the C/O ratio, varying by less than
0.1~mag between the C/O=1.1 and 1.4 cases. These differences become
$\la0.01$~mag in the $J$-band. Just for the bluest optical pass-bands
the choice between these two C/O values causes significant differences
the bolometric corrections, amounting for instance to $\ga0.4$~mag in
$B$. Anyway the $\Teff\le3200$~K C-stars are much fainter (more than
4~mag in a Vega system) at these blue wavelengths than at the
$I$-band, and therefore these cases are of little practical
interest. Hence, we adopt by default the C/O=1.1 C-star spectra, for
all \Teff. A more elaborated approach will be followed after spectra
for a wider range of C/O values become available (Aringer et al., in
preparation).

Finally, as part of the TRILEGAL project (Girardi et al. 2005), we
have included spectra for DA white dwarfs from Finley et al. (1997)
and Homeier et al. (1998) for \Teff\ between 100\,000 and 5\,000~K,
and $\log g$ between 7 and 9.

The Castelli \& Kurucz (2003) and Allard et al. (2000) tabulated
spectra extend well into the far-IR, until 160 and 971~$\mu$m
respectively. In the other cases, the available spectra were tabulated
just up to the mid-IR, namely until 12.5~$\mu$m for Fluks et
al. (1994) and 2.5~$\mu$m for Loidl et al. (2001). In all cases we
have extended the flux to the very far-IR using a scaled
Rayleigh-Jeans law starting from the point of maximum tabulated
$\lambda$. This is clearly a crude approximation. van der Bliek et
al. (1996), for instance, find that black-body fluxes in the far-IR
present discrepancies of the order of 20 per cent with respect to the
solar flux. Decin \& Eriksson (2007) recently discussed the status of
the predicted spectral energy distributions in the IR, to whom we
refer for a detailed discussion of this topic. It turns out that 
our predicted BC for the mid- and far-IR will be uncertain
by up to a few tenths of a magnitude for the coolest M and C
stars. However, we expect that the errors in colours 
will be somewhat smaller, and they will become irrelevant in stars with
significant mass-loss, which are exactly those with the most
prominent fluxes in the mid- and far-IR. In fact, in these cases the
emission at mid- and far-IR wavelengths is dominated by the circumstellar dust
shells and not by the stellar photosphere (see Sect.~\ref{sec_dust}).

\subsection{Filter sets so far considered}
\label{sec_photsys}

Table~\ref{tab_photsys} presents all the photometric systems we have
considered so far. They include several traditional ground-based
photometric systems (Johnson-Cousins-Glass, Str\"omgren-Crawford,
Washington, BATC), systems belonging to some influential wide-field
surveys (DENIS, 2MASS, OGLE-II, SDSS) and cameras (WFI, UKIDSS), and
then to some of the most successful satellite-based cameras of the
recent past (Tycho, ACS) or still in operation (WFPC2, NICMOS,
Spitzer, AKARI).  Other systems are presently being added to the
list. 

Note that the transformations to some of these photometric systems are
described in other papers (e.g. Girardi et al. 2007 for WFPC2 and ACS;
Vanhollebeke et al. 2007 for OGLE-II); they are listed in
Table~\ref{tab_photsys} just for the sake of completeness and no
further detail on these systems will be provided here. 
All systems in the Table are already available at the web
interface (see Sect.~\ref{sec_datatables}).

\begin{sidewaystable*}
\small
\caption{Photometric systems for which our new isochrones are made
available (updated to November 2007).}
\label{tab_photsys}
\begin{tabular}{llllll}
\hline
Photometric system & Filter names & Central & kind of &
references & additional information \\ & & wavelength range & zero-points
& for filters & \\
\hline
Johnson  & $UBV$ & 3642 -- 5502 \AA  & Vega$^{(\rm a)}$ & Ma\'{\i}z-Apell\'aniz (2006) & supersedes Girardi et al. (2002) \\
Bessell \& Brett & $U_{\rm X}B_{\rm X}BV(RI)_{\rm C}$ & 3627 \AA -- 4.72~$\mu$m & Vega & Bessell (1990) & \\
               & $JHKLL'M$ &                        &  & Bessell \& Brett (1988) & \\
$UBVRIJHK$$^{(\rm b)}$ & $UBV(RI)_{\rm C}JHK$ & 3642 \AA -- 2.15~$\mu$m & Vega & Ma\'{\i}z-Apell\'aniz (2006) & $^{(\rm a)}$ \\
           &            &                         &      & Bessell \& Brett (1988), Bessell (1990) &  \\
Str\"omgren-Crawford & $uvby$ & 3474 -- 5479 \AA & Vega$^{(\rm a)}$ & Ma\'{\i}z-Apell\'aniz (2006) & \\
                     & H$\beta_{\rm w}$H$\beta_{\rm n}$ &                  &      & Moro \& Munari (2000) & $^{(\rm f)}$ \\
Tycho & $V_{\rm T}B_{\rm T}$ & 4250 -- 5303 \AA & Vega$^{(\rm a)}$ & Ma\'{\i}z-Apell\'aniz (2006) & \\
SDSS & $ugriz$ & 3587 -- 8934 \AA & AB$^{(\rm c)}$ & Fukugita et al. (1996) & details in Girardi et al. (2004) \\
OGLE-II & $UBVI$ & 3695 -- 8321 \AA & Vega & OGLE website & details in Vanhollebeke et al. (2007) \\
ESO/WFI & wide filters & 3469 -- 8590 \AA & Vega & ESO/WFI website & supersedes Girardi et al. (2002) \\
Washington & $CMT_1T_2$ & 3982 -- 8078 \AA & Vega & Geisler (1996) & details in Girardi et al. (2002) \\
DENIS & $IJK_{\rm s}$ & 7927 \AA -- 2.15~$\mu$m & Vega & Fouqu\'e et al. (2000) & details in Cioni et al. (2006) \\
2MASS & $JHK_{\rm s}$ & 1.23 -- 2.16~$\mu$m & Vega$^{(\rm e)}$ & Cohen et al. (2003) & supersedes Bonatto et al. (2004) \\
UKIDSS & $ZYJHK_{\rm s}$ & 8823 \AA -- 2.20~$\mu$m & Vega & Hewett et al. (2006) & supersedes files distributed since 2006 \\
BATC & all 15 filters & 3383 -- 9722 \AA & AB & Yan et al. (2000) & supersedes files distributed since 2006 \\
HST/WFPC2 & all wide filters & 1700 -- 9092 \AA & Vega+ST+AB & Holtzman et al. (1995) & details in Girardi et al. (2007) \\
HST/ACS & all WFC+HRC & 2645 -- 9130 \AA & Vega & Sirianni et al. (2005) & details in Girardi et al. (2007) \\
HST/NICMOS & $JHK$-equivalent & 1.11 -- 2.04~$\mu$m & Vega+ST+AB & D. Figer, priv. comm. & details in Girardi et al. (2002) \\
Spitzer & all MIPS+IRAC & 3.52 -- 154~$\mu$m & Vega,AB$^{(\rm d)}$ & Spitzer website & details in Groenewegen (2006) \\
AKARI & all filters &  2.32 -- 160~$\mu$m & Vega,AB$^{(\rm d)}$ & AKARI website & details in Groenewegen (2006) \\
\hline
\end{tabular}
\\
$^{(\rm a)}$ Zero-points are also taken from Ma\'{\i}z-Apell\'aniz
(2006). \\ $^{(\rm b)}$ Combines $UBV$ from Ma\'{\i}z-Apell\'aniz (2006) with
$(RI)_{\rm C}JHK$ from Bessell \& Brett (1988) and Bessell
(1990). This combination is provided as being the most useful for the
analysis of ground-based data.  Supersedes Girardi et al. (2002). \\
$^{(\rm c)}$ We adopt the traditional definition of magnitudes, using Pogson's
formula, and not Lupton et al.'s (1999) one; see Girardi et al. (2004)
for the motivations of this choice. \\ $^{(\rm d)}$ A Vegamag zero-point
reproduces the photometry quite accurately for most IR systems (see
Reach et al. 2005). AB magnitudes, however, may be convenient because
then the magnitudes can be easily converted to the flux in
milli-Jansky by means of a single formula, i.e. $f_\nu\,{\rm (mJy)} =
-0.4\,m_{\rm AB}\,{\rm (mag)} - 6.56$. \\ $^{(\rm e)}$ We use zero-points
determined by Ma{\'{\i}}z Apell{\'a}niz (2007) using Bohlin's (2007)
Vega spectrum. As a consequence, the 2MASS zero-point offsets measured
by Cohen et al. (2003) do not apply. \\ $^{(\rm f)}$ For the H$\beta$ filters,
we simply assume $\beta={\rm H}\beta_{\rm wide}-{\rm H}\beta_{\rm
narrow}=2.873$ for Vega (cf.\ Hauck \& Mermilliod 1998). The
zero-point is arbitrarily set as ${\rm H}\beta_{\rm wide}{\rm
(Vega)}=0$.
\end{sidewaystable*}

\section{Reprocessing by circumstellar dust}
\label{sec_dust}

Circumstellar dust reprocesses a substantial fraction of the photons
emitted by the coolest evolved stars. Since, on one hand, the efficiency
of dust extinction decreases quickly above $\sim 1 \mu$m and, on the
other hand, dust does not survive at temperatures above 1000--2000~K,
the general effect of the presence of dust in an astrophysical source
is to shift part of the intrinsic emitted power from the
optical-ultraviolet to the mid- and far-infrared region.

A serious difficulty in taking into account in detail the effects
of dust is that the optical properties (i.e. the efficiency of
absorption and scattering as a function of wavelength) differ
significantly in different environments. This is particularly true for
the envelopes of AGB stars, which are believed to be one of the two
main producers of interstellar grains, the other being the supernovae
blast waves (e.g. Dwek 1998; Todini \& Ferrara 2001; Morgan \& Edmunds
2003; Calura et al. 2007). As for the dust composition, the general
wisdom is that the envelopes of C-type stars are composed by
carbonaceous grains while those of M-type stars are dominated by
silicates, however several recent detailed studies, mainly based on
Spitzer data (e.g. Lebzelter et al. 2006; Verhoelst et al. 2006) and
in more detailed modelling (e.g. Ireland \& Scholz 2006; Woitke 2006;
Ferrarotti \& Gail 2006) demonstrated that the situation may be more
complex. Moreover, the grain size distribution, which together with
the composition determines the optical properties, is expected to be
significantly different from that inferred in the general interstellar
medium.

A further big difficulty in the understanding of AGB envelopes is 
the occurrence of strong mass loss and the possible relation with
the type of dust that can condensate in their outflows.  The massive
winds in AGB stars are commonly ascribed to the coupled effect between
large amplitude pulsations and radiation pressure on dust
grains. According to detailed dynamical studies, in C-rich stars the
radiation pressure on dust can in fact be strong enough to explain the
observed outflows (see e.g. Winters et al. 2000; H\"ofner et
al. 2003). Instead in O-rich stars the possible type of grains are
either not stable or opaque enough near $\sim 1 \mu$m to explain the
outflows (e.g. Ireland \& Scholz 2006; Woitke 2006). 

Furthermore, recent spectroscopic analyses of oxygen-rich AGB
stars based on ISO data for the LMC (Dijkstra et al. 2005) and the
Galactic Bulge (Blommaert et al. 2006) have shown a clear correlation
between the dust composition and the mass-loss rates. For instance,
the dust features displayed by oxygen-rich stars are mainly ascribed
to Al$_2$O$_3$ for low $\dot{M}$, while going to higher $\dot{M}$ they
are interpreted with an increasing amount of silicates. These findings
are consistent with the latest results from Spitzer data of AGB stars
belonging to the globular cluster 47~Tuc (Lebzelter et al. 2006). The
dust mineralogy actually changes as the star evolves at increasing
luminosities and pulsation periods. On the other hand, for C stars
unveiling the dust-condensation sequence is more problematic due to
its complex dependence on a number of factors (Speck et al. 2006; 
Thompson et al. 2006).
 
What emerges is that, despite the conspicuous progress in both
observational and theoretical grounds, the issue of dust formation and
evolution in AGB stars is still affected by large uncertainties. For
these reasons in this paper we will explore just a few possible kinds
of circumstellar dust, essentially following the prescriptions by
Groenewegen (2006, G06) and Bressan et al. (1998, B98). Alternative
and more complex descriptions will be examined in subsequent papers.

\subsection{Method}
\label{sec_dustmethod}

This section outlines the adopted procedure to include the effect
of dust in our isochrones.
  
In general, for a given star of $(L,\Teff)$, losing
mass at a rate $\dot{M}$, with dust expansion velocity $v_{\rm exp}$
and dust-to-gas ratio $\Psi$, radiation transfer (RT) calculations
provide the dust optical depth $\tau$ and its corresponding
attenuation/emission spectrum. The optical depth $\tau$ scales with
the other quantities via the approximated relation
\begin{equation}
\tau(1\mu{\rm m}) \simeq A_{\rm d}\,\dot{M}\,\Psi\,v_{\rm exp}^{-1} 
L^{-0.5} \,\,\,\,,
\label{eq_scaling}
\end{equation}
(see G06 and references therein) where $A_{\rm d}$ is a quantity that is set 
mainly by the dust composition. 

For practical convenience
we have pre-computed a number of reference RT models
which are parametrized as a function of a basic set of parameters, namely:
$\tau(1\mu{\rm m})$, \Teff, $L$, dust condensation temperature $T_{\rm c}$,  
and for different dust mixtures, as 
specified  in  Sects.~\ref{sssec_g06}  and \ref{sssec_b98} 
for G06 and B98 models, respectively.
Then, each star along the isochrone is assigned its proper dusty 
envelope by using 
Eq.~(\ref{eq_scaling}) with the corresponding values of 
$\dot{M}$, $\Psi$, $v_{\rm exp}$, $L$, $T_{\rm eff}$, and $A_{\rm d}$.

Each  determination of  $\tau(1\mu{\rm m})$ corresponds to specific dust
attenuation and emission functions, which in turn fix the ratio
between the flux derived from RT calculations
and the dust-free intrinsic flux, i.e. $F_{\lambda}/F_{\lambda, 0}$.
Finally, bolometric corrections
are derived, i.e. 
$\Delta{\rm BC}_{\lambda}=-2.5\log(F_{\lambda}/F_{\lambda, 0})$.

In summary, for each AGB star along the isochrone, 
the general scheme proceeds through the following steps:
\begin{itemize}
\item We first distinguish between O- and C-rich envelopes, using the 
photospheric C/O ratio. The dust composition is selected among those
for which RT calculations have been performed.
 \item $v_{\rm exp}$ and $\Psi$ are computed as a function of $L$, \Teff, 
C/O and $\dot{M}$, using the formalism described in
Sects.~\ref{ssec_vexp} and \ref{ssec_dtog} below.
\item The proper $\tau$ is calculated via 
Eq.~(\ref{eq_scaling}).
\item The corresponding $\Delta{\rm BC}_\lambda$ relation is 
interpolated at the effective wavelengths $\lambda_{\rm eff}$ 
of all filters considered in our
isochrones. The resulting $\Delta{\rm BC}_{\lambda_{\rm eff}}$ 
values are added to the
absolute magnitudes computed previously via Eq.~(\ref{eq_bcfinal}).
\end{itemize}

\subsection{Expansion velocity}
\label{ssec_vexp}

The wind terminal velocity or expansion velocity, $v_{\rm exp}$,
is calculated according to the formalism presented by Elitzur
\& Ivezi\'c (2001), to which the reader should refer for the
details of the underlying theoretical analysis.  In brief, following
Elitzur \& Ivezi\'c (2001) the dust-wind problem is solved by means of
a suitable scaling approach, so as it is possible to express
$v_{\rm exp}$ as a function of mass-loss rate, stellar luminosity and
various dust parameters (dust-to-gas ratio, chemical composition,
condensation temperature, etc.) in the form
\begin{equation}
v_{\rm exp}=\displaystyle\left(A \dot{M}_{-6}\right)^{{1}/{3}}
\left(1+B\displaystyle\frac{\dot{M}_{-6}^{{4}/{3}}}{L_4}\right)^
{-{1}/{2}}
\label{eq_vexp}
\end{equation}
where $v_{\rm exp}$ is expressed in km s$^{-1}$; $\dot{M}_{-6}$ is the
mass-loss rate in units of $10^{-6} M_{\odot}$~yr$^{-1}$; $L_{4}$ is
the stellar luminosity in units of $10^{4}\, L_{\odot}$.

The dependence of $v_{\rm exp}$ on the dust-to-gas ratio is included
in the definition of the parameters $A$ and  $B$ (through the quantity
$\sigma_{22}$, see below):
\begin{equation}
A=3.08\:10^{5} T_{\rm c3}^4\, Q_{\star}\, \sigma_{22}^2\, \chi_{0}^{-1}
\end{equation}
\begin{equation}
B=\left[2.28\frac{Q_{\star}^{1/2}\,\chi_{0}^{1/4}}{Q_V^{3/4}\,
\sigma_{22}^{1/2}\,T_{\rm c3}}\right]^{-4/3}
\end{equation}
which in turn depend on several quantities quoted here below.  The
condensation temperature in units of $10^3$~K is denoted with $T_{\rm
c3}$.  The parameter $Q_{\star}$ is the Planck average of the
efficiency coefficient for radiation pressure, evaluated at the
effective temperature, while $Q_V$ is the efficiency factor for
absorption at visual.  The quantity $\chi_{0}$ is the ratio:
\begin{equation}
\chi_{0}=\displaystyle\frac{Q_{\rm P}(T_{\rm eff})}{Q_{\rm P}(T_{\rm c})},
\end{equation}
where  $Q_{\rm P}(T)$ is the Planck average of the absorption
efficiency at the temperature $T$.

We define $\sigma_{22}=\sigma_{\rm gas}/(10^{-22}\,\,{\rm cm}^{2})$,
where $\sigma_{\rm gas}$ (in cm$^{2}$)
 is the dust cross-sectional area per gas particle
at condensation
\begin{equation}
\sigma_{\rm gas} = \pi a^2 \displaystyle\frac{n_{\rm dust}}{n_{\rm gas}}
\,\,\,\,\,.
\label{eq_sigmag}
\end{equation}
In the above equation $a$ is the mean size (in cm) of the dust grains,
$n_{\rm dust}$ and ${n_{\rm gas}}$ are the number densities (in
cm$^{-3}$) of dust and gas particles, respectively.
Similarly to Elitzur \& Ivezi\'c (2001), here it is assumed that all
dust grains have the same size.

The dust-to-gas mass ratio is defined as
\begin{equation}
\Psi = \frac{\rho_{\rm dust}}{\rho_{\rm gas}}
\label{eq_psi}
\end{equation}
where $\rho_{\rm dust}$ and $\rho_{\rm gas}$ correspond to the
density (in g cm$^{-3}$) of the matter in the form of dust and gas,
respectively.

Assuming that the dust grains have a
specific density $\rho_{\rm grain}$ (in g cm$^{-3}$)
the explicit form of the dust density reads
\begin{equation}
\rho_{\rm dust} = \displaystyle\frac{4}{3}\pi a^3\rho_{\rm grain}\,n_{\rm dust}\,\,\,\,\,.
\label{eq_rhod}
\end{equation}
With the approximation that all the gas is made up of hydrogen and
helium, with abundances (in mass fraction) $X_{\rm H}$ and $X_{\rm
  He}$ respectively, then
\begin{equation}
\rho_{\rm gas} = A_{\rm gas} m_{\rm H}   n_{\rm gas}
\label{eq_rhog}
\end{equation}
where  $A_{\rm gas}\simeq 4/(4 X_{\rm H} + X_{\rm He})$
is the mean molecular weight of the gas in units of the mass
of the H atom, $m_{\rm H}=1.674\, 10^{-24}$~g, and
$n_{\rm gas}$ is the number density of the gas particles
(in cm$^{-3}$).

Finally, inserting Eqs.~(\ref{eq_rhod}) and (\ref{eq_rhog}) into
Eq.~(\ref{eq_psi}), we can re-write Eq.~(\ref{eq_sigmag}) to show
the dependence on the dust-to-gas ratio
\begin{equation}
\sigma_{\rm gas} = \displaystyle\frac{3}{4}
\frac{A_{\rm gas}\, m_{\rm H}}{a\, \rho_{\rm grain}}\,\Psi\,
\end{equation}
It is worth noting that in the Elitzur \& Ivezi\'c (2001) formalism the
dust-to-gas mass ratio is indeed a free parameter. In
Sect.~\ref{ssec_dtog} we detail our scheme to specify $\Psi$ as a
function of basic stellar parameters like $\dot{M}$ and the C/O ratio.

Moreover, given the dust chemical type (either amorphous carbon or
silicates) in Elitzur \& Ivezi\'c (2001) the quantities $Q_{\rm P}$
and $Q_{\star}$ are assigned fixed values corresponding to a
particular choice of the dust condensation temperature ($T_{\rm
c}=800$ K) and the stellar effective temperature ($T_{\rm eff} = 2500$
K).  We have carried out additional runs of DUSTY (Ivezi\'c et
al. 1999) in order to evaluate $Q_{\rm P}$ and $Q_{\star}$ for various
values of $T_{\rm c}$ and $T_{\rm eff}$, and for different dust
compositions.  The results are presented in Table~\ref{tab_dustpar}
and are used in our calculations by interpolating in
\Teff\ and $T_{\rm c}$ and allowing for different dust mixtures. In the
present paper we adopt $T_{\rm c}=1000$~K when C/O$<1$ and $T_{\rm
c}=1500$~K when C/O$>1$.  The grain size is set $a=0.1\,\,\mu$m in all cases.
The other input parameters, i.e.
effective temperature and dust chemical type, are just those
predicted by our stellar isochrones.

It should be also mentioned that Eq.(~\ref{eq_vexp}) is valid as long
as  dust condensation is  efficient enough to drive an
outflow. As discussed in Elitzur \& Ivezi\'c (2001) 
this condition corresponds  to the minimum mass-loss rate:
\begin{equation}
\dot{M}_{\rm min} = 3\, 10^{-9} \displaystyle\frac{M^2}
{Q_{\star}\sigma_{22}^2 L_4 T^{1/2}_{k3}}\quad\quad\quad\quad\quad\quad
(M_{\odot}\,{\rm yr}^{-1})
\label{eq_mlmin}
\end{equation}
where $T_{\rm k3}$ is the kinetic temperature in units of $10^3$~K at
the inner boundary of the dust shell. For simplicity we assume that
the kinetic temperature $T_{\rm k}$ coincides with the condensation
temperature $T_{\rm c}$ of the dust.  
Interestingly we have verified that $\dot{M}_{\rm min}$ given
by Eq.~(\ref{eq_mlmin}) is in good agreement with the results from
hydrodynamical wind calculations, i.e. $\dot{M}_{\rm min} \approx 3-5\,
10^{-7}$ $M_{\odot}$ yr$^{-1}$ as predicted by Wachter et al. (2002) 
for C-stars.

For $\dot{M} < \dot{M}_{\rm min}$, while dust is not a driving mechanism 
to produce a wind, it may be still formed passively 
in the circumstellar ejecta.
To handle this case we have simply evaluated  
the expansion velocity setting $\dot{M}=\dot{M}_{\rm min}$ in the
corresponding Eq.~(\ref{eq_vexp}). In this way $v_{\rm exp}$ is seen to 
decline smoothly with the luminosity to the lowest measured values and,
perhaps more importantly, by using 
$v_{\rm exp}=v_{\rm exp}(\dot{M}_{\rm min})$ 
together with the actual $\dot{M}$ in Eq.~(\ref{eq_scaling}), 
we can deal with the very optically-thin circumstellar envelopes 
($\tau(1\mu{\rm m}) \la 10^{-2} - 10^{-3}$) 
expected at low $\dot{M}$.

One important feature of Eq.(~\ref{eq_vexp}) is that it naturally
predicts a maximum value for the expansion velocity, which is attained
at large $\dot{M}$ as a consequence of the weakening of radiative
coupling in optically thick winds: $v_{\rm max} \simeq v_{\rm
exp}(\tau_V = 1.3)$.  This aspect may bring along significant
implications for the interpretation of the observed data
(Sect.~\ref{ssec_obsvexp}; see Fig.~\ref{fig_vexp}).

\begin{table*}
\caption{Characteristic quantities used in the dust models,
following the formalism developed by Elitzur \& Ivezi\'c (2001).  They
are defined in Sect.~\ref{ssec_vexp}.}
\begin{tabular}{cccclllrl}
\hline
dust composition & $Q_{V}$ & $\rho_{\rm grain}$ (gr cm$^{-3}$) & a
($\mu$m) &
$T_{\rm eff}$ & $Q_{\star}(T_{\rm eff})$ & $Q_{P}(T_{\rm eff})$
& $T_{\rm c}$ &
$ Q_{P}(T_{\rm c})$\\
\hline
100\% AlOx &  0.370  & 3.2 & 0.1 & 2000 & 0.0189 &  0.0087  & 1500 & 0.00811 \\
           &         &     &     & 2500 & 0.0344 &  0.0105  & 1200 & 0.00897 \\
           &         &     &     & 3000 & 0.0599 &  0.0128  & 1000 & 0.0107  \\
           &         &     &     & 3500 & 0.0958 &  0.0157  &  800 & 0.0145 \\
\hline
100\% Silicate & 0.779 & 3.0 & 0.1 & 2000 & 0.0688 &  0.0531 & 1500 & 0.0396 \\
           &         &     &       & 2500 & 0.106  &  0.0696 & 1200 & 0.0333 \\
           &         &     &       & 3000 & 0.159  &  0.0884 & 1000 & 0.0303 \\
           &         &     &       & 3500 & 0.231  &  0.110  &  800 & 0.0291 \\
\hline
100\% Silicon Carbide & 5.17 & 3.2 & 0.1 &
                                 2000 & 0.148  &  0.0825  & 1500 & 0.0576 \\
           &         &     &   & 2500 & 0.293  &  0.120   & 1200 & 0.0438 \\
           &         &     &   & 3000 & 0.527  &  0.176   & 1000 & 0.0348 \\
           &         &     &   & 3500 & 0.847  &  0.249   &  800 & 0.0264 \\
\hline
100\% AMC &   2.53  & 1.9 & 0.1 &   2000 & 0.170 &  0.132 & 1500 & 0.0955 \\
           &         &     &    &   2500 & 0.274 &  0.180 & 1200 & 0.0802 \\
           &         &     &    &   3000 & 0.425 & 0.242  & 1000 & 0.0730 \\
           &         &     &    &   3500 & 0.615 & 0.313  &  800 & 0.0686 \\
\hline
\end{tabular}
\label{tab_dustpar}
\end{table*}

\subsection{Dust-to-gas mass ratio}
\label{ssec_dtog}

Under  the assumptions of steady-state,
spherically symmetric, radial outflow with complete coupling 
between dust and gas (i.e. both components sharing the same velocity),
from the mass-conservation law $\dot M = 4\, \pi\, r^2\,\rho\,v_{\rm exp}$,
the dust-to-gas mass ratio
$\Psi$ -- defined by Eq.~(\ref{eq_psi}) -- can be expressed
\begin{equation}
\Psi=\frac{\dot{M}_{\rm dust}}{\dot{M}_{\rm gas}}
\end{equation}
where $\dot{M}_{\rm dust}$ and $\dot{M}_{\rm gas}=\dot{M}-\dot{M}_{\rm
dust}$ denote the mass-loss rates in the form of dust and gas,
respectively. For any given combination of $\dot M$ and C/O,
$\dot{M}_{\rm dust}$ is calculated on the basis of the formalism
presented by Ferrarotti (2003) and Ferrarotti \& Gail (2006, and
references therein), to whom we refer for many details. The general
scheme is the following:
\begin{equation}
\dot{M}_{\rm dust} = \sum_i \dot{M}_{{\rm dust},i}
\end{equation}
where the summation includes the contribution of various dust species,
each one being expressed
\begin{equation}
\dot{M}_{{\rm dust},i} = \dot{M}\,X_{\rm seed}
	\left( \frac{A_{{\rm dust},i}}{A_{\rm seed}} \right)
	f_{{\rm dust},i}
	\,\,\,,
\label{eq_fdust}
\end{equation}
where $X_{\rm seed}$ is the total mass fraction of the
seed\footnote{Following Ferrarotti \& Gail (2006), with seed or key
element we denote the least abundant element among those required to form
the dust species under consideration; e.g. Si is the seed element of
the silicate compounds in O-rich stars.} element in the circumstellar
envelope, $A_{\rm seed}$ denotes its atomic weight, whereas $A_{{\rm
dust},i}$ corresponds to the mean molecular weight of the dust
species.  The adopted values are taken from Ferrarotti
\& Gail (2002).  The quantity $f_{{\rm dust},i}$ is the
condensation degree of the dust species, being defined as the fraction
of the seed element condensed into dust grains.  In our calculations
the condensation degrees are estimated as a function of the total
mass-loss rate and C/O ratio by means of the the analytic fits
provided in Ferrarotti (2003).

The types of dust species that may condensate in the AGB circumstellar
envelopes depend critically on the available abundances of the main
dust forming elements, i.e. C, O, Mg, Si, S, and Fe.  We recall that
in the present calculations the chemical mixture of metals at the ZAMS
is assumed to be scaled solar according to the Grevesse \& Noels
(1993) compilation. While the surface abundances of C and O may change
in the course of the stellar evolution (mainly due to dredge-up
episodes and HBB), those of Mg, Si, S, and Fe remain practically
unaltered.

Following the results of non-equilibrium dust-formation calculations
carried out by Ferrarotti \& Gail (2006), it is useful to define two
critical carbon abundances, namely $Y_{\rm C,1}= Y_{\rm O} - 2 Y_{\rm
Si}$ and $Y_{\rm C,2}=Y_{\rm O} - Y_{\rm Si}+Y_{\rm S}$, where $Y=X/A$
is the abundance in mol$\,{\rm g}^{-1}$.  They define sharp boundaries
between three different domains of dust compounds, as described in the
following.

Adopting the solar O abundance of Grevesse \& Noels (1993), and under
the assumption that the surface abundance of O is not affected during
the evolution, then $Y_{\rm C,1}$ and $Y_{\rm C,2}$ define two
critical C/O ratios, namely (C/O)$_1=0.90$ and (C/O)$_2=0.97$.  
The invariance of O is a common assumption, since
standard predictions of detailed AGB nucleosynthesis indicate that the
material dredged-up at thermal pulses comprises only a small amount of
oxygen, i.e. $X_{\rm O} \sim 0.5-4 \% $ (Boothroyd \& Sackmann 1988;
Izzard et al. 2004). However one should consider that this result is
highly dependent on the treatment of convective boundaries (see
e.g. Herwig 2000 for high O inter-shell abundances); in addition the
surface abundance of O may be even depleted as a consequence of
efficient HBB in low-metallicity high-mass AGB stars (Ventura et
al. 2002).

Since in our calculations the oxygen variations are negligible in most
cases, we adopt the initial O abundance to define three main regimes
of dust composition as a function of (C/O)$_1$ and (C/O)$_2$, which
broadly correspond to the three main spectral classes of AGB stars (M,
S, C).  Details are given in Sects.~\ref{sssec_dustM} to
\ref{sssec_dustC}.

When applied to our isochrones, the formalism just outlined produces
$\Psi$ values typically up to 0.004, 0.002, 0.0001 for O-rich stars,  
and up to 0.01, 0.02, 0.02 for C-rich stars,
with  $Z=0.019, 0.008, 0.001$, respectively. These are  
maximum predicted values, and they are not meant to be representative of 
AGB stars at varying $Z$. Detailed population synthesis models are
needed to explore this issue, which is postponed to future analysis.
  
Anyhow, from comparing the upper values given above we derive
some interesting trends.  
While for O-rich stars 
we find a direct proportionality between $\Psi$ and $Z$, for 
C-rich stars the relation becomes more complex and non-monotonic. 
In the case of O-rich envelopes the relation $\psi \propto Z$ is
explained by considering that 
the maximum amount of dust
that can be formed is controlled by the abundances of 
the seed elements (i.e. Si, Fe), which scale with the initial 
metallicity.
 
In the case of C-rich envelopes one key parameter in the formation of dust
is the excess of carbon with respect to oxygen, and this quantity 
does not trace the initial metallicity as it is crucially affected by
the third dredge-up and HBB during the TP-AGB phase. 
It follows that the large $\Psi$ values predicted for C-rich stars even at 
low $Z$  reflect their relatively
high C/O ratios as predicted by our TP-AGB models at decreasing metallicity.

It is worth noticing that these theoretical expectations are in line
with the observational evidence presented in van Loon (2000) and
Marshall et al. (2004) for the proportionality of the dust-to-gas ratio
on $Z$ in oxygen-rich stars; and van Loon et al. (1999) for the enhanced
abundance of free carbon available for molecule and dust formation in
metal-poor carbon stars. More observational data on this issue has been
recently provided by e.g. 
Speck et al. 2006, Thompson et al. 2006, 
Sloan et al. 2006, Zijlstra et al. 2006, Lagadec et al. 2007, 
Matsuura et al. 2007.

\subsubsection{Stars with C/O $< 0.90$}
\label{sssec_dustM}
In this interval there are sufficient oxygen atoms to allow the
formation of the silicate-type dust. These conditions roughly
correspond to the domain of M stars.  Two main dust
contributions are considered:
\begin{equation}
\frac{d{M}_{\rm dust}}{dt} = \displaystyle
\frac{d{M}_{\rm sil,M}}{dt}+\frac{d{M}_{\rm iro,M}}{dt}
\end{equation}
that correspond to silicate dust and iron dust, respectively.  These
can be explicited
\begin{eqnarray}
\displaystyle\frac{d{M}_{\rm sil,M}}{dt} & = & \dot{M} X_{\rm Si}
\frac{A_{\rm sil}}{A_{\rm Si}} f_{\rm sil} \\
\displaystyle\frac{d{M}_{\rm iro,M}}{dt} & = & \dot{M} X_{\rm Fe}
\frac{A_{\rm iro}}{A_{\rm Fe}} f_{\rm iro}
\end{eqnarray}
where $A_{\rm Si}$ ($A_{\rm Fe}$) is the atomic weight of Si (Fe),
$A_{\rm Si}$ ($A_{\rm iro}$) is the effective molecular weight and
$f_{\rm sil}$ ($f_{\rm iro}$) is the condensation degree of the
silicate (iron) dust.

The condensation degree of silicate dust, $f_{\rm sil}$ accounts for
the possible contributions of olivine-type dust, pyroxene-type dust,
and quartz-type dust:
\begin{eqnarray}
f_{\rm sil}=f_{\rm ol}+f_{\rm py}+f_{\rm qu}
\end{eqnarray}
with clear meaning of the symbols.
The effective molecular weight of the silicate dust mixture is expressed
\begin{eqnarray}
A_{\rm sil}=(f_{\rm ol}\,A_{\rm ol}+f_{\rm py}\,A_{\rm py}+f_{\rm qu}\,A_{\rm qu})/ f_{\rm sil}
\end{eqnarray}

From the analysis of Ferrarotti \& Gail (2001), the amount of silicate
dust in form of olivine, pyroxene and quartz depends critically on the
$Y_{\rm Mg}/Y_{\rm Si}$ ratio, which is $\sim 1.07$ according to the
chemical mixture adopted in this work (Grevesse \& Noels 1993).  At
this value the non-equilibrium dust condensation calculations by
Ferrarotti \& Gail (2001) predict that the dust mixture is mainly
composed by olivine and pyroxene, with a small fraction of quartz.
Following the results for an M-star model with $\dot M= 10^{-5}\,
M_{\odot}$ yr$^{-1}$ and $Y_{\rm Mg}/Y_{\rm Si} \simeq 1.07$, the
condensation degrees are assumed to obey the ratios $f_{\rm ol}/f_{\rm
py}=4$ and $f_{\rm ol}/f_{\rm qu}=22$.

Finally, $f_{\rm sil}$ and $f_{\rm iro}$ are evaluated with the
analytic fits provided by Ferrarotti (2003):
\begin{eqnarray}
\displaystyle f_{\rm sil} & = & 0.8\,\displaystyle
\frac{\dot{M}_{-6}}{\dot{M}_{-6}+5} \, \displaystyle
\sqrt{\frac{Y_{\rm C,1}-Y_{\rm C}}{Y_{\rm C,1}}}\\
\displaystyle f_{\rm iro} & = & 0.5\,\displaystyle
\frac{\dot{M}_{-6}}{\dot{M}_{-6}+5}
\end{eqnarray}

\subsubsection{Stars with $0.90\le$ C/O $ < 0.97$}
\label{sssec_dustS}
In this interval a small amount of silicates can be produced because of
the deficit in O and the dust composition is iron-dominated.  These
conditions may roughly include the class of S stars.
Two main dust contributions are considered:

\begin{equation}
\frac{d{M}_{\rm dust}}{dt} = \displaystyle
\frac{d{M}_{\rm sil,M}}{dt}+\frac{d{M}_{\rm iro,M}}{dt}
\end{equation}
that correspond to silicate dust and iron dust, respectively.
These can be explicited
\begin{eqnarray}
\displaystyle\frac{d{M}_{\rm sil,M}}{dt} & = & \dot{M} X_{\rm Si}
\frac{A_{\rm sil}}{A_{\rm Si}} f_{\rm sil} \\
\displaystyle\frac{d{M}_{\rm iro,M}}{dt} & = & \dot{M} X_{\rm Fe}
\frac{A_{\rm iro}}{A_{\rm Fe}} f_{\rm iro}
\end{eqnarray}
with the same meaning of the symbols as in Sect.~\ref{sssec_dustM}.

Following Ferrarotti (2003) the condensation degrees are approximated with
\begin{eqnarray}
\displaystyle f_{\rm sil,S} & = & 0.1\,\displaystyle
\frac{\dot{M}_{-6}}{\dot{M}_{-6}+5} \\
\displaystyle f_{\rm iro,S} & = & 0.5\,\displaystyle
\frac{\dot{M}_{-6}}{\dot{M}_{-6}+5}
\end{eqnarray}

We note that for given $\dot{M}$ the total condensation degree
of silicate dust is lower than in the case of M stars (because of the
lower abundance of available O).  The relative ratios between $f_{\rm
ol}$, $f_{\rm py}$, and $f_{\rm qu}$ are assumed the same as those for
M-stars, which is a reasonable assumption (see Ferrarotti \& Gail
2002).

\subsubsection{Stars with C/O $\ge 0.97$}
\label{sssec_dustC}
In this interval the dust mixture becomes carbon-dominated.
These conditions apply to the class of C stars.  Two dust
contributions are included:

\begin{equation}
\frac{d{M}_{\rm dust}}{dt} = \displaystyle
\frac{d{M}_{\rm sic,C}}{dt}+\frac{d{M}_{\rm car,C}}{dt}
\end{equation}
that correspond to silicon carbide dust and solid
carbon dust, respectively. These can be explicited
\begin{eqnarray}
\displaystyle\frac{d{M}_{\rm sic,C}}{dt} & = & \dot{M} X_{\rm Si}
\frac{A_{\rm sic}}{A_{\rm Si}} f_{\rm sic} \\
\displaystyle\frac{d{M}_{\rm car,C}}{dt} & = & \dot{M} X_{\rm C} f_{\rm car}
\end{eqnarray}
$X_{\rm Si}$, and $X_{\rm C}$ represent the element abundances (in
mass fraction), which are given by the initial values on the main
sequence for Fe and Si, while the current atmospheric value is adopted
for C.  The quantities $A_{\rm Si}$ and $A_{\rm sic}$ are the atomic
weight of Si, and the effective molecular weight of the silicate dust
mixture, respectively.

Finally, $f_{\rm sic}$, and $f_{\rm car}$ denote the degree of
condensation of the elements Si and C in the corresponding dust
species. Following Ferrarotti (2003) they are evaluated with
\begin{eqnarray}
\displaystyle f_{\rm sic} & = & 0.5\,\displaystyle
\frac{\dot{M}_{-6}}{\dot{M}_{-6}+5}\\
\displaystyle f_{\rm car} & = & 0.5\,\displaystyle
\frac{\dot{M}_{-6}}{\dot{M}_{-6}+5}\, \left(\frac{\rm C}{\rm O}-1\right)
\end{eqnarray}

\subsection{Radiative transfer calculations}
\label{sec_radiative}

Bolometric corrections are provided from two independent 
RT calculations, which are updates
and extensions of those originally described by B98 and G06. In both
cases the models have been calculated with 1-dimensional codes that
solve the RT equation and the thermal balance equation in a
self-consistent way (see Granato \& Danese 1994; Groenewegen 1993, and
Groenewegen 1995).

Although they differ in several minor details which are related to the
structure of the original codes -- as described in Sects.~\ref{sssec_g06}
and \ref{sssec_b98} --  B98 and G06 calculations 
produce very similar results, which proves
the robustness of both RT models.   
For instance, whereas B98 tabulate
the complete output $F_\lambda$ spectra with a regular spacing in
$\log\lambda$ (here set to be $\Delta\log\lambda=0.04$~dex), G06 code
provides directly the integrated $F_{\lambda_{\rm eff}}$ for a given
set of input pass-bands. In fact, we
have estimated that the differences in $\Delta{\rm BC}_{\lambda_{\rm eff}}$ 
coming from these two different
approaches, for the broad-band filters we are using, are of the order
of just $\sim0.05$~mag for $\tau(1\mu{\rm m})=1$. Considering the
overall uncertainties, these differences are small enough to be
safely ignored.

Much more important, instead, are the differences associated with the
assumed dust compositions, and the coverage of stellar
parameters, as detailed below.

\subsubsection{Extended G06 calculations}
\label{sssec_g06}

The dust models that are incorporated in the isochrones are an
extension of the work described in G06. The models have been
calculated for (arbitrary) values of $L = 3000 L_{\odot}$, $d =
8.5$~kpc, $v_{\rm exp} = 10\,{\rm km\,s}^{-1}$, $\Psi=0.005$.  In G06,
only few effective temperatures were considered. Here we calculated
models for spectral types M0,1,$\ldots$,10 ($T_{\rm eff}$ = 3850,
3750, 3650, 3550, 3490, 3397, 3297, 3129, 2890, 2667, 2500~K) from
Fluks et al. (1994) for O-rich stars, and $T_{\rm eff}$ = 3400, 3200 ,
3000, 2800, 2650 K from Loidl et al. (2001) for C-rich stars (with a
C/O ratio of 1.1). The O- and C- stellar input spectra are for solar
metallicities.

Several types of dust are considered (see G06 for a discussion). For
O-rich stars, (1) 100\% Aluminium Oxide (AlOx; amorphous porous
Al$_2$O$_3$), with optical constants from Begemann et al. (1997), (2) a
combination of 60\% AlOx and 40\% Silicate (optical constants from
David \& P\'egouri\'e 1995), (3) 100\% silicate.
A condensation temperature of $T_{\rm c} = 1500$~K was assumed in the
first two cases, 1000~K for the pure silicate dust.

For C-stars, we consider either a combination of 85\% Amorphous Carbon
(AMC) and 15\% Silicon Carbide (SiC) with optical constants from,
respectively, Rouleau \& Martin (1991) for the AC1 species and
$\alpha$-SiC from P\'{e}gouri\'{e} (1988), and assuming a $T_{\rm c}$
of 1200~K, or 100\% AMC with a $T_{\rm c}$ of 1000~K.

\begin{figure}
\resizebox{\hsize}{!}{\includegraphics{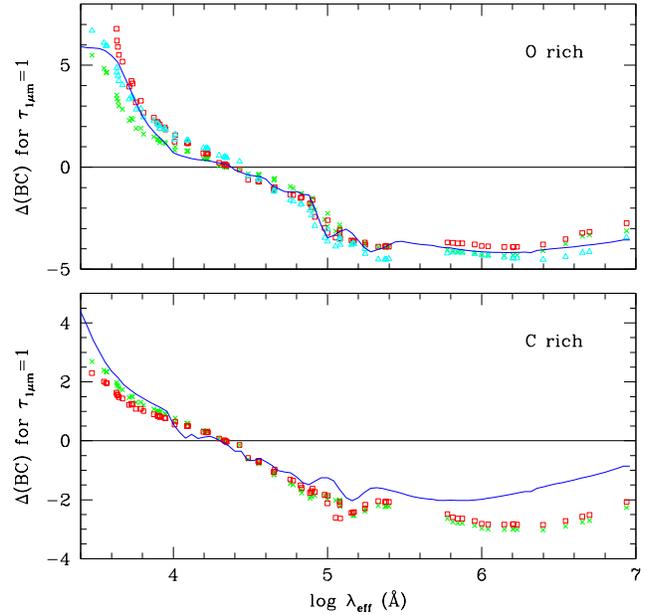}}
\caption{The absorption/emission properties of the dust mixtures
considered in this work, in the form of $\Delta{\rm BC}$ values and
limited to the case of $\tau_{1\mu {\rm m}}=1$ and input stellar
spectra of $\Teff=3000$~K.  {\em Top panel:} For O-rich stars, the
dust compositions are 100\% AlOx (crosses), 100\% Silicates 
(triangles), 60\% AlOx + 40\%
Silicate (squares) following the G06 approach, and Silicates
(continuous line) following B98.  {\em Bottom panel:} For C-rich
stars, the dust compositions are 100\% AMC (crosses) and 85\% AMC +
15\% SiC (squares) following the G06 approach, and Graphites
(continuous line) following B98. }
\label{fig_groenlambda}
\end{figure}

The results of such computations consist in the apparent stellar
magnitudes for several values of $\dot{M}$ (and hence $\tau$) ranging
from $10^{-10}$ to $4\times10^{-5}\,\Msun\,{\rm yr}^{-1}$, and for
each spectral type and filter under consideration. Over this range of
mass loss $\tau$ increases up to $\sim\!3$ for AMC (at 11.33~$\mu$m)
and $\sim\!45$ for AlOx (at 11.75~$\mu$m). A total of 90 filters were
considered in these specific computations. These magnitude tables were
converted into $\Delta{\rm BC}(\tau)$ values that directly give the
difference between the magnitudes of non-dusty ($\tau=0$) and dusty
stars\footnote{The use of $\Delta{\rm BC}(\tau)$ ensures that any
residual difference in the zero-points between the G06 magnitudes, and
those used in our new isochrones, will be erased.}.

\subsubsection{B98 calculations}
\label{sssec_b98}

In B98 for each dust type (either O-rich or C-rich)
the reference grid of AGB dusty envelopes consists of
models calculated for $L=10000$, $T_{\rm eff}=3000 K$ 
while varying the optical depth $\tau(1\mu{\rm m})$ from $5 \,
10^{-4}$ to $200$
in 50 equal logarithmic steps. The
condensation temperature is  assumed $T_{\rm c} = 1000$ K for O-rich
envelopes, and $T_{\rm c} = 1500$ K for C-rich envelopes.
Additional tests have shown that,
 for the adopted dust mixtures, the resulting spectral shapes
are little affected by changes in $L$ and $T_{\rm eff}$ within the
typical ranges for AGB stars, for fixed values of  $\tau(1\mu{\rm m})$.

The radiative transfer of the photospheric stellar spectrum
through the dust is computed with the Granato \& Danese (1994) code
(originally built for AGN tori), assuming spherical symmetry and the
radial dust density distribution decreasing as $r^{-2}$. The adopted
dust grain model is that of Rowan-Robinson et al. (1986)
\footnote{The dust model by Rowan-Robinson et al. (1986)
 is based on observations of circumstellar dust
shells around late-type C- and M-type stars, that appear
to require absorption efficiencies decreasing as $\lambda^{-1}$
in the far-IR.
This wavelength dependence is ascribed to grains with a disordered, damaged
or amorphous structure.
See the original paper for more details.},
which
considers six different kinds of grains: AMC of radius $a=0.1\mu{\rm
m}$ and graphites of $0.03$ and $0.01\mu{\rm m}$ for C-rich mixtures,
and amorphous silicate of $a=0.1\mu{\rm m}$ and silicates of $0.03$
and $0.01\mu{\rm m}$ for O-rich mixtures.

Although we updated in several aspects the B98 model, in particular in the
dust grain properties, the detailed exploration with Spitzer data of these
updates is still underway and will be presented in a forthcoming paper.
Therefore here we adopt the already well established B98 model in its
original
form. The dust composition we consider is carbon for C-type mass-losing
stars, and silicates for M-type ones.

The relation between
$\tau(1\mu{\rm m})$ and the mass loss is given by Eq.~(\ref{eq_scaling})
with $A_{\rm d}=1.5\times10^{11}$ for silicates and 
$A_{\rm d}=6.6\times10^{11}$ for
the carbonaceous mix (with $v_{\rm exp}$ in ${\rm km}\,{\rm s}^-1$,
$L$ in solar units). 
The quantity $A_{\rm d}$ is slightly affected by the spectrum of the
illuminating star. 
However, B98 have checked  that the errors in
$\tau(1\mu{\rm m})$ estimated from Eq.~(\ref{eq_scaling}) 
with the above values 
of $A_{\rm d}$ are kept within 5\% for spectra with $T_{\rm eff}$
in the range 2500-4000~K. This is a fair approximation because
models with $\tau(1\mu{\rm m})$ differing by less than 5-10\% produce almost
identical spectra.

\subsubsection{Overview of RT models}
Figure~\ref{fig_groenlambda} illustrates the behaviour of $\Delta{\rm
BC}$ as a function of wavelength, for the dust compositions considered
in this work, and for the specific case of $\tau(1\mu{\rm m})=1$ and
$\Teff=3000$~K. It can be noticed that all curves agree very well in
the case of O-rich mixtures, and differ significantly only for the
C-rich mixtures at $\lambda>10\mu{\rm m}$. These differences become
higher for higher values of $\tau$, and can be ascribed to the use of
different dust mixtures and spectral libraries.

In summary, we consider several possible choices of dust
composition, namely: four for O-rich envelopes and 
 three for C-rich envelopes. Alternatively,
we can ignore dust setting $\tau=0$. All these cases are allowed in
our isochrone calculations.

\section{A preliminary comparison with observations}
\label{sec_obs}

Gi00 isochrones have been extensively used in the literature for the
study of resolved stellar populations. Most of their use has regarded
the optical photometry of stars in the most populated evolutionary
phases like the main sequence, sub-giant branch, RGB, red clump and
early-AGB. Now the addition of the detailed TP-AGB evolution together
with the infrared photometry open new perspectives for the application
of these new stellar models, 
that we briefly illustrate in the following via comparisons
with Galactic Disk and Magellanic Cloud data.

Most of the comparisons are intended to illustrate the locus of our
new isochrones in different diagrams, and do not consider crucial
aspects of the host galaxies such as their age--metallicity relations
and star formation histories. A detailed comparison between present
models and the observed star counts in several parts of these
diagrams, is left to future papers.

\begin{figure}
        \resizebox{\hsize}{!}{\includegraphics{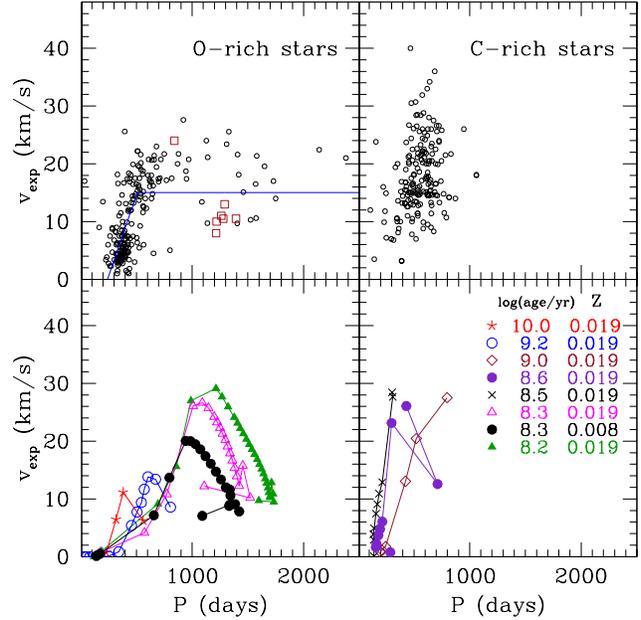}}
\caption{Expansion velocities of circumstellar flows
against pulsation periods of variable AGB stars, for both O-rich (left
panels) and C-rich objects (right panels).  Upper panels: Galactic
observations (empty circles) are taken from the compilations assembled
by Groenewegen et al. (1998) for O-rich stars and Groenewegen et
al. (2002) for C-rich stars. Observed data for LMC variables (Marshall
et al. 2004) is also plotted (empty squares). The fit relation
proposed by Vassiliadis \& Wood (1993) is shown by the solid line.
Bottom panels: Predicted expansion velocities for a few selected
isochrones with $Z=0.019$ and $Z=0.008$.  Ages are indicated in the
figure. See the text for more details.}
\label{fig_vexp}
\end{figure}

\subsection{Expansion velocities of variable AGB stars}
\label{ssec_obsvexp}

As first check of our models we compare the predicted expansion
velocities (Sect.~\ref{ssec_vexp}) as a function of pulsation periods
with observations of variable AGB stars in the Galactic Disk and
Magellanic Clouds. The results, displayed in Fig.~\ref{fig_vexp}, are
consistent with the empirical estimates.

A relevant remark should be made in respect to the O-rich stars. While
for $P < 500-600$ days a positive correlation between periods and
expansion velocities seems to exist in the observed data, this trend
does not proceed at longer periods, where a considerable scatter of
$v_{\rm exp}$ is detected. It should be noticed that while reproducing
the mean observed trend at shorter periods, the widely-used relation
of Vassiliadis \& Wood (1993; solid line) fails to describe the data
at longer periods, where $v_{\rm exp}$ is assumed to be constant ($15$
km s$^{-1}$).

In the bottom panels of Fig.~\ref{fig_vexp} we show the results of our
calculations for a few selected isochrones with $Z=0.019$ and
$Z=0.008$. 
 Regardless of age all isochrones describe a similar non-monotonic evolution on
the $v_{\rm exp}-P$ diagram. At increasing periods, first the
expansion velocity increases, reaches a maximum and eventually
declines.  This bell-shaped behaviour is particularly well developed
for the youngest isochrones (with ages of $\sim 1.6\, 10^{8}$ yr and
$\sim 2\, 10^8$ yr), which correspond to massive AGB stars (with
initial masses of $4.5\, M_{\odot}$ and $4.1\,M_{\odot}$) and it
appears to explain the existence of variables with high periods and
relatively low expansion velocities. 

To this respect it is also interesting to compare
the two isochrones with $\log({\rm age}/{\rm yr})= 8.3)$ and different
metallicities ($Z=0.019$ and $Z=0.008$). The one with  
$Z=0.008$ reaches, on average, lower $v_{\rm exp}$. Moreover, 
the few OH/IR stars in the LMC for which
$v_{\rm exp}$ has been measured (empty squares; Marshall et al. 2004)
fall just in correspondence of the declining branches of the $v_{\rm
exp}$ curves.  This fact suggests that the low expansion velocities of
LMC OH/IR stars compared to those of Galactic OH/IR stars with the
same $P$, may not reflect a pure effect of metallicity (i.e. a linear
$Z$-dependence of the dust-to-gas ratio $\Psi$ such that the lower $Z$
the lower $v_{\rm exp}$), as commonly invoked (see the discussion in
Marshall et al. 2004); rather it may be the result of a more complex dependence
of $v_{\rm exp}$ as a function of stellar and dust parameters ($L,\,
\dot{M},\, M,\, \Psi$). A deep investigation of this issue will be
performed in future work. 

\subsection{The LMC and SMC near-infrared photometry}

TP-AGB stars become striking objects in galactic surveys conducted at
pass-bands redder than $I$. The best available data for the AGB
population in galaxies is likely the one obtained from the extensive
near- and mid-infrared surveys of the Magellanic Clouds.

\begin{figure*}
\begin{minipage}{0.48\hsize}
\resizebox{\hsize}{!}{\includegraphics{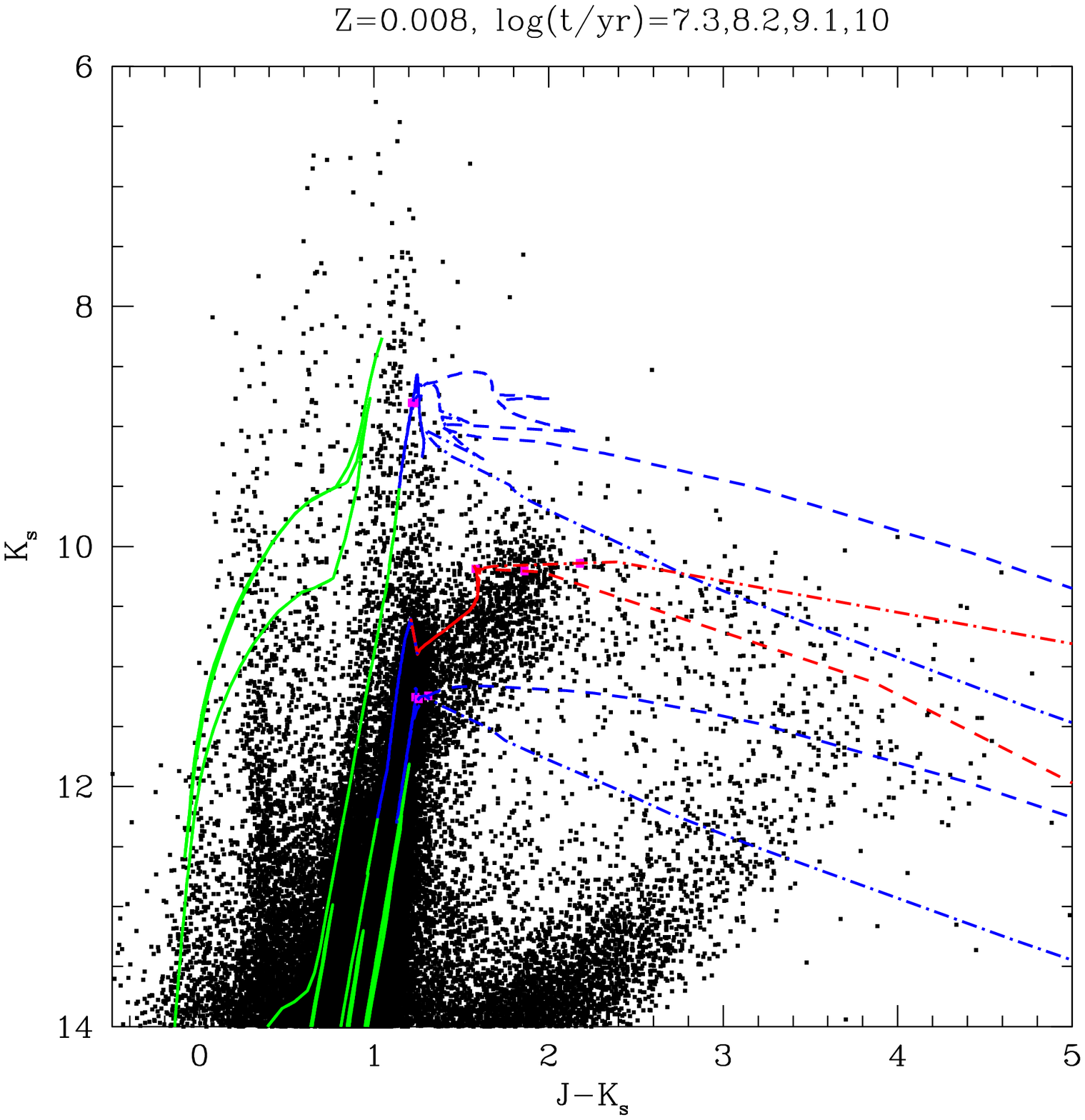}}
\end{minipage}
\hfill
\begin{minipage}{0.48\hsize}
\resizebox{\hsize}{!}{\includegraphics{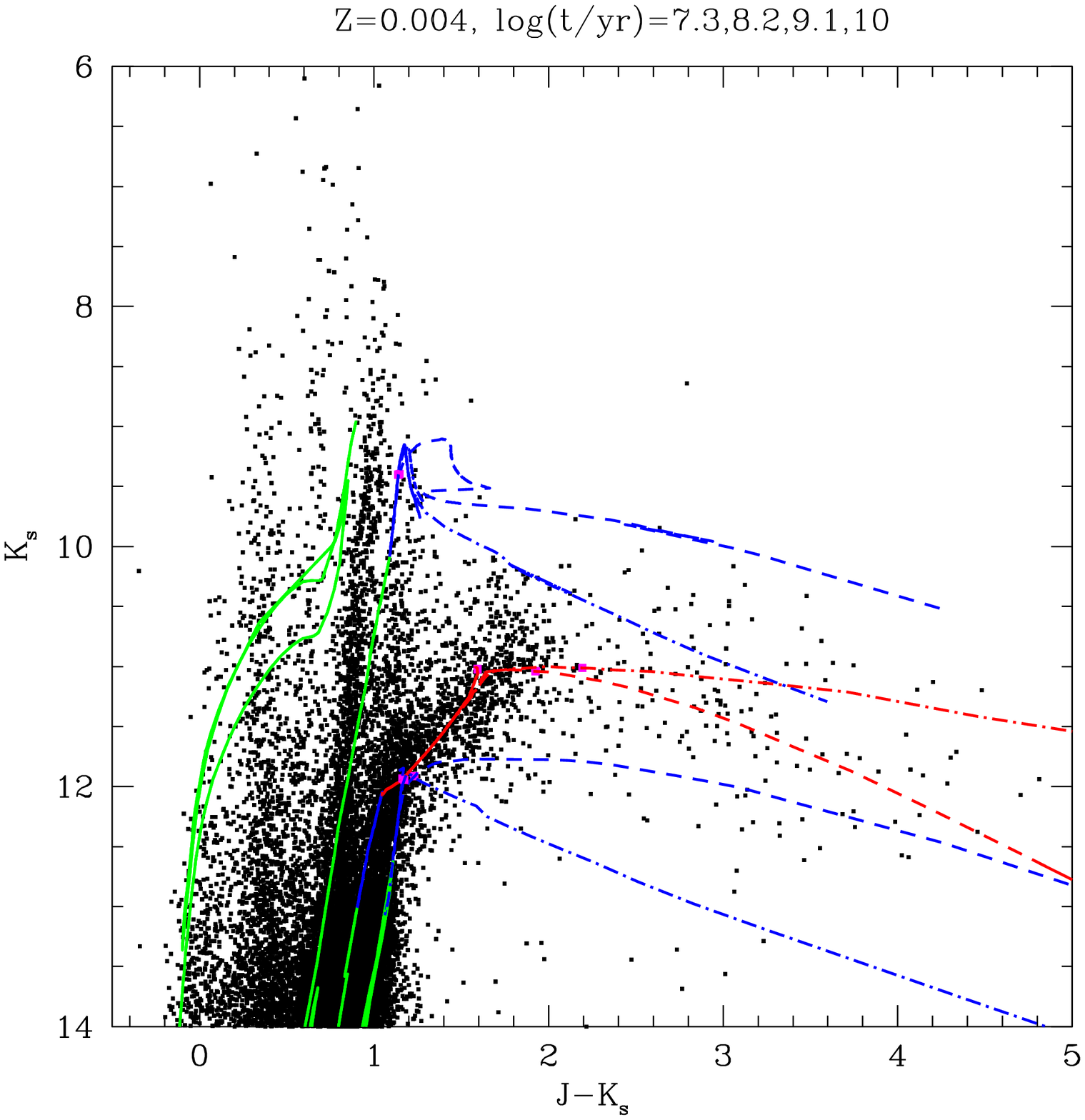}}
\end{minipage}
\caption{The 2MASS \ks\ vs. \jks\ diagram for the inner
LMC (left panel) and for the SMC (right). Over-imposed are isochrones
with $Z=0.008$ shifted by a distance modulus of $18.5$, and with
$Z=0.004$ shifted by $19.0$, for the LMC and SMC cases,
respectively. The isochrones are drawn in different colours according
to their evolutionary status: pre-TP-AGB stars (green), and TP-AGB
stars O-rich (blue) and C-rich (red). The selected ages are:
$\log(t/{\rm yr})=7.3$ (no AGB present), 8.2 (well developed O-rich
TP-AGB), 9.1 (well-developed C-rich TP-AGB), and 10.0 (short O-rich
TP-AGB), for 3 different choices of dust-obscuration: the no-dust case
(continuous lines), cf. Bressan et al. (1998; dot-dashed lines) and
c.f. Groenewegen (2006; the 0.85\,AMC+0.15\,SiC mixture for C-stars,
and 0.6\,AlOx+0.4\,Silicate for O-rich stars; dashed lines). In order
to limit confusion in this plot, the isochrones are drawn up to their
point of maximum mass-loss rate on the TP-AGB, i.e. ignoring their
final post-AGB stages in which the isochrones cross back to the blue
part of the diagram. Moreover, a magenta dot mark points where the
mass loss increases above $10^{-6}~\Msun\,{\rm yr}^{-1}$; above this
indicative limit the stellar lifetimes are necessarily short, and few
stars are expected to occupy isochrone sections to the red of these
points.}
\label{fig_kjk}
\end{figure*}

In Fig.~\ref{fig_kjk} we plot the 2MASS \ks\ vs. \jks\ diagram of the
inner LMC and SMC. The data are taken directly from the SAGE and
S$^3$MC catalogs (Blum et al. 2006, and Bolatto et al. 2007,
respectively), and hence correspond to objects that have counterparts
in the Spitzer pass-bands.

We start discussing the LMC case depicted in the left panel. The
circular area of $\pi$~sqrdeg centered on the LMC bar at
$\alpha_{2000}=5^{\rm h}23\fm5, \delta_{2000}= -69\degr45\arcmin$, is
large enough to include thousands of AGB stars brighter than the TRGB,
and small enough to somewhat limit the presence of features that are
not due to the LMC. These latter features are however still evident in
the plot, and consist of (1) the vertical fingers in the top left part
of the CMD, at $\ks<11$~mag and $\jks<1.0$~mag, caused by foreground
stars in the Milky Way disk and halo, and (2) the plume of objects at
the bottom right part of the CMD, roughly at $\ks>12$~mag
vs. $\jks>1.6$~mag, which largely consists of background galaxies (see
Nikolaev \& Weinberg 2000, his Region L). Outside of these two CMD
regions, the bulk of objects indeed consists of stars located in the
LMC.

On this diagram, we over-plot our isochrones for the initial
metallicity $Z=0.008$ (which are adequate to describe the youngest LMC
populations) and at four selected ages:
\begin{itemize}
\item At $\log(t/{\rm yr})=7.3$, AGB stars are not present and the most
prominent near-IR stars are red super-giants burning He in their cores.
These stars draw a diagonal line which coincides with sequence H of
Nikolaev \& Weinberg (2000; their Fig.~3).
\item At $\log(t/{\rm yr})=8.2$, the TP-AGB phase is well developed and
populated by luminous O-rich stars undergoing HBB. These stars draw a
second diagonal sequence in the CMD, corresponding to sequence G of
Nikolaev \& Weinberg (2000) and joining their sequence F at fainter
magnitudes which is populated by early-AGB stars. These features are
also well described by our new isochrones.
\item At $\log(t/{\rm yr})=9.1$, HBB does not occur and the TP-AGB phase
is dominated by a well-populated red tail of C-rich stars (sequence J
in Nikolaev \& Weinberg 2000), which departs from the O-rich sequence
at $\ks=10.3$~mag, $\jks=1.2$~mag and reaches \jks\ as red as
2.0~mag. This feature is overall well described by the $\log(t/{\rm
yr})=9.1$ isochrones without dust, apart from the fact that in this
case the red tail just reaches $\jks=1.6$~mag, which is the colour of
all C-star spectra cooler than $\sim3000$~K in the Loidl et al. (2001)
library.
\item At $\log(t/{\rm yr})=10.0$, the third dredge-up does not occur, and
the only TP-AGB stars are O-rich. They correspond to sequence F in
Nikolaev \& Weinberg (2000), located right above the TRGB.
\end{itemize}

The above items mainly describe the behaviour of the isochrones at the
regime of weak mass-loss, in which colours and magnitudes are affected
by circumstellar dust at the level of just $\la0.2$~mag. What happens
in the regime of high mass-loss rates (say
$|\dot{M}|>10^{-6}\,\Msun\,{\rm yr}^{-1}$) is also well evident in the
figure, when we look at the isochrones with dust obscuration: the
upper fraction of their TP-AGB phase departs to very red colours at
increasing \ks-band magnitudes, reaching $\jks>4$~mag. The
location and slope of this extended red tail (sequence K in Nikolaev
\& Weinberg 2000) is also well described by our isochrones, and
especially by the $\log(t/{\rm yr})=9.1$ one which runs over the most
densely populated part of this extended red tail.

\begin{figure*}
\begin{minipage}{0.48\hsize}
\resizebox{\hsize}{!}{\includegraphics{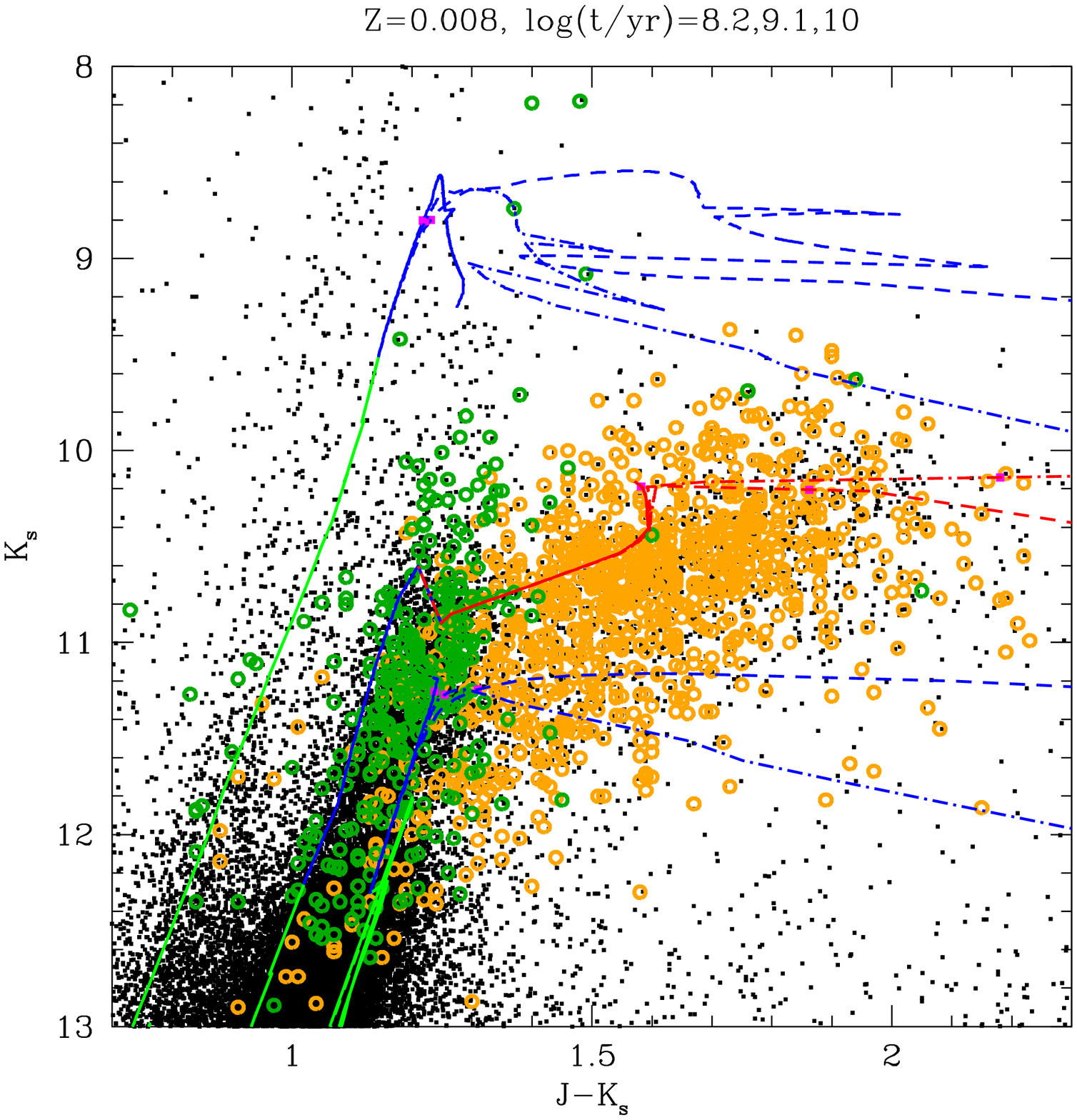}}
\end{minipage}
\hfill
\begin{minipage}{0.48\hsize}
\resizebox{\hsize}{!}{\includegraphics{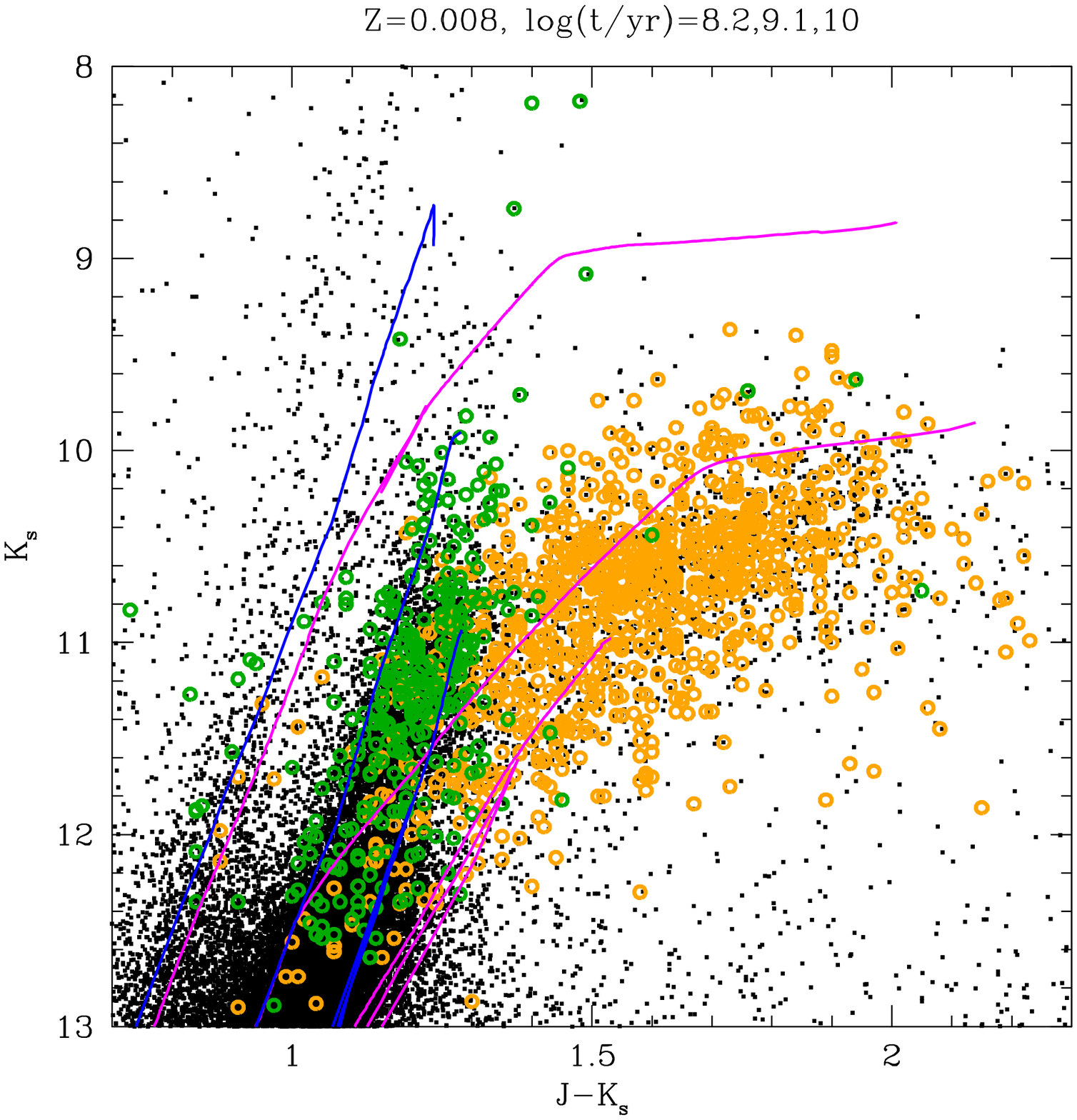}}
\end{minipage}
\caption{{\bf Left panel}: The 2MASS \ks\ vs. \jks\ diagram for the
LMC, now detailing the CMD region that corresponds to the bulk of
TP-AGB stars, and limiting the isochrones to the ages of $\log (t/{\rm
yr})=8.2$, $9.1$, and $10.0$. Additional LMC data are also plotted:
The small circles are LPVs in the LMC that have been spectroscopically
classified as being of type M (dark-green) or C (orange), according to
the data compilation from Groenewegen (2004). The meaning of the
different lines along the isochrones are the same as in
Fig.~\ref{fig_kjk}. The {\bf right panel} shows the same data as
compared to Gi00 (blue lines) and Cordier et al. (2007; magenta
lines). Notice that these latter sets of isochrones do not contain a
C-star phase.
}
\label{fig_kjk_zoom}
\end{figure*}

The left panel of Fig.~\ref{fig_kjk_zoom} details the region of the
\ks\ vs. \jks\ diagram that includes the bulk of the TP-AGB
stars. In addition to the 2MASS data for the inner LMC, we plot the
Groenewegen (2004) database of LMC stars spectroscopically classified
as being of types M and C. The latter are LPVs selected from the OGLE
database for the entire LMC.  Although they are not distributed evenly
over the CMD, they clearly illustrate that the bulk of M- (dark-green)
and C-type (orange) stars are well separated in the CMD; in particular
they show the striking red tail of C stars extending from
$\jks\sim1.3$~mag up to $\jks\sim2.0$~mag, and the confinement of M
stars at the bluest colours ($\jks\la1.3$~mag).

Then, the figure evidences how the new isochrones reproduce the
general appearance of the data, and in particular the red tail of C
stars which is well described by the $\log(t/{\rm yr})=9.1$
isochrone. It is also remarkable how well the isochrones predict the
neat separation between O-rich (M-type) and C-rich (C-type) stars in
the diagram.  In the $\log(t/{\rm yr})=9.1$ isochrone, the red tail
reaches $\jks\sim1.6$ even in absence of circumstellar dust. Dust
obscuration in stars with $|\dot{M}|\la10^{-6}~\Msun\,{\rm yr}^{-1}$
provides an additional $\Delta(\jks)\sim0.2-0.3$~mag that improves the
comparison at the reddest part of the red tail, at $\jks$ between 1.6
and 2.0~mag. As thoroughly discussed in Marigo et al. (2003a), the main
physical effect driving the appearance of the red tail of C stars are
the cool \Teff\ caused by changes in molecular opacities as the third
dredge-up events increase the photospheric C/O ratio.

The right panel of Fig.~\ref{fig_kjk_zoom} over-plots isochrones from
Gi00 and Cordier et al. (2007) over the same data. These isochrone sets
contain the complete TP-AGB phase but computed in a rather crude way,
i.e. without considering the third dredge-up events and HBB
nucleosynthesis, and using $\Teff(L,M,M_{\rm core})$ relations derived
from scaled-solar mixtures.  These isochrones fail in reproducing the
basic features indicated by 2MASS and spectroscopic data, although in
different ways:
\begin{itemize}
\item
The Gi00 isochrones describe well the locus of O-rich AGB stars, as
expected for TP-AGB models that do not consider the third dredge up
process. These isochrones are confined to $\jks\la1.3$~mag, and do not
cross the red tail of C stars between $\jks\sim1.3$ and 2.0~mag.
\item
The Cordier et al. (2007) isochrones, instead, do produce a sort of
red tail, reaching $\jks\sim2.1$~mag, but they do this {\em while still on
an O-rich phase}, and for all ages younger than about 1~Gyr. This is
completely at odds with observations, which show that only C-rich
stars (with very few exceptions) populate the red tail at $\jks>1.3$~mag.
\end{itemize}
The reasons why Cordier et al. (2007) O-rich TP-AGB models do reach so
red \jks\ colours -- so as to mimic C-type stars, though they are
not -- likely reside in both the extremely low $\Teff$ -- as cool as
2400~K at the TP-AGB end -- they reach using the Wagenhuber (1996)
formula, and on the use of a colour-\Teff\ relation derived from
C-type stars (from Bergeat et al. 2001). These choices are highly
questionable. First, independent integrations of model atmospheres
(e.g. Marigo 2002) suggest that O-rich stars can hardly reach
\Teff\ values lower than $\simeq3000$~K, even at solar
metallicities\footnote{Note that the bulk of M giants in the
Magellanic Clouds are of spectral types earlier than M4, and in
general much hotter than the typical M giants in the Milky Way. A
handful of cool M giants and supergiants of spectral type later than
M6 is also known in the Magellanic Clouds (e.g. Groenewegen \&
Blommaert 1998; van Loon et al. 1998, 2005). In order to explain them,
and also the more common giants of spectral type later than M8
observed in the solar neighbourhood (with \Teff\ as small as 2400~K,
see Fluks et al. 1994), our atmosphere models require super-solar
metallicities.}. Second, it has long been known that (\jk)--\Teff\
relations for C-stars cannot be applied to O-rich stars, and
vice-versa, since these two kinds of stars have very different
near-infrared spectra.

From the failure of both the G00 and Cordier et al. (2007) isochrone
sets in describing a red tail made of C stars, it follows that such
models are not suitable for the interpretation of near-infrared
photometry in the age interval in which C stars form. For LMC
metallicities, this interval is quite ample, going from a few
$10^8$~yr up to about 5~Gyr (see also Frogel et al. 1990).

Finally, the 2MASS \ks\ vs. \jks\ diagram for the area within the SMC
covered by the S$^3$MC survey (right panel of Fig.~\ref{fig_kjk}) can
be described in quite a similar way to the LMC one. The most obvious
difference between the two panels is in the $\sim0.5$~mag shift of all
features coming from the LMC--SMC difference in distance, that we also
took in consideration when plotting the isochrones. Also, due to the
particular selection of data the SMC plot presents less contamination
from background galaxies than the LMC one.

\subsection{The LMC and SMC mid-infrared photometry}

\begin{figure*}
\begin{minipage}{0.48\hsize}
\resizebox{\hsize}{!}{\includegraphics{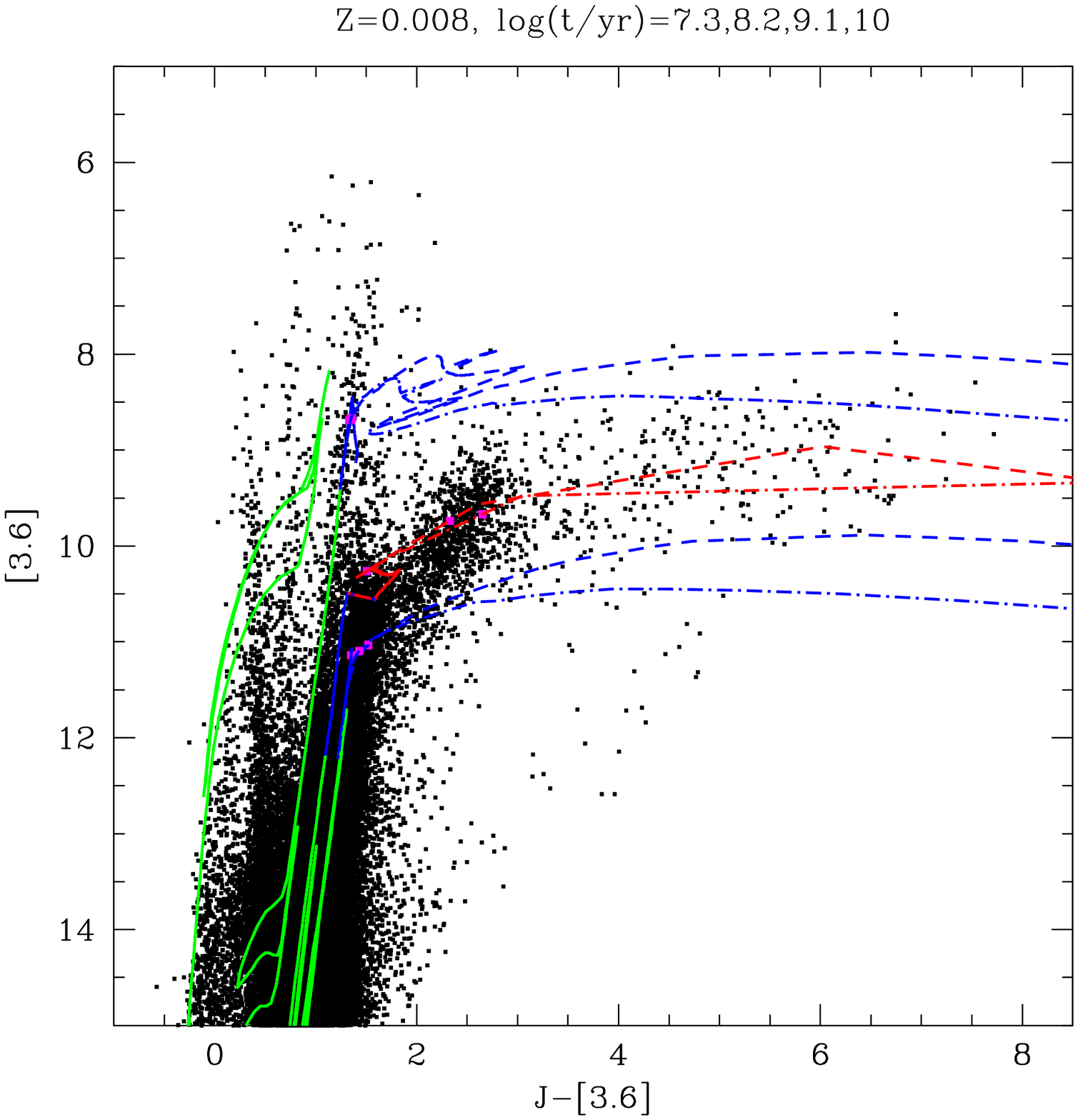}}
\end{minipage}
\hfill
\begin{minipage}{0.48\hsize}
\resizebox{\hsize}{!}{\includegraphics{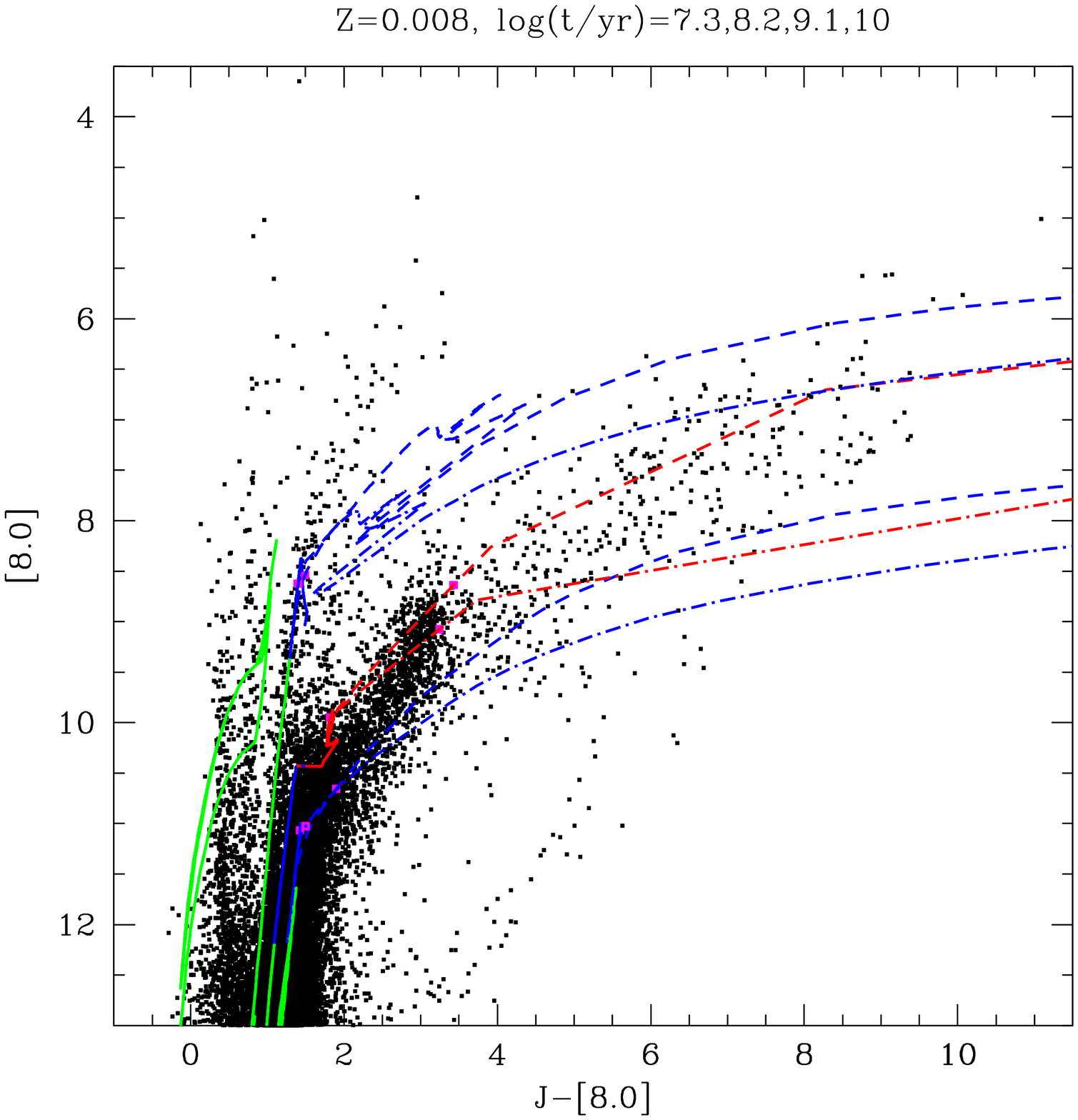}}
\end{minipage}
\begin{minipage}{0.48\hsize}
\resizebox{\hsize}{!}{\includegraphics{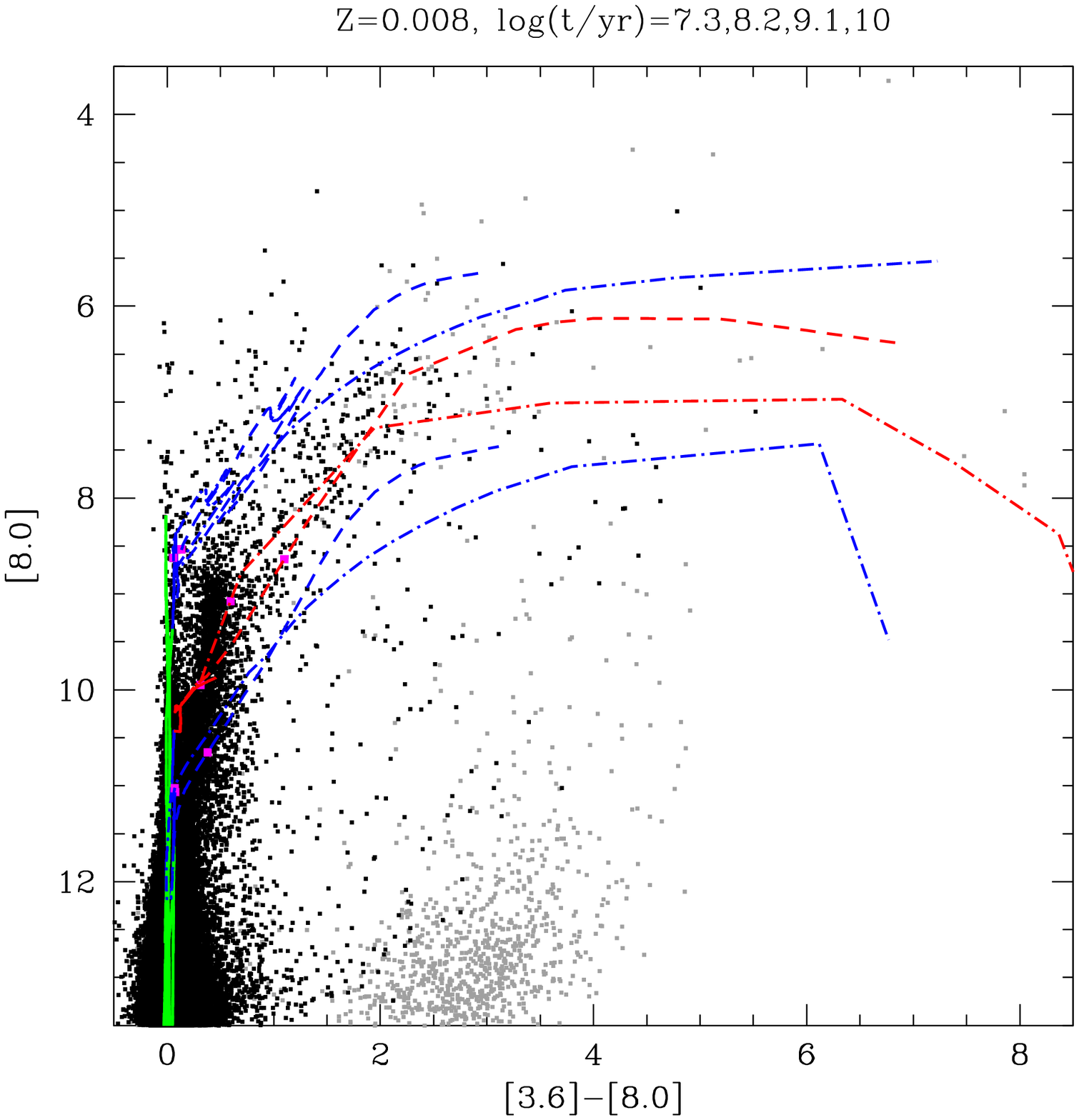}}
\end{minipage}
\hfill
\begin{minipage}{0.48\hsize}
\resizebox{\hsize}{!}{\includegraphics{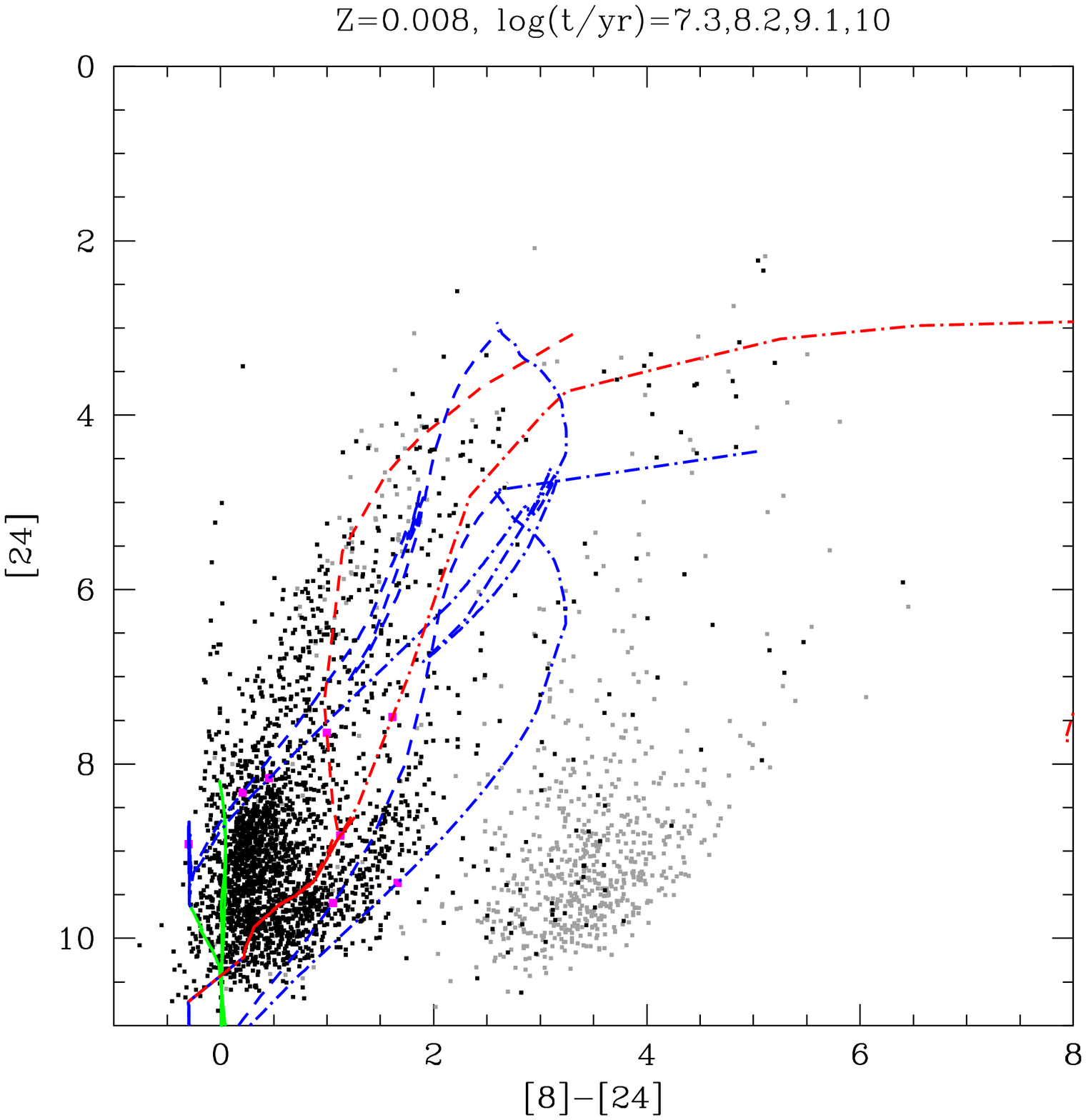}}
\end{minipage}
\caption{A comparison between the new isochrones and the near- to
mid-infrared data for the LMC. The isochrones are the same as plotted
in the left panel of Fig.~\ref{fig_kjk}.  The data (dots) come from
the SAGE epoch~1 point source catalog (Blum et al. 2006) and combine
IRAC+MIPS with 2MASS data inside a circle of 1 degree radius in the
inner LMC. The panels show the 2MASS+IRAC [3.6] vs. $J-[3.6]$,
2MASS+IRAC [8.0] vs. $J-[8.0]$, IRAC [8.0] vs. $[3.6]-[8.0]$, and
IRAC+MIPS [24] vs. $[8.0]-[24]$ diagrams. The gray dots correspond to
objects not detected by 2MASS, most of which are expected to
correspond to background galaxies and young pre-main sequence stars,
which are not discussed in this paper.}
\label{fig_sage}
\end{figure*}

\begin{figure*}
\begin{minipage}{0.48\hsize}
\resizebox{\hsize}{!}{\includegraphics{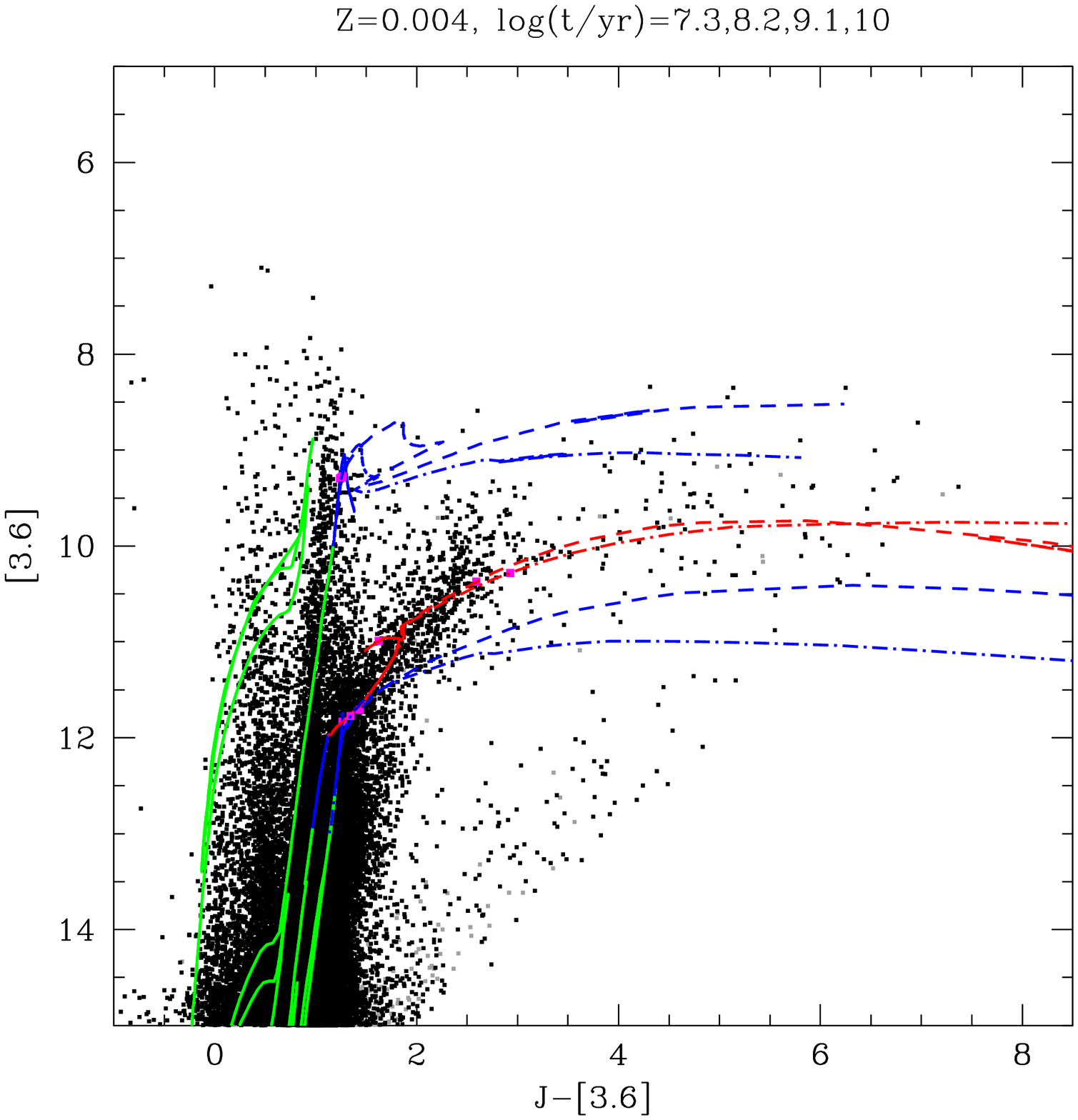}}
\end{minipage}
\hfill
\begin{minipage}{0.48\hsize}
\resizebox{\hsize}{!}{\includegraphics{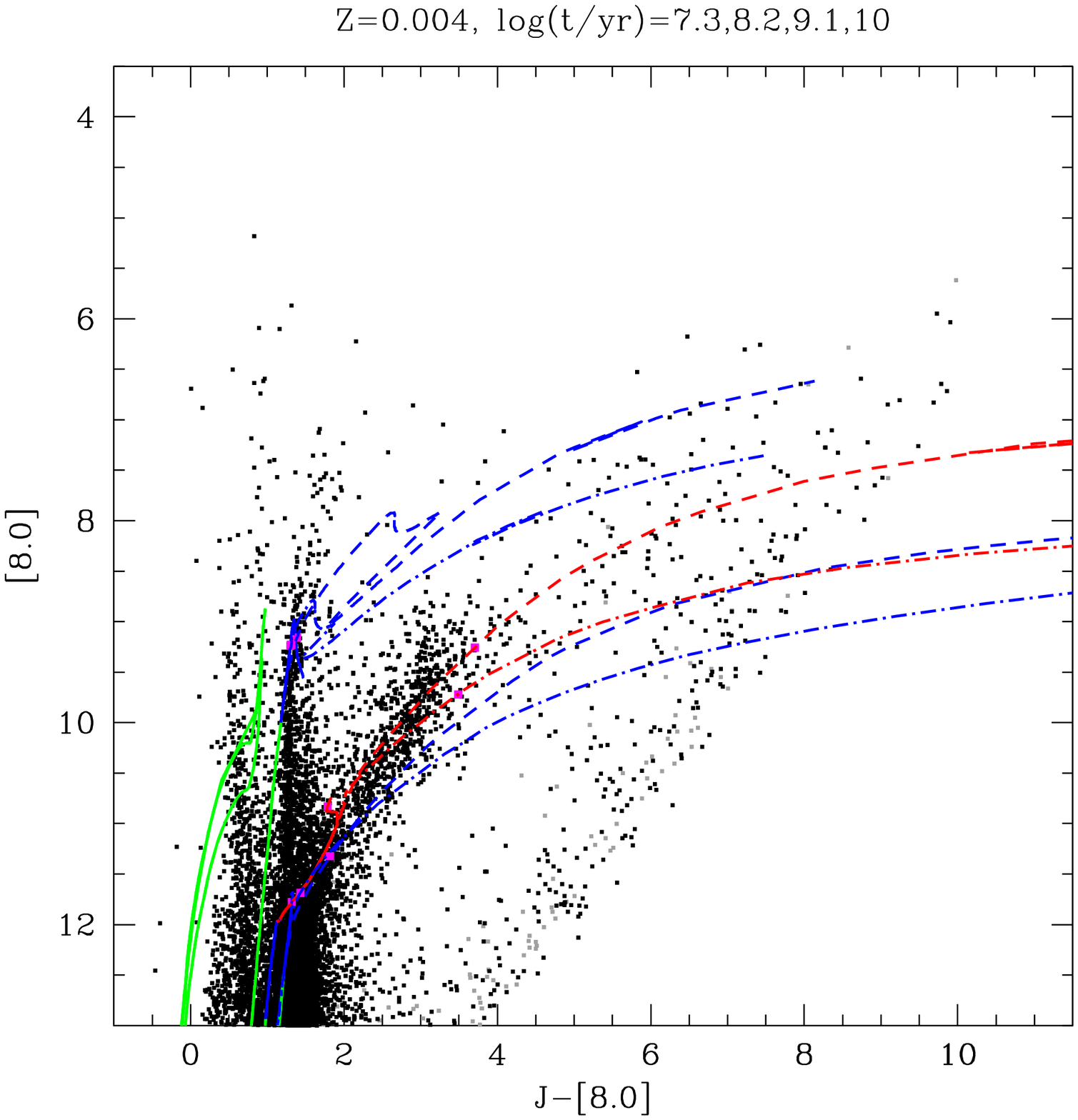}}
\end{minipage}
\begin{minipage}{0.48\hsize}
\resizebox{\hsize}{!}{\includegraphics{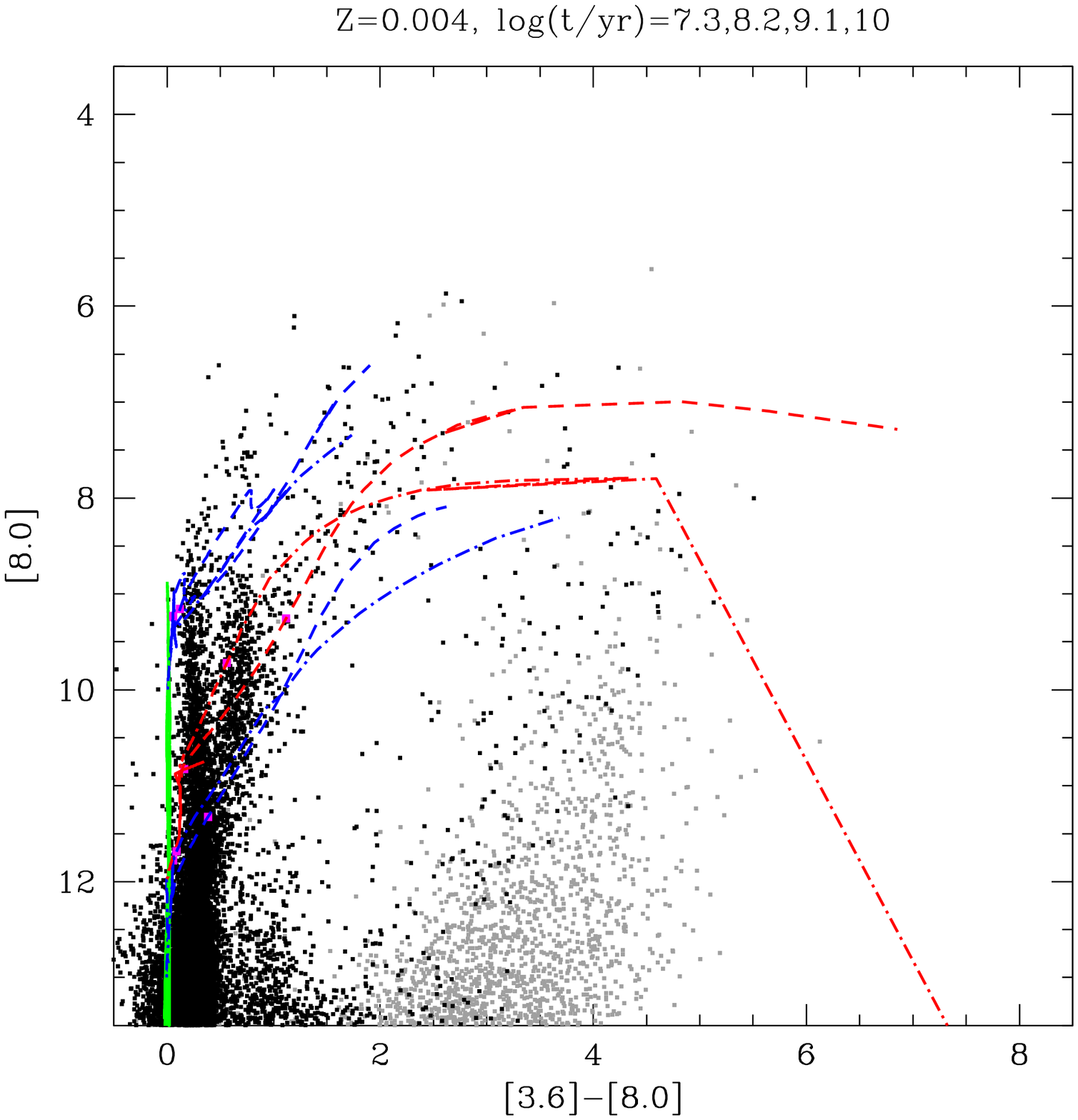}}
\end{minipage}
\hfill
\begin{minipage}{0.48\hsize}
\resizebox{\hsize}{!}{\includegraphics{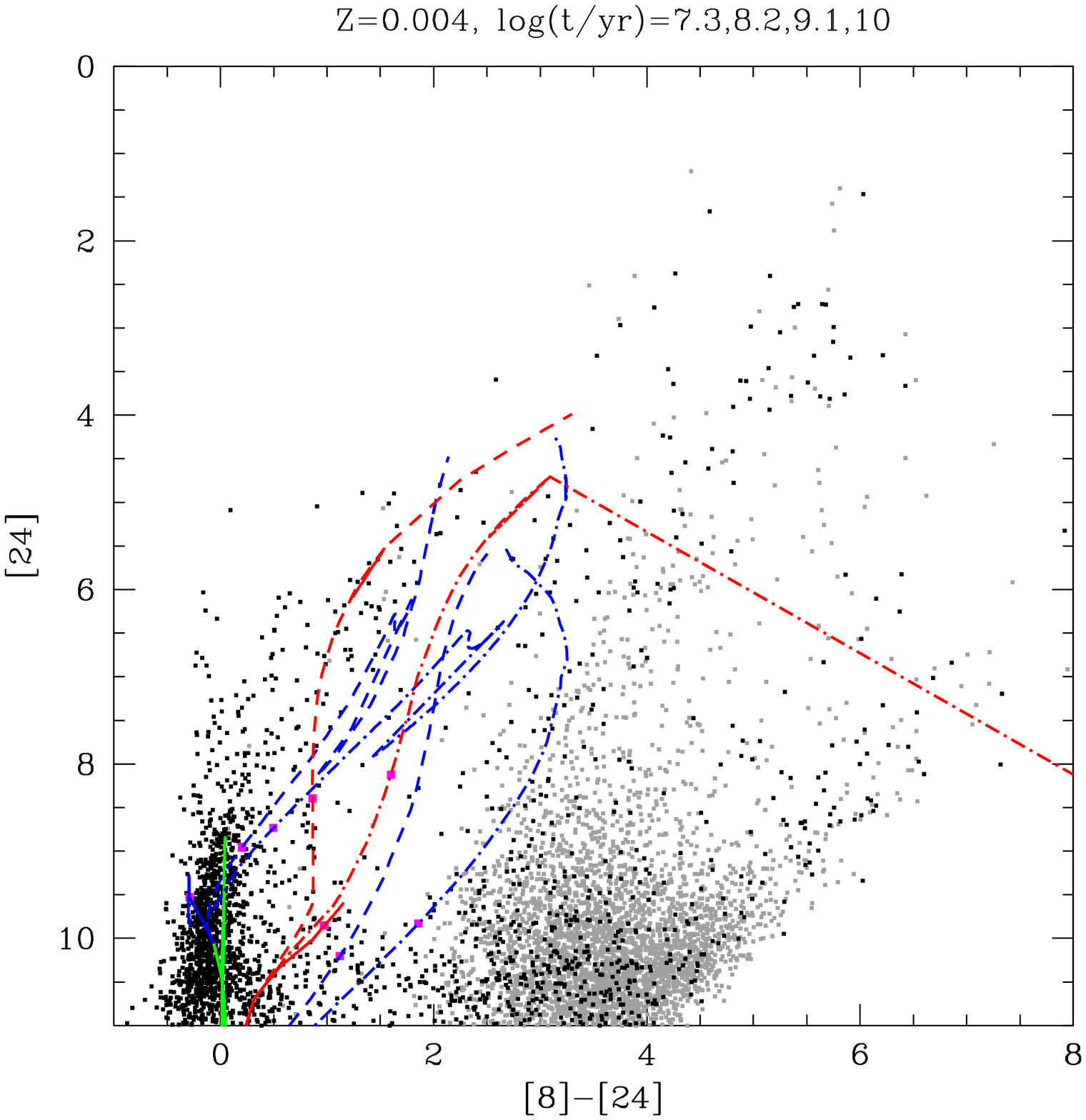}}
\end{minipage}
\caption{As Fig.~\ref{fig_sage} but for
the SMC. The data (dots) come from the S$^3$MC catalog (Bolatto et
al. 2007). The isochrones are the same as in the right panel of
Fig.~\ref{fig_kjk}. }
\label{fig_s3mc}
\end{figure*}

Figures~\ref{fig_sage} and \ref{fig_s3mc} show a series of CMDs
involving the near- and mid-IR photometry of the Magellanic Clouds,
comparing the new isochrones with 2MASS+SAGE data for the inner LMC
(Blum et al. 2006) and with 2MASS+S$^3$MC data for the SMC
(Bolatto et al. 2007)\footnote{The S$^3$MC data, originally in flux
units, have been converted to magnitudes using the zero-points from
Cohen et al. (2003) for $JH\ks$ and from Bolatto et al. (2007)
for Spitzer pass-bands.}. In all these plots, the effect of
circumstellar dust becomes crucial in describing the observed
photometry. The features present in these CMDs (except for the
IRAC+MIPS [24] vs. $[8.0]-[24]$ diagram, for which the data is poorer)
are essentially the same as those present in the \ks\ vs. \jks\
diagram. In particular, all of them show contamination from foreground
Milky Way stars in the form of a narrow vertical finger confined to
the left part of the CMDs, and contamination from background galaxies
at their bottom-right corner. The gray dots correspond to objects not
detected by 2MASS, most of which are expected to correspond to
background galaxies and young pre-main sequence stars, which are not
of interest in this paper.

The LMC population in these plots again shows O-rich sequences (core
He burning red super-giants plus O-rich TP-AGBs) that progressively
merge into an almost-vertical blue sequence when colours involving
redder pass-bands are considered; plus an inclined C-rich sequence that
can be identified as the main body of the $\jks<2.0$~mag red
tail. From the extremity of this more populated red tail a very
extended and less populated plume departs, that corresponds to
the stars with high mass-loss rates. Our isochrones reproduce well the
position of these plumes in the several CMDs, and moreover predict
that they are populated by both O- and C-rich stars.

\subsection{Integrated broad-band colours}
\label{sec_vkevol}
Broad band colours of Magellanic Cloud clusters represent the basic
observable against which any population synthesis model should be
compared for testing its predicting performance.

\begin{figure*}
\begin{minipage}{0.47\hsize}
        \resizebox{\hsize}{!}{\includegraphics{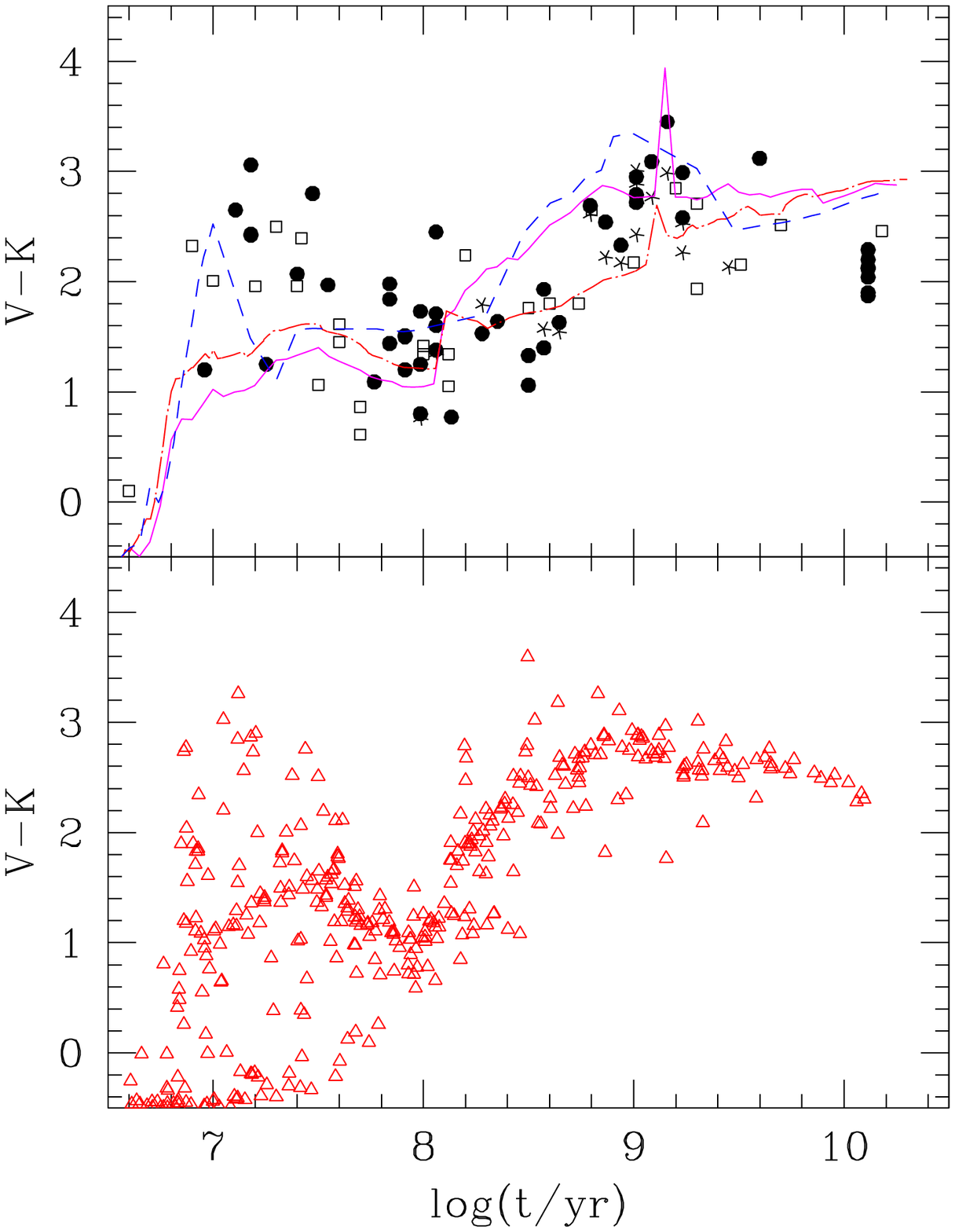}}
\end{minipage}
\hfill
\begin{minipage}{0.47\hsize}
        \resizebox{\hsize}{!}{\includegraphics{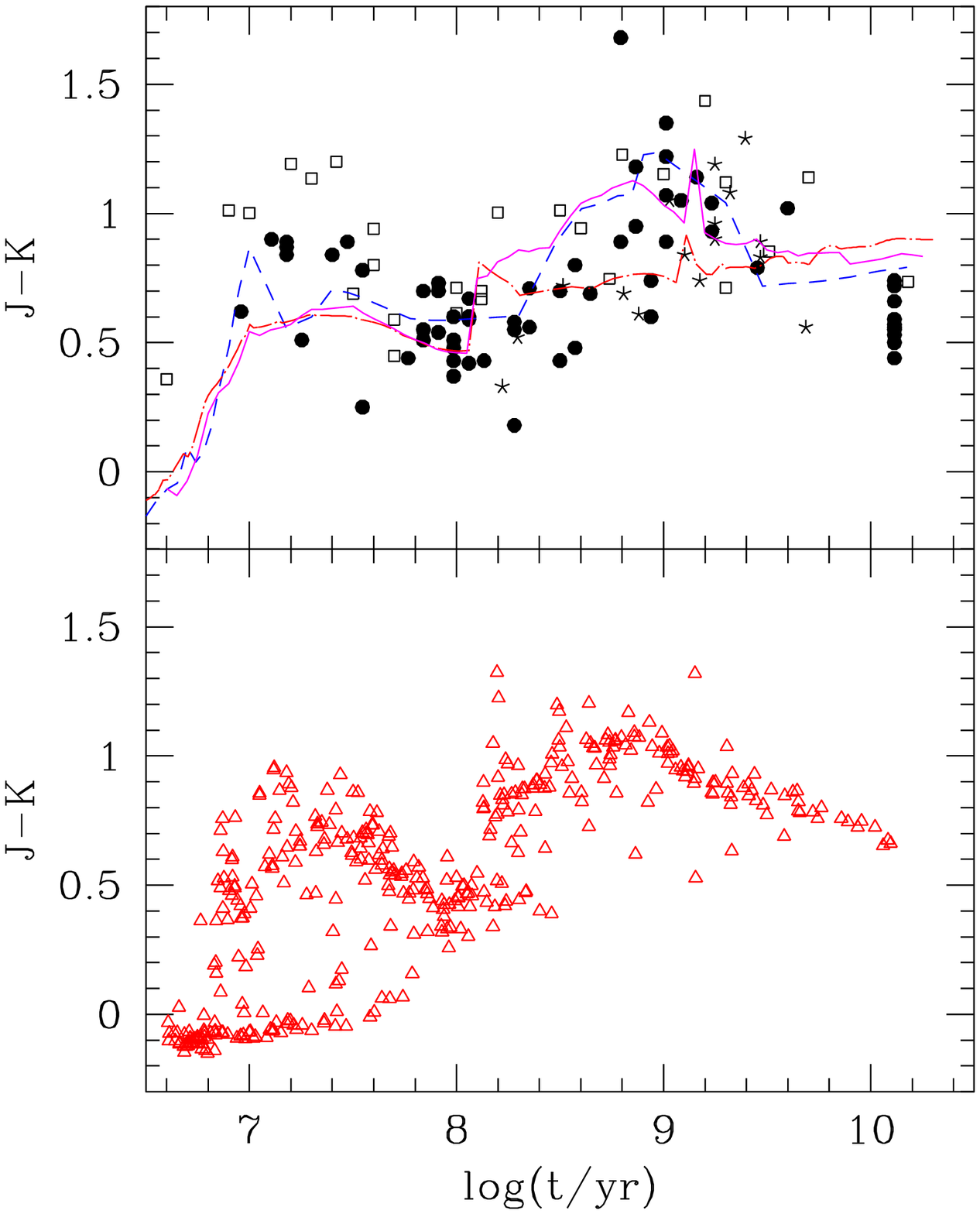}}
\end{minipage}
\caption{Broad-band $V-K$ and $J-K$ colours of LMC
clusters as a function of cluster age.  {\bf Top panels}: Observed
photometric data taken from various compilations: Persson et al (1983,
filled circles); Kyeong et al. (2003, empty squares); and the $V$-band
photometry by Goudfrooij et al. (2006) adopting an aperture radius of
50 arcsec, combined to the $JHK_{\rm s}$ photometry from Pessev et
al. (2006; starred symbols).  Clusters are assigned ages by adopting
the $S$-parameter--age calibration by Girardi et al. (1995).
Predicted SSP colours from various authors are superimposed for
comparison: this work (solid line, $Z=0.008$), Maraston (2005; dashed
line, $Z=Z_{\odot}/2$), Bruzual \& Charlot (2003; dot-dashed line,
$Z=0.008$).  We note that the spike in the colour evolution at
$\sim10^9$~yr is real and is explained in detail by Girardi \&
Bertelli (1998). {\bf Bottom panels}: Simulations of stellar clusters
based on the new isochrones. They roughly follow the distribution of
typical sizes (or initial total masses) and metallicities of LMC
clusters. The age distribution is assumed to be constant. The
simulated sample has a limiting magnitude of $M_V^{\rm lim}=-6.5$,
which roughly corresponds to the one of the observed samples in the
top panel.}
\label{fig_sspcolor}
\end{figure*}

Figure~\ref{fig_sspcolor} displays the results for two integrated
colours, $V-K$ and $J-K$, that are expected to be significantly
affected by the existence of TP-AGB stars.  By looking at
Fig.~\ref{fig_sspcolor} (top panels) we actually see that the observed
data for both $V-K$ and $J-K$ colours show a rising trend starting
from $t\sim 10^8$~yr towards older ages, which is currently well
explained by our current single-burst stellar populations (SSP) models
(solid line) as due to the development of the TP-AGB phase.  The
reader is referred to Appendix~\ref{sec_revagb} for a discussion on
the behaviour of $V-K$ at ages $t\sim 10^8$~yr and its relation to the
evolution of the most massive AGB stars.  Integrated magnitudes are
calculated accounting for the luminosity variations (i.e. the
long-lived luminosity dips in low-mass stars) driven by thermal
pulses, and the effect on the emitted spectrum by circumstellar dust.

For comparison we also show the results of other two widely-used SSP
models, namely Charlot \& Bruzual (2003; dot-dashed line) and Maraston
(2005; dashed line).  A few points should be noticed in respect to the
age interval $t \ga 10^8$~yr, when AGB stars are present.  While
Charlot \& Bruzual (2003) colours tend to be bluer than the observed
ones because of a likely underestimation of the TP-AGB phase (see
Bruzual 2007 for a recent revision of this aspect in their models),
Maraston (2005) tends to form a sort of upper envelope to the data,
especially in $V-K$. Our results are intermediate between the two.

The bottom panels of Fig.~\ref{fig_sspcolor} illustrate the same
colour evolution, by means of simulated clusters. The cluster
simulations are performed by randomly adding stars to the isochrones,
with the probability of a given initial stellar mass being given by
the Chabrier (2001) log-normal initial mass function.  The addition of
stars to a simulated cluster is stopped when its initial total mass
reaches a given value $M_{\rm T}$. The entire cluster sample is
assumed to follow a $M_{\rm T}^{-1.5}$ distribution, which is derived
from the luminosity function of LMC clusters (Elson \& Fall 1985; see
also Girardi et al. 1995, sect. 8.2). The cluster metallicities follow
the LMC age-metallicity relation from Pagel \& Tautvai\v{s}ien\`e's (1998)
bursting model, which has been shown to reproduce the observed
trend. The distribution of cluster ages is assumed to be constant;
although this is not a realistic assumption (see Girardi et al. 1995),
it allows us a good illustration of the colour evolution over the
complete age range. A posteriori, we impose a limiting magnitude of
$M_V^{\rm lim}=-6.5$~mag to the simulated sample, which roughly
corresponds to the one of the observed samples in the top panels of
Fig.~\ref{fig_sspcolor}. The results present the intrinsic colour
dispersion expected for the cluster data due to their small numbers of
evolved stars (stochastic fluctuations), and in this aspect they look
clearly more similar to the observations than the continuous narrow
SSP lines depicted in the top panel. The main discrepancy between
models and simulations is at ages younger than $10^8$~yr, where a
bunch of very blue clusters is predicted but absent from the
observed sample\footnote{Isochrones in this age range were taken from Bertelli
et al. (1994).}. 
The blue clusters are essentially 
those for which no red super-giant was
predicted, and derive naturally from the fact that the red super-giant 
phase is short-lived. In order to recover the observed colours of the youngest
clusters the red super-giant phase has to be significantly 
longer and cooler 
than in the Bertelli et al. (1994) isochrones. 
On one hand, a longer duration could be
attained, for instance, by invoking a larger efficiency of overshoot,
but this may in turn inhibit the development of the blue loops
of massive stars (see the discussion in Alongi et al. 1991).
On the other hand, the effective temperatures of 
red super-giants depend very much on the
treatment of the density inversion that is predicted to take 
place in the outermost atmospheric layers. A deeper analysis of this issue
is beyond the scope of the paper and it is postponed to future work.

Another questionable point involves the ages between $10^8$~yr and
$\sim4\times10^8$~yr, after the AGB phase develops: in this age
interval, our models become on the mean redder than the data. As
mentioned in the Appendix, removing this feature from the models
poses a theoretical difficulty as it would as well prevent the significant
enrichment of helium and nitrogen in the most massive AGB stars, 
which is instead required to explain the observed abundances 
in planetary nebulae of type I. Finally, at older ages 
the adoption of an age--metallicity
relation allows us to reproduce the colours of the oldest LMC
clusters, which have ages and metallicities typical of old Galactic
globular clusters.

In short, these isochrones offer a valid alternative for the modelling
of the integrated light of star clusters and galaxies, since they pass
the LMC cluster test performing similarly well as the widely used
Charlot \& Bruzual (2003) and Maraston (2005) SSP models, when we
look at the mean colour evolution (top panels of
Fig.~\ref{fig_sspcolor}). Moreover, the isochrones we provide do allow
the simulation of resolved stellar populations, which is essential for
their further testing -- as illustrated by the simulations in the
bottom panels of Fig.~\ref{fig_sspcolor}. The isochrones are based on
internally consistent sets of tracks, which naturally obey the fuel
consumption theorem, and provide other quantities necessary for a
consistent modelling of galaxies -- namely the chemical yields and
remnant masses (see Marigo \& Girardi 2001 for a discussion on this
latter point). Having all these components together allows us the
simultaneous comparison of the model results with many different
observables, indicating hidden problems (like the young blue clusters
predicted in Fig.~\ref{fig_sspcolor}) and hopefully pointing the way
to future improvements in the models.

\section{Data tables and final remarks}
\label{sec_conclu}

\subsection{Data tables}
\label{sec_datatables}

The data presented in this paper are provided in two different ways:

(1) As a repository of static tables down-loadable from the VizieR
Catalogue Service at CDS ({\tt http://vizier.u-strasbg.fr}) and from
the URL {\tt http://stev.oapd.inaf.it/dustyAGB07}. This database
includes not only the isochrones, but also bolometric corrections,
extinction coefficients, dust attenuation curves, etc.

(2) Via an interactive web interface at {\tt
http://stev.oapd.inaf.it/cmd}, that allows users to build isochrones
(or sequences of them) for different values of age and metallicity,
photometric systems, and dust properties. The same web form provides
access to a series of other services, like the computation of
luminosity functions, integrated magnitudes, and simulated star
clusters.

All information about these data are provided in the form of either
{\tt ReadMe} files in case (1), or HTML help pages in case (2). Most
of the information available will be of interest only to the people
who actually access the data, and therefore will not be repeated
here. Suffice it to recall that isochrones are provided for any metal
content in the range $0.0001\le Z\le0.03$ (corresponding to
populations with $-2.31\le\mh\le+0.22$, for scaled-solar initial
compositions), and ages in the range $7.8\le\log (t/{\rm yr})\le10.2$
(i.e. from 0.063 to 15.8~Gyr). The initial mass ranges from 0.15 to
7~\Msun.

Moreover, the {\tt http://stev.oapd.inaf.it/cmd} URL allows the mass
range to be extended from 7 to 100~\Msun, using the tracks from
Bertelli et al. (1994) isochrones. This allows the age range to be
extended down to about 4~Myr for complete isochrones -- or even down
to $t=0$ in the case one is interested in obtaining an extended
zero-age main sequence. The only problem with this is that the tracks
for massive stars are not available for $Z_\odot\le Z\le0.03$, and in
this case have to be taken from the $Z=0.02$ set of Bertelli et
al. (1994).

The basic information provided along isochrones consists on the
luminosity, effective temperature, surface gravity, and absolute
magnitudes in the photometric system under consideration. For the
TP-AGB phase, this information is complemented with the core mass,
surface C/O ratio, pulsation mode and period, and mass-loss rate. The
TP-AGB section of an isochrone can be easily identified by the
information about the core mass, whereas the C-star phase is
identified by $\co>1$. These latter variables are also useful to
simulate, a posteriori, the luminosity and \Teff\ variations caused by
pulse cycles on the TP-AGB, as we have done in a series of papers
(e.g. Marigo et al. 2003; Cioni et al. 2006; Girardi \& Marigo 2007b).

Finally, we remark that the {\tt http://stev.oapd.inaf.it/cmd} web
interface is going to be continuously expanded, either with new
extensions to the tracks (e.g. with the planetary nebulae and white
dwarf sequences), new photometric systems, new spectra from which to
derive bolometric corrections, and different dust
compositions. Interstellar extinction will be included in the
isochrones in a consistent way (Girardi et al., 2007, in prep.). In
any case, the original database, as here described, will remain
available as well.

\subsection{Final remarks}
Users of the new isochrones should be aware that:
\begin{itemize}
\item The isochrones are expected to perform reasonably well for
AGB populations in the Milky Way disk and in the Magellanic Clouds,
against which data they have already been compared or calibrated. The
quality of the results at very low metallicities ([Fe/H]$\la-1.5$) is
yet to be verified.
\item Present-day model atmospheres for cool stars are still far from
satisfactory. Molecular line lists are acknowledged to be incomplete,
whereas for LPVs dynamical model atmospheres would probably provide
more realistic spectral energy distributions than the static models we
have adopted for both M- and C-type stars. The template spectra used
in or models do not cover well enough the expected range of
metallicities, effective temperatures, and surface gravities of AGB
populations in resolved galaxies.
\item Dust properties may differ from the compositions assumed in this
work, and they likely change along the TP-AGB evolution as a function
of parameters other than those here considered.  For a given dust
composition, RT models have been calculated for a large range of
optical depths, even if particular combinations of dust and mass-loss
rate are unlikely. For example, pure AlOx is preferentially seen in
stars with low mass-loss rates only, and the combination of AlOx and
silicates is most appropriate for intermediate mass-loss rates. For
the Galaxy, the mass-loss rate below which the pure AlOx models should
probably be used, and above which pure silicates are likely to be most
appropriate are, respectively, about 10$^{-7}$ $M_{\odot}$ yr$^{-1}$
and 10$^{-6}$ $M_{\odot}$ yr$^{-1}$.  At different metallicities these
ranges may be different and therefore the full range of models has
been calculated for all compositions here considered.  An extended
list of other caveats associated with dusty circumstellar envelopes is
presented in G06.
\item Mass-loss prescriptions from evolved stars are still very uncertain,
and probably none of the presently available formalisms can be applied
with confidence. For O-rich AGB stars, even the basic physical process
driving mass loss is being debated (see Woitke 2006).
\item In constructing our theoretical isochrones, we have made 
all possible effort to ensure internal consistency, but this has not
always been possible. For instance, the dust condensation
prescriptions here adopted are likely inconsistent with those assumed
in the hydrodynamic calculations from which mass-loss formulas were
derived. On the other hand, the uncertainties resulting from such
inconsistencies are no more important than those resulting from a
fundamental lack of understanding of certain aspects of stellar
evolution and mass loss.
\end{itemize}
Work is in progress to verify or improve on the points above; in
particular, we are performing an extensive study of the Magellanic
Cloud data (from $I$ to $[24]$, in preparation), and checking the
results at low metallicities using data for dwarf galaxies and
globular clusters (Gullieuszik et al. 2007ab, and work in
preparation). Model atmospheres of cool giants are being extended by
Aringer et al. (in prep.).

Despite the caveats, we believe that our new isochrones constitute a
significant progress with respect to the past. We recall that the
simple discussion of the above mentioned points would have been
meaningless in the context of the more crude TP-AGB isochrones listed
in Table~1.

In fact, we believe this is the first extended set of isochrones that
includes the TP-AGB evolutionary phase with the level of completeness
and detail it deserves. Crucial effects such as the third dredge up
(which makes C-type stars), hot bottom burning (which largely prevents
the most massive AGB stars from becoming C-type), and variable
molecular opacities (which causes the cool red tail of C stars in
near- and mid-IR CMDs) have been considered. Mass-loss rates are
described with a formalism that discriminates between C- and M-type
stars, and leads to the attainment of the super-wind regime during the
latest stages of the TP-AGB evolution.  The transition from the first
overtone to the fundamental mode is predicted, which helps triggering
higher mass-loss rates. Parameters associated with the third dredge-up
are calibrated so as to reproduce the C-star luminosity functions
observed in the Magellanic Clouds, and the lifetimes of C- and M-type
stars derived from star counts in Magellanic Cloud clusters. Many
photometric systems are considered, properly taking into account the
different spectra of O-rich and C-rich stars, including the stellar
obscuration by circumstellar dust at optical to near-IR wavelengths,
and its dust emission in the mid- and far-IR.  The expansion velocity
of the wind and the dust-to-gas ratio are made to vary consistently as
a function of stellar parameters ($\dot M$, $T_{\rm eff}$, $L$, C/O,
$Z$), in place of simpler parametrizations often used in the
literature.  Dust reprocessing of radiation makes the most evolved AGB
stars to disappear in optical pass-bands, and appear at a very
extended red tail in near- and mid-IR CMDs. All these features make
the new isochrones potentially very useful in the study of systems
containing AGB stars, both resolved and unresolved.

The new models have been included in the TRILEGAL code (Girardi et
al. 2005; Girardi \& Marigo 2007b)\footnote{{\tt
http://trilegal.ster.kuleuven.be/}\\ or {\tt
http://stev.oapd.inaf.it/trilegal}} for the simulation of resolved
stellar populations in the Milky Way and in external galaxies, and in
the GRASIL code (GRAphite and SILicate; Silva et al. 1998, Panuzzo et
al. 2005)\footnote{
{\tt
http://adlibitum.oat.ts.astro.it/silva/default.html} or {\tt
http://web.oapd.inaf.it/granato/grasil/grasil.html}} for computing the
integrated spectrum of galaxies from the X-ray to the radio
domain taking into account the effects of dust. Forthcoming papers will present applications of both codes,
including the extensive testing of our TP-AGB tracks and dust models
against observational data.

\begin{acknowledgements}
This study was supported by the University of Padova
(Progetto di Ricerca di Ateneo CPDA052212), COFIN INAF 2005, and 
contract ASI-INAF I/016/07/0. We
acknowledge Rita Gautschy-Loidl for providing us the C-star spectra,
and Bernhard Aringer for the help in implementing them. We thank Anil
Seth, Ben Williams, Enrico Held, Marco Gullieuszik, Leandro Kerber,  
and Mauro Barbieri
for pointing out some bugs and problems in preliminary versions of
these isochrones, and Jacco van Loon for his very useful suggestions.
\end{acknowledgements}


\appendix
\section{Revised evolution of massive TP-AGB models}
\label{sec_revagb}

\begin{figure}
\resizebox{\hsize}{!}{\includegraphics{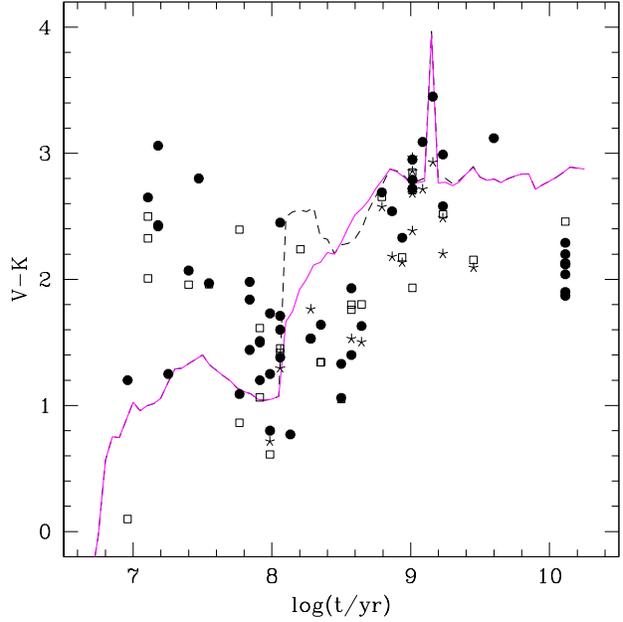}}
\caption{Integrated $V-K$ colours as a function of SSP age, derived
from the original Paper~I TP-AGB tracks (dashed line), and after the
revision applied to the TP-AGB models with initial masses
$\ge 3.0\, M_{\odot}$ (solid line). The adopted metallicity is
$Z=0.008$.  Observed colours of LMC clusters are taken from the
compilations by Persson et al. (1983; filled circles), Kyeong et
al. (2003; empty squares), and Pessev et al. (2006, starred symbols).
}
\label{fig_sspold}
\end{figure}

The TP-AGB evolution of more massive stars, with initial masses $3.0
\le M/M_{\odot} \le 5.0$, has been modified  as follows.
The starting point that has stimulated the revision is the behaviour
of the integrated colours (e.g. $V-K$, $J-K$, $H-K$) as a function of
age as displayed by the Magellanic Clouds' clusters of age
$t\ga10^8$~yr. As shown in Fig.~\ref{fig_sspold}, the comparison with
SSP colours, derived from the isochrones including the original TP-AGB
tracks from Paper~I, would indicate a sizable excess in the predicted
$V-K$ colours at ages around $1-4\times 10^8$~yr. The sudden jump to
$V-K\sim 2.5$ at $t \sim 10^8$~yr is due to the occurrence of the AGB
phase in models with $\mini\sim 5$~\Msun, which corresponds to the
maximum initial mass (for our set of models) required to develop an
electron-degenerate C-O core. In the $1-4\times 10^8$ yr age
range, while the integrated $V$-band luminosity is essentially
produced by less evolved stars still populating the region close to
the main-sequence turn-off (see e.g. Charlot \& Bruzual 1991), a
significant contribution in the $K$-band comes from the most massive
AGB stars, with initial masses in the range $3.0 M_{\odot} \le \mini \le
5.0 M_{\odot}$.

It follows that the $V-K$ colour jump at $t\sim 10^8$~yr should be
ascribed to an overestimation of the TP-AGB phase of more massive AGB
models, that translates into either a) too high luminosities and/or b)
too long lifetimes.  Of course these aspects are inter-related as they
are both highly sensitive to model details dealing with two crucial
processes that affect the evolution of these stars, namely HBB and
mass loss by stellar winds.

Given the high degree of uncertainty that affects the theoretical
predictions for $\dot{M}$ along the AGB phase, 
the theoretical discrepancy is likely attributed to an 
underestimation of the mass-loss efficiency in massive AGB stars.
A simple remedy 
could be obtained artificially just introducing a
multiplicative factor in front of our mass-loss law (similarly to
the $\eta$ parameter of the Reimers law) for $M_{\rm i} \ga 4.0\,
M_{\odot}$, but in this case we would 
introduce a discontinuity in our adopted formalism for $\dot M$, as
the correction factor should then vanish at lower $M_{\rm i}$. 
 
After a general re-consideration of the many prescriptions of our
synthetic TP-AGB model, we have found that a more physically-sound
way to obtain higher mass-loss rates may come from an improved 
treatment of the pulsation
properties of the most massive AGB stars, compared to the original 
description given in Paper~I.

\begin{figure}
\resizebox{\hsize}{!}{\includegraphics{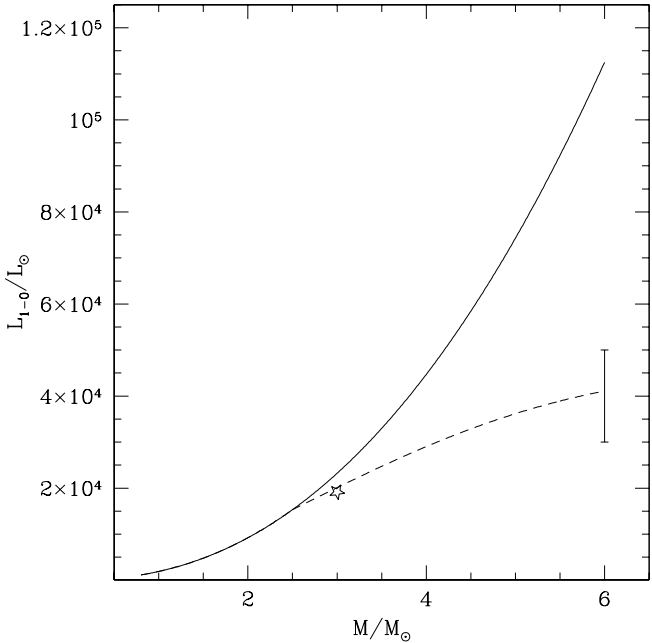}}
\caption{Transition luminosity defining the switching from fundamental
mode to first-overtone pulsation mode, as a function of the current
stellar mass, assuming $Z=0.008$ and $\log T_{\rm eff} =3.52$.  The
original relation of Paper~I is shown as a solid line
(Eq.~\ref{eq_p1p0old}), while the modified relation at higher stellar
masses (Eq.~\ref{eq_p1p0}) is drawn with a dashed line.  The starred
symbol corresponds to a pulsation model  by Fox \& Wood (1982)
with $M_{\rm i} = 3.0\, M_{\odot},\, Z=0.019,\, \log(T_{\rm
  eff})=3.49$), while the vertical bar roughly defines the 
range of critical luminosity for a few models 
calculated Fox \& Wood (1982) with $M_{\rm i} = 6\, M_{\odot}$.}
\label{fig_lp1p0rev}
\end{figure}

Specifically, following the results of non-adiabatic pulsation models
of Fox \& Wood (1982) for stellar masses $3-6\, M_{\odot}$, it seems
likely that in Paper~I the switching of the pulsation mode from the
first overtone to the fundamental one has been assumed to occur at too
high luminosities in the most massive TP-AGB models.

We recall that
the critical luminosity, $L_{1-0}$, is defined as the luminosity at
which the growth rates of the first overtone mode and the fundamental
mode are equal. In Paper~I the critical luminosity has been derived from the
pulsation models calculated by Ostlie \& Cox (1986) in the mass range
$0.8 - 2.0\, M_{\odot}$ and the fitting relation reads
\begin{eqnarray}
\label{eq_p1p0old}
\log \left(L_{1-0}/L_{\odot}\right)_{\rm MG07} & = &
-14.516+2.277\log (M/M_{\odot}) \\
\nonumber
& & +5.046\log T_{\rm eff}-0.084\log(Z/0.02)\,\, .
\end{eqnarray}

In Figure~\ref{fig_lp1p0rev} we compare $L_{1-0}$ as predicted by the
original Paper~I prescription (solid line), extended over the entire
mass range of interest, with the results from Fox \& Wood
(1982). We see that while at $M = 3\, M_{\odot}$ 
Eq.~(\ref{eq_p1p0old}) is fully consistent with the results 
from Fox \& Wood (1982), at  $M = 6\, M_{\odot}$,
$\left(L_{1-0}/L_{\odot}\right)_{\rm MG07}$ exceeds significantly the range
of critical luminosities derived from the Fox \& Wood (1982) data.
Based on these results we apply a rough
correction $\Delta \log L_{\rm cor}$ to
Eq.~(\ref{eq_p1p0old}) in order to decrease the transition luminosity
for $M \ga 3.0\, M_{\odot}$, so that we write
\begin{equation}
\label{eq_p1p0}
\log \left(L_{1-0}/L_{\odot}\right) = \log \left(L_{1-0}/L_{\odot}\right)_{\rm
  MG07} + \Delta\log L_{\rm cor}
\end{equation}
with
\begin{equation}
 \Delta\log L_{\rm cor} = -0.125\, (M/M_{\odot} -2.5) \,\,\,\,.
\end{equation}
The revised relation for $L_{1-0}/L_{\odot}$ is also plotted in
Fig.~\ref{fig_lp1p0rev} (dashed line). We see that it coincides with
the old relation for $M \le 2.5\, M_{\odot}$, while it flattens to
lower luminosities for larger masses.
It should be observed that, given the paucity of the 
 theoretical  pulsation  models presently available, 
this revision is meant to be 
just a crude attempt to improve our treatment of $L_{1-0}$.
Indeed, new pulsation models over 
wider ranges of the key parameters (i.e. mass, metallicity,
C/O ratio, etc.) are urgently required to allow for  
a real progress in modelling the pulsation properties of AGB stars.

The new formula for the critical luminosity $ L_{1-0}$ is used to
re-calculate the TP-AGB evolution of models with initial masses
$3.0\le M/M_{\odot} \le 5.0$ for all choices of the metallicity.
Massive models start to pulsate in the fundamental mode at lower
luminosities than assumed in Paper~I, which favours the attainment of
larger mass-loss rates with consequent earlier termination of the AGB
phase. This effect is more pronounced at larger masses ($M \ge 4\,
M_{\odot}$) shortening the TP-AGB lifetimes by a factor of $2-5$.  The
impact on the integrated $V-K$ colours is good as we see in
Fig.~\ref{fig_sspold}. The jump at $t\sim 10^8$ yr is reduced and a
more gradual increase of the colour proceeds at larger ages, allowing
for the seeked consistency with the observations.  Similar results are
obtained for other integrated colours, like $J-K$, and $H-K$.

It should also be emphasized that the TP-AGB fuel at ages
$t\ga10^8$~yr cannot be arbitrarily cut down since otherwise we would
run into other problems, related to the observed properties of the
most luminous AGB stars and their progeny.  For instance, we have
verified that even invoking an extreme shortening of the TP-AGB phase
for these stars the predicted integrated $V-K$ and $J-K$ colours are
still consistent with the observed ones, but in this case we cannot
fulfill the nucleosynthesis constraint set by type I planetary
nebulae. These latter are characterized by high He and N abundances,
which are currently interpreted as the result of hot-bottom burning in
more massive AGB stars (e.g. Marigo et al. 2003b; Pottasch \&
Bernard-Salas 2006). A too short activation of the CNO cycle at the
base of the convective envelope would not allow to reach the He and N
surface enrichment measured in type I planetary nebulae.  In other
words, the nuclear fuel burnt during the TP-AGB phase cannot be
constrained only on the base of the integrated luminosities, but it
should also be checked against indications provided by abundance
analyses in the stellar ejecta. In fact, as extensively discussed by
Marigo \& Girardi (2001) a precise fraction of the nuclear fuel
consumption (emitted light) in post-main sequence stars ends in the
form of chemical (mainly He,N,O) yields. One cannot produce large
amounts of He and N in the most massive AGB stars (as observed in 
their PN descendants),
without adding a marked contribution of these same stars to the
integrated light.

\end{document}